\documentclass[preprint2]{aastex}

\usepackage{graphicx}
\usepackage{natbib}

\def\ga{\hspace{1ex} ^{>} \hspace{-2.5mm}_{\sim} \hspace{1ex}}
\def\la{\hspace{1ex} ^{<} \hspace{-2.5mm}_{\sim} \hspace{1ex}}
\newcommand{\BibTeX}{ \textrm{B\kern-.05em\textsc{i\kern-.025em b}\kern-.08em
    T\kern-.1667em\lower.7ex\hbox{E}\kern-.125emX} }
\usepackage{color}
\newcommand{\tdif}[2]{\frac{{\rm d}#1}{{\rm d}#2}}
\newcommand{\pdif}[2]{\frac{\partial {#1}}{\partial {#2}}}

\slugcomment{}

\shorttitle{Thermal Evolution of Super-Earths}
\shortauthors{Tachinami, Senshu and Ida}

\begin{document}

\title{Thermal evolution and lifetime of intrinsic magnetic fields of Super Earths in habitable zones}

\author{C. Tachinami\altaffilmark{1} }
\affil{Department of Earth and Planetary Sciences, Tokyo Institute of Technology,
    Meguro, Tokyo 1528551 (Japan)}
\email{ctchnm@geo.titech.ac.jp}

\author{H. Senshu\altaffilmark{2}}
\affil{Planetary Exploration Research Center, Chiba Institute of Technology, 2-17-1 Tsudanuma, Chiba 2750016 (Japan)}

\and

\author{S. Ida\altaffilmark{1}}
\affil{Department of Earth and Planetary Sciences, Tokyo Institute of Technology,
    Meguro, Tokyo 1528551 (Japan)}

\begin{abstract}
We have numerically studied the thermal evolution of
various-mass terrestrial planets in habitable zones,
focusing on duration of dynamo activity to generate
their intrinsic magnetic fields, which may be one of
key factors in habitability on the planets.
In particular, we are concerned with super-Earths,
observations of which are rapidly developing.
We calculated evolution of temperature distributions
in planetary interior, using Vinet equations of state, Arrhenius-type formula
for mantle viscosity, and the astrophysical mixing length theory 
for convective heat transfer modified for mantle convection.
After calibrating the model with terrestrial planets in
the Solar system, we apply it for 0.1--$10M_{\oplus}$
rocky planets with surface temperature of $300~\mbox{K}$ (in habitable zones) 
and the Earth-like compositions.
With the criterion for heat flux at the CMB
(core-mantle boundary), the lifetime of the magnetic fields
is evaluated from the calculated thermal evolution.
We found that the lifetime slowly increases with 
the planetary mass ($M_p$) independent of
initial temperature gap at the core-mantle boundary ($\Delta T_{\rm CMB}$)
but beyond a critical value $M_{c,p}$ ($\sim O(1)M_{\oplus}$)
it abruptly declines by the mantle viscosity enhancement due to
the pressure effect.
We derived $M_{c,p}$ as a function of $\Delta T_{\rm CMB}$ 
and a rheological parameter (activation volume, $V^*$).
Thus, the magnetic field lifetime of super-Earths with $M_p > M_{p,c}$ 
sensitively depends on $\Delta T_{\rm CMB}$, 
which reflects planetary accretion, and $V^*$, 
which has uncertainty at very high pressure.
More advanced high-pressure experiments and first-principle simulation
as well as planetary accretion simulation
are needed to discuss habitability of super-Earths.

\end{abstract}


\section{Introduction}

Many of exoplanets so far detected may be gas giants with
masses $\ga 100 M_{\oplus}$, because massive planets are more easily
to be detected.
However, recently 
several super-Earths with masses of a few to ten $M_{\oplus}$ 
have been discovered by improved radial velocity measurements
\citep[e.g.,][]{Udry2007} or microlensing observations
\citep[e.g.,][]{Beaulieu2006}.
On-going radial velocity \citep{Mayor2009} and
microlensing \citep{Gould2010} surveys 
and theoretical studies \citep[e.g.,][]{Ida2004,Ida2008,Ida2010} 
strongly suggest ubiquity of super-Earths in extra solar planetary systems.

Space transit surveys such as CoRoT and Kepler will also detect
many super-Earths.
In fact, CoRoT have detected the minimum mass 
transiting planet (CoRoT-7b) \citep[]{Leger2009+CoRoT7b,Queloz2009}.  
With an assumed composition,
a mass-radius relationship of the super-Earths 
gives planetary masses from transit observational data.
On the other hand, if the planetary masses are 
obtained by follow-up radial velocity
observations, the mass-radius relationship 
can be used to estimate the planetary composition, although there 
is some ambiguity depending on the amount of H$_2$O \citep{Sotin2007}.
\citet{Valencia2006} used Birch-Murnaghan equation of state (EOS) 
for rocks and metals 
to obtain a mass-radius relationship of various-mass terrestrial 
planets under some conditions (core ratio, surface temperature, and etc.). 
\citet{Sotin2007} also considered ocean planets 
which contain 50\% of H$_2$O. 
They used the EOS including thermal pressure to describe $P-V-T$ relations for ices 
under extremely high pressure and obtained a mass-radius relationship 
for both terrestrial and ocean planets.

Because super-Earths should exist also in habitable zones,
the aspects related to the habitability of super-Earths
are being discussed.
Planetary habitability is often discussed in terms of the stability of
liquid water on the planetary surface \citep{Kasting1993}.
Assuming planets that are massive enough to maintain dense atmosphere,
a range of orbital radius in which liquid water is stable
is called a "habitable zone."

In addition to the existence of liquid water,
evolution of amount and composition of planetary atmosphere
may also be an important factor for habitability.
It is believed that most fraction of the
present atmosphere of the Earth
was formed by impact degassing \citep[e.g.,][]{Abe1985}
and it consisted of CO$_2$ and H$_2$O with more than 100 bars.
The plate tectonics on the Earth
has removed huge amount of CO$_2$ from the Earth's atmosphere
on Gyr timescales \citep[]{Tajika1992}.  

\citet{Valencia2007plate} and \citet{ONeill2007} investigated 
the possibility of plate tectonics on the surface of super-Earths. 
The plate tectonics would significantly affect amount and composition of
planetary atmosphere through carbonate-silicate cycle with degassing and weathering.
It also has stabilizing effect of planetary surface temperature, since temperature dependence of weathering rate of carbonate 
works as a negative feedback mechanism for surface temperature change \citep{Tajika1992}. 
\citet{Valencia2007plate} argued that super-Earths could invoke 
plate tectonics, because terrestrial planets larger than Earth may have a thinned surface thermal boundary 
layer and increased yield stress.  
On the other hand, \citet{ONeill2007} showed that 
super-Earths would have stagnant-rid style mantle convection without 
plate tectonics. 
It is noted that these studies \textit{a priori} assumed 
thermal structure of super-Earths,
as the studies on a mass-radius relation did.

Although the thermal effect on the mass-radius relation
may be negligible, plate tectonics should depend on
planetary thermal evolution history.
Dynamo activity to generate a magnetic field
also depends on the thermal evolution.
Planetary magnetic field prevents stellar winds 
from splitting the planetary atmospheres.
The magnetic field also prevents cosmic rays from penetrating 
to the planetary surface.
Thus, it may be one of the most important 
factors for land-based life to be maintained.
It is widely accepted that the Earth's magnetic field is attributed to a
dynamo effect \citep[e.g.,][]{Glatzmaier1995,Kuang1997,Kageyama1997}
in the metallic core. 
The intrinsic magnetic field may be sustained if convective fluid
motions in the core is vigorous enough, in other words, the heat flux
through the core surface is large enough. 
Thus, detailed study on thermal evolution of the core is needed 
to evaluate generation of the magnetic field.

Using the box model (see below), 
\citet{Papuc2008} modeled the thermal evolution of 
the various-mass terrestrial planets on geological 
timescales to discuss the evolution of planetary surface activity, 
i.e., the plate velocity and the degassing rate. 
Their results showed that the super-Earths may have dense atmosphere 
in their early history, since larger planets have higher degassing rates.
Since they were concerned with planetary surface activity, 
they focused on treatment of heat generation of radio
activity in the mantle and surface heat flow,
leaving treatment of cores simple.
We will show that careful treatment of the core
such as the effects of 
inner core nucleation and increase in heat capacity
due to high compression and gravitational energy stored in the core,
which \citet{Papuc2008} neglected, are  
important for the study of dynamo activity
(Note that the evaluation of the surface activity is hardly affected
by the careful treatment of the core).

In a series of papers, Schubert, Stevenson and their colleagues
developed a "box" model for thermal evolution 
of the terrestrial planets in our Solar system
\citep[e.g.,][]{Schubert1979,Stevenson1983}.
In their model, the thermal structure is described by
two boxes that correspond to the mantle and the core.
The temperature variation in each box is neglected and
the temperature distributions in the Earth are
represented by three distinct temperatures of
the core, the mantle, and the planetary surface.
The heat flow is evaluated by
thermal conduction through the thermal boundary layers 
at CMB and the planetary surface
with the thermal Boundary Layer Theory (BLT)
\citep{Stevenson1983}.
In the BLT, the heat flux is 
determined by the thickness of the thermal
boundary layer, which is given by local 
Rayleigh number.
Because this model is easily treated, it provides
a powerful tool to explore general trends of thermal evolution
of terrestrial planets.

For Earth's thermal evolution, we can use observational
constraints such as surface heat flow and inner core size. 
With the calibrated model, unknown parameters such as
initial temperature distribution of Earth's interior 
\citep[][also see section 2.7]{Yukutake2000} or 
potassium abundance in the core
\citep{Nimmo2004} can be constrained.
Rheological parameters and impurity abundance in the core
are also estimated (see section 3.3).
The existence of the magnetic field for Mercury
and early decay of the magnetic field for Venus and Mars 
are consistent with calculations
with reasonable choice of initial temperature
or impurity abundance in the cores 
by BLT \citep{Stevenson1983} and MLT (Appendix B).

\citet{Gaidos2010} simulated thermal evolution of various sized super-Earths 
by the BLT
to evaluate the magnetic activity of super-Earths.
They concluded that massive rocky planets ($ > 2.5 M_{\oplus}$) 
can not sustain magnetic field,
because they found an inversion of gradients of the melting 
and adiabatic curves in the core under such high pressure.
Since the outer core is solid in the inverse state, the cooling of the 
inner liquid core, in which dynamo operates, is inhibited.
Although the possibility of the inversion raised by the paper 
is a very important, the conclusion depends on high pressure material properties 
such as meting and adiabatic curves that need to be confirmed.
In the present paper, we point out another important factor to
inhibit dynamo activity in super-Earths, drastic increase in the mantle viscosity
due to pressure effect.
Even if the inversion in the core does not occur, the enhanced
mantle viscosity quickly terminates dynamo activity in the core. 

In the present paper, we are concerned with lifetime
of magnetic fields of super-Earths.
There is no observational constraint 
for super-Earths to calibrate parameters for a model 
and initial/boundary conditions.
Here, clarifying key quantities for generation of 
the magnetic field, we derive planetary mass dependence 
of lifetime of magnetic activity and how it depends on
the model parameters and initial/boundary conditions.
To reduce unknown parameters, we consider super-Earths
in nearly circular orbits in habitable zones.
The analysis on the key rheological parameters
will provide new motivations for high pressure experiments
and first principle simulations, 
since state-of-arts high pressure experiments have already 
reached the pressure at the bottom of Earth's mantle.
We will point out that initial temperature distribution
sensitively affects the lifetime, which gives 
new motivations for theories of planet accretion from planetesimals
and core formation.

Here, we develop a thermal evolution
model to discuss the existence of the intrinsic magnetic field in
terrestrial planets with various masses. 
Since we are concerned with heat flux across the CMB,
we calculate detailed radial temperature distribution
in both core and mantle.
We use the Mixing Length Theory (MLT) to calculate
the heat flow.
The MLT is commonly used to study stellar interior.
We use the modified version for 
low Reynolds number flow in solid planets that have
radial discontinuities in their interior
\citep[][ and references therein]{Sasaki1986,Abe1995,Senshu2002,Kimura2009}.
The modified MLT is useful for calculations of
super-Earths that may have additional higher-pressure phase transitions
such as a post-post perovskite transition (although we
do not consider it in the present paper) and
early-stage planets that may have convection-barriers at
upper/lower mantle boundary \citep[e.g.,][]{Honda1993}
or density crossover at melt/solid boundary 
\cite{Labrosse2007}.
We will also show the results for the
two-layer convection case in which convective flow does not penetrate
the spinel-perovskite transition at upper/lower mantle boundary, 
while most of calculations are
in the cases of one-layer convection.
We compare the MLT with the conventional BLT 
in details and show that they are in good agreement 
with each other in the case of one-layer convection (Appendix A).

We will show that the lifetime is
rather shorter for super-Earths than for Earth-mass planets
for nominal parameters of solid state.
The mechanism to suppress dynamo activity
in super-Earths found in this paper is independent of that in 
\citet{Gaidos2010}, so that it may be likely that 
super-Earths are magnetically inactive.
We also point out that the choice of initial conditions 
and rheological parameters highly affect
the thermal evolution of the planets.
In section 2, we explain our numerical model. The
numerical results will be shown in section 3. Finally we 
discuss the habitability for terrestrial planets in view of
the intrinsic magnetic field.

\section{Numerical Model}

We follow thermal evolution of a planet 
that is a cooling process from 
a hot early state due to accretion from planetesimals
and core-mantle differentiation, for various-mass terrestrial
planets by using a one-dimensional spherically symmetric model. 

As described below, there are unknown parameters for
rheological properties and initial conditions of planets.
Furthermore, even in our Solar system, the terrestrial planets
have different compositions (a water-rock-iron ratio) and
surface temperature that is regulated by orbital radius.
In extrasolar planetary systems, more variations in
the water-rock-iron ratio and surface temperature 
should exist due to different metal abundance of 
host molecular clouds as well as different
pressure/temperature state of disks and planetary atmosphere and
planetary formation processes.
These unknown variations could make the study on the 
mass dependence meaningless. 

In order to reduce the uncertainty,
in a ``nominal'' case (see section 2.8),
we consider planets with surface temperature 
($T_{\rm surf}$)
of 300K and the mantle/core mass ratio ($\zeta_{\rm m/c}$) of $7:3$,
which is the same as that for the Earth.
The planets may correspond to extrasolar terrestrial planets 
in nearly circular orbits in habitable zones.
In the nominal case, 
the other unknown parameters in mantle rheology, core impurity 
and initial temperature
are calibrated with data of the Earth.

The restriction to the nominal case enables us to
derive clear planetary mass dependences.
We also discuss how these dependences change by
different choice of the parameters. 
In Appendix B, we also carry out the runs with different
$T_{\rm surf}$ and $\zeta_{\rm m/c}$
to calculate thermal evolution of Mercury, Venus and Mars.
The results show that our model can be applied for
these planets too, if we use reasonable non-nominal parameters.
Systematic survey for thermal evolution of super-Earths 
in the non-nominal cases is left for future works.

The thermal evolution is calculated by the following methods:
\begin{enumerate}
\item The radial density distribution is calculated by using 
VINET equation of state taking into account pressure dependence (section 2.1).
Since this distribution is almost independent of 
evolution of temperature distribution and the inner core growth as 
explained below, we use 
the distribution calculated in the initial state ($t = 0$)
throughout the entire thermal evolution.
\item Given a temperature distribution in the interior at time $t$,
the radius of the inner solid core is
calculated by the melting temperature with pressure and
composition dependences (section 2.2).
Sulfur is considered as the impurity in the core and
its concentration in the outer liquid core is self-consistently
calculated with the condensation of the inner core (section 2.2).
\item The heat transfer throughout the mantle is calculated by 
the astrophysical mixing length theory modified for
solid planets (we discuss its validity and usefulness
in section 2.3). 
The mantle viscosity in the heat transfer equation
is estimated by using Arrhenius type formulation (section 2.5).
By subtracting the energy loss during the timestep ($\Delta t$), 
we obtain a new temperature distribution 
at $t + \Delta t$ and go back to step 2.
Time evolution is calculated by iterations of step 2 and 3.

\end{enumerate}
In the following subsections, we will explain 
each step in more detail.

\subsection{Density profile}

The hydrostatic stratification is calculated by 
\begin{equation}
\label{eq:HSE}
\tdif{P}{r}=\rho(r) g(r); \; \tdif{r}{m}=\frac{1}{4\pi r^2 \rho(r)},
\end{equation}
where $P,r,\rho,g,$ and $m$ are pressure, radius, density, gravitational
acceleration, and mass inside the radius $r$, respectively. 
Vinet EOS\citep[]{Vinet1987} is given by
\begin{equation}
\label{eq:VINETEOS}
P=3K_0\frac{1-x}{x^2}\exp(\phi(1-x)),
\end{equation}
where
\begin{equation}
\phi=\frac{3}{2}(K_0^{\prime}-1),
\end{equation}
$x=\rho_0/\rho$, and $\rho_0, K_0$ and $K_0^{\prime}$ are density, bulk
modulus, and its pressure differentiation at zero pressure,
respectively.

Birch-Murnaghan EOS is "finite strain" EOS, in which pressure
is expressed by Taylor series expansions of finite strain,
and it is often used for the calculation 
of interior of solid planets.
However, the finite strain EOS do not accurately represents the
volume variation under very high compression
(if pressure exceeds the bulk modulus of zero pressure, 
the expansion never converges).
Vinet EOS is derived from a general inter-atomic potential energy function. 
For simple solids, Vinet EOS provides more accurate representations of the 
volume variations with pressure, under very high pressure.
Since we are concerned with interior of super-Earths under very high pressure,
we adopt Vinet EOS rather than Birch-Murnaghan EOS, following
\citep{Valencia2007radius}.

Since thermal contraction is small enough
(it is less than 1\% in physical size for 100K change in the average temperature of Earth's interior), 
we neglect the temperature dependence in Vinet EOS.
The density profile in the planet is calculated by 
numerically solving equations
(\ref{eq:HSE}) and (\ref{eq:VINETEOS}).

\begin{figure}[htbp]
\begin{flushleft}
\includegraphics[scale=0.5, angle=0]{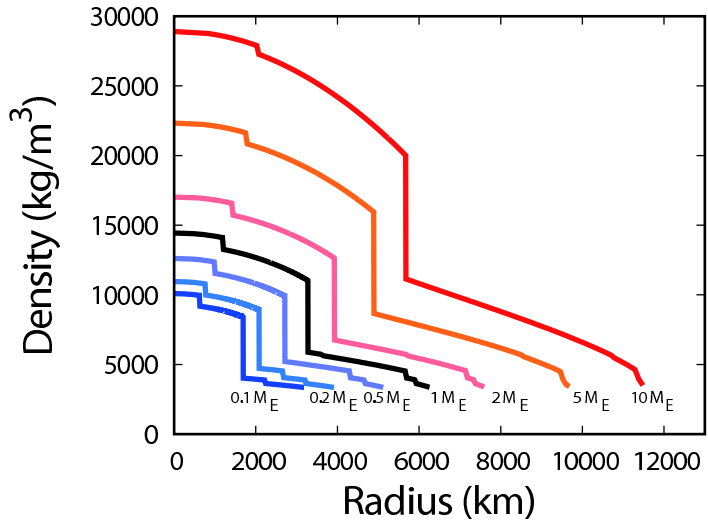}
\end{flushleft}
\caption{Radial density profiles for 0.1, 0.2, 0.5, 1, 2, 5, 10 $M_{\oplus}$ planets 
(with inner core of 6 wt\% of each core) obtained by our model in the nominal case.}
\label{fig:density}
\end{figure}

The compositions are assumed as
olivine and $\gamma$-spinel for upper mantle, 
perovskite and post-perovskite for lower mantle, 
Fe and FeS for outer core, and Fe for inner core
with the properties given in Table \ref{ta:PP}.
As temperature decreases, the inner solid core grows and sulfur
moves from the inner core to the outer core
(see section \ref{sec:thermal evolution of the core}).
In general, the inner core growth changes the volume of
the whole core, because the parameters,
$\rho_0, K_0$ and $K_0^{\prime}$, are different between
Fe and FeS.
However, since we numerically found that the volume change
 of the whole core is very small, we neglect it.
Figure \ref{fig:density}
shows the result of numerical calculation
of the density profile for 0.1 to 10 Earth-mass planets. 
This result is different in the radii of the core and the mantle from
the results by \citet{Valencia2006} and \citet{Sotin2007}
by a few \%, which may be due to different choice of
parameter values in the EOS.
But, this difference does not affect the thermal evolution.

\subsection{Thermal evolution of the core}
\label{sec:thermal evolution of the core}

The temperature distribution of the core is determined as follows:
\begin{enumerate}
\item The inner solid core: We assume that each part of
the inner core memorizes the temperature at which it solidified,
because of inefficient heat transfer due to conduction in the
solid core. 
\item The outer liquid core: 
We assume that the liquid core has an adiabatic temperature
distribution by vigorous convection. 
\item Time evolution:
The radius of the inner core and the temperature at the CMB is
determined by total energy of the inner and outer cores
as described below, and the total energy is given as a function of time
with integrating heat flux at the CMB.
\end{enumerate}

The adiabatic temperature gradient in the outer core is
given by \citep{Soul1997,Yukutake2000,Valencia2006}
\begin{equation}
\label{eq:ATG}
\pdif{T}{r}=\frac{\rho g\gamma_G}{K_s}T,
\end{equation}
where $\gamma_G$ and $K_s$ 
are Gr\"{u}neisen parameter and bulk modulus of
the liquid core. 
Depth variation of $\gamma_G$ is calculated as 
$\gamma_G=\gamma_{G0}(\rho_0/\rho)^q$ 
(the parameter values used are summarized in Table \ref{ta:PP}).
The density at 0 pressure $\rho_{\rm 0OC}$, 
bulk modulus $K_{\rm 0OC}$, and its pressure derivation 
$K'_{\rm 0OC}$ of the outer core 
are given by impurity concentration $x_{{\rm S}}$ as:
\begin{equation}
x_{{\rm FeS}}= x_{{\rm S}}\frac{Z_{{\rm Fe}}+Z_{{\rm S}}}{Z_{{\rm S}}}
\end{equation}
\begin{equation}
{\rho_{0}}_{{\rm OC}}=\left(\frac{1-x_{{\rm FeS}}}{\rho_{{\rm Fe}}}+\frac{x_{{\rm FeS}}}{\rho_{{\rm FeS}}}\right)^{-1}
\end{equation}
\begin{equation}
{K_{0}}_{{\rm OC}}=\frac{1}{{\rho_0}_{{\rm OC}}}\frac{1}{\frac{1-x_{{\rm FeS}}}{\rho_{{\rm Fe}}}\frac{1}{K_{{\rm Fe}}}+\frac{x_{{\rm FeS}}}{\rho_{{\rm FeS}}}\frac{1}{K_{\rm FeS}}}
\end{equation}
\begin{equation}
 {K^{\prime}_{0}}_{{\rm OC}}=-1+{\rho_0}_{{\rm OC}}{K_0}_{{\rm OC}}\left(\frac{1-x_{{\rm FeS}}}{\rho_{{\rm Fe}}}\frac{1+K^{\prime}_{{\rm Fe}}}{K^2_{{\rm Fe}}}+\frac{x_{{\rm FeS}}}{\rho_{{\rm FeS}}}\frac{1+K^{\prime}_{{\rm FeS}}}{K^2_{{\rm FeS}}}\right),
\end{equation}
where $x_{\rm Fe}$, $x_{{\rm FeS}}$, $Z_{{\rm Fe}}$, and 
$Z_{{\rm S}}$ are mass fraction of Fe and FeS, molar weights of Fe 
and S, respectively.

The inner core nucleation decelerates cooling of the core
by release of gravitational energy due to
the change in the density distribution and
by release of latent heat \citep{Stevenson1983, Gubbins2004}.
The light elements are kicked into the outer core, 
resulting in depression of a melting point of the outer core
\citep{Stevenson1983,Yukutake2000}. 
The boundary between inner and outer cores is located at
the intersection between adiabatic and melting curves in the core. 
We use a Lindeman's equation for the melting curve of pure iron, 
\begin{equation}
\Gamma(\rho)=\Gamma_0\left(\frac{\rho_0}{\rho}\right)^{2/3}\exp\left\{\frac{2\gamma_0}{q}\left[1-\left(\frac{\rho_0}{\rho}\right)^q\right]\right\}.
\end{equation}
We also consider the depression of a melting point by concentration of light elements. We define the melting point of Fe-FeS alloy as
\begin{equation}
T_{\rm melt}=(1-2 x_{\rm S})\Gamma(\rho),
\end{equation}
and the factor $(1-2 x_{\rm S})$ expresses 
the depression of the melting point due to dissolution of light elements
\citep{Usselman1975,Stevenson1983}.
Assuming that the outer
core is well mixed by convection,
\begin{equation}
x_{\rm S}=x_{\rm 0S}\frac{M_{{\rm c}}}{M_{{\rm c}}-M_{{\rm ic}}},
\end{equation}
where $M_{{\rm ic}}$ and $M_{\rm c}$ are the inner core mass
and total mass of the inner and outer cores and
$x_{\rm 0S}$ is the initial impurity concentration.  
In the nominal case, we adopt $x_{\rm 0S}=0.1$.

Given the inner core radius, we can calculate the total energy of
the core ($E_{\rm core}$) which is sum of 
the gravitational energy ($E_g$), latent heat ($E_l$), and
thermal energy ($E_{th}$). 
As described above, the temperature at the CMB is given as
a function of the radius of the inner core. 
As a result, we can obtain $E_{\rm core}$
as a function of the temperature at the CMB.  
Conversely, the radius of the inner core and 
the temperature at the CMB are given as a function of $E_{\rm core}$.

The energies are given by
\begin{equation}
\begin{array}{ll}
E_g & 
{\displaystyle
=-\int^{r_{{\rm ic}}}_{0}4\pi r^3\rho_{{\rm ic}}(r)g_{{\rm ic}}(r)dr-\int^{r_{{\rm c}}}_{r_{{\rm ic}}}4\pi r^3\rho_{{\rm oc}}(r)g_{{\rm oc}}(r)dr, } \\
E_l & =LM_{ic}, \\
E_{th} & 
{\displaystyle
=\int^{r_c}_04\pi r^2\rho(r)C_p(r)T(r)dr, }
\end{array}
\end{equation}
where $L$ is the latent heat released by solidification of 
unit mass of iron, which 
is assumed to be constant of $1.2\times 10^{6}$ J/kg \citep{Anderson1997},
and not to depend on the impurity concentration in the outer core, and
$C_p$ is specific heat with constant pressure. 
Both the gravitational energy and the latent heat 
are released after the inner core starts to solidify.
Gravitational energy is also released 
by thermal contraction, which will be discussed in section 2.6. 
The total energy $E_{\rm core}$ decreases with the rate that is equal to
the heat flux at the bottom of the mantle (see section 2.3).
Detailed calculations of the energies are given in Appendix C.

\subsection{Heat transfer throughout mantle}

The mantle is cooled by irradiation from
the planetary surface and heated by heat flow
from the core and internal radioactivity (see below). 
The heat transfer equation is: 
\begin{equation}\label{TT}
\rho C_p\pdif{T}{t}=
\frac{1}{r^2}\pdif{{}}{r}\left\{r^2k_c\left(\pdif{T}{r}\right)+r^2k_v\left[\left(\pdif{T}{r}\right)-\left(\pdif{T}{r}\right)_s \right]\right\}
+\rho Q,
\end{equation}
where $k_c$ is thermal diffusion coefficient,
$Q$ is radioactive heat production rate, 
$(\partial T/\partial r)_s$ is the adiabatic temperature gradient, and
the first and second terms in the right hand side represent 
conductive and convective fluxes.

To evaluate the convective flux in the mantle, 
we use the astrophysical
mixing length theory (MLT) modified for solid planets 
\citep[][and references therein]{Sasaki1986,Abe1995,Senshu2002,Kimura2009},
rather than the conventional parameterized convection model 
\citep[PCM; e.g.,][]{Sharpe1979}
or the commonly used 
boundary layer theory \citep[BLT; e.g.,][]{Stevenson1983}.

The PCM is very simple (Appendix A).
However, it uses the values of
$k_c$ and Rayleigh number $Ra$ that represents the whole mantle,
which are difficult to evaluate for real mantle because of
huge spatial variation of the mantle viscosity.
As a result, although the PCM can be applied to study overall 
trend of thermal evolution, it may not be accurate enough for
evaluation of heat flux across the CMB ($F_{\rm CMB}$),
which we are concerned with in the present paper.
In the BLT, since the heat flux is expressed by quantities 
only in the thermal boundary layer (Appendix A),
the BLT has better resolution for evaluation of $F_{\rm CMB}$.
Since the modified MLT also uses local values of physical quantities,
it quantitatively agrees with the BLT for wide range of parameters,
while the PCM does not agree with the BLT and the MLT 
for the cases in which viscosity variation is large in the mantle,  
as shown in Appendix A.
As explained below, since the MLT is more easily to be applied
for super-Earths, we use the MLT.

In the early Earth, the upper/lower mantle boundary
could have worked as a barrier for convection \citep{Honda1993}.
Tentative stagnancy at the upper/lower mantle boundary 
is also suggested for some subduction slabs 
in the present Earth \citep{Wortel2000}. 
The density overturn at melt/solid boundary in deep magma ocean
in the early Earth may have also worked as the barrier \citep{Labrosse2007}.
In super-Earths, post-post perovskite transition 
in deep mantle at high pressure could also work
as a barrier \citep{Umemoto2006}.

As explained below, the modified MLT
is easily applied for mantle convection with barriers,
without tuning of parameters for each barrier.
Although most of our calculations in the present paper
only consider the surface boundary and CMB
(in some runs we consider the upper/lower mantle boundary 
at spinel-perovskite transition as well),
we use the modified MLT for future extensions of calculations
with various convection barriers. 
In the following papers, we will consider the effects of
other barriers.

In the MLT, the coefficient for convective heat transfer is given by
\begin{eqnarray}
k_v=\left\{ \begin{array}{cr}
\displaystyle 0 & \displaystyle {\rm for} \; \left(\pdif{T}{r}\right)>\left(\pdif{T}{r}\right)_s\\ \\
\displaystyle \frac{\rho^2 C_p \alpha g
 \ell^4}{\eta}\left[\left(\pdif{T}{r}\right)-\left(\pdif{T}{r}\right)_s \right]\  & 
\displaystyle {\rm for} \; \left(\pdif{T}{r}\right)<\left(\pdif{T}{r}\right)_s\\
\end{array} \right.
\label{eq:k_v}
\end{eqnarray} 
where $\eta$ and $\ell$ are viscosity and the mixing length,
respectively. 
Here, the velocity of fluid blobs is evaluated by
Stokes velocity rather than free fall velocity
in the original MLT, in order to apply the model
to low Reynolds number flow in the mantle. 
In the astrophysical context such as stellar interior, 
the density scale height is usually adopted as $\ell$.
For calculation of thermal evolution of the Earth, 
it is proposed that a distance ($D$) from the closest barrier 
such as the CMB or the top of the mantle layer
is appropriate for $\ell$
\citep[][and references therein]{Sasaki1986,Abe1995,Senshu2002,Kimura2009}.
The detailed comparison with the PCM and BLT in Appendix A
shows that $\ell = 0.82D$ is the best choice.
We adopt $\ell = 0.82D$ for all the runs in the present paper.

With this choice, as approaching a barrier, $k_v$ rapidly 
decreases in proportion to $\ell^4$ and the conductive term
dominates in Eq.~(\ref{TT}).
As a result, thermal boundary layers, in which the conductive 
heat transfer dominates, are automatically represented.
Thereby, the modified MLT is easily applied for
calculation for thermal evolution of the proto-Earth or super-Earths.

The mantle viscosity is discussed in section 2.5.
The heat flux and temperature at the CMB determined from 
the calculation of the mantle heat transfer is used as a boundary
condition for thermal evolution of the mantle.
The total energy in the core is interpolated from the result of section 2.2.
It decreases with time, according to the calculated heat flux at the CMB.

\subsection{Internal heat source}

For thermal evolution of terrestrial planets on geological timescales,
long-lived radiogenic elements ($^{40}$K, $^{232}$Th,
$^{235}$U, and $^{238}$U) are important heat sources in the mantle. 
The estimated amounts of these elements in the Earth are compiled in table
\ref{ta:radiogenic heat} \citep{VanSchmus1995}. 
Here we assume the same abundances of the radiogenic elements
in the mantle in super-Earths as those in the Earth
and that the elements are distributed uniformly throughout the mantle.
The heat production rate at time $t$, $Q(t)$, is
given by $Q(t)=HU\exp(- \lambda (t - t_{\oplus}))$
where $U$, $H$, $t_{\oplus}$ and $\lambda$ are the abundance, heat
production rate, and decay constant of the element, respectively.

\subsection{Temperature and pressure dependency of mantle viscosity}

The mantle viscosity is one of the most important physical parameters 
to simulate thermal evolution,
since it determines the heat transfer efficiency 
in the mantle (see Eqs. [\ref{TT}] and [\ref{eq:k_v}]).
The viscosity sensitively depends on temperature and pressure, 
and both of temperature and pressure widely vary throughout the mantle. 
Here, we adopt Arrhenius type formulation for temperature-
and pressure-dependent viscosity model \citep{Ranalli2001}: 
\begin{equation}
\eta(T,P)=\frac{1}{2}\left[\frac{1}{B^{1/n}}\exp\left(\frac{E^*+PV^*}{nRT}\right)\right]\dot{\epsilon}^{(1-n)/n},
\label{eq:viscosity}
\end{equation}
where $R$, $\dot{\epsilon}$, $n$, $B$, $E^*$ and
$V^*$ are universal gas constant, strain rate, creep index, 
Barger coefficient, activation energy,
and activation volume of mantle, respectively.
We use different values of these parameters for upper and lower mantles. 
The mineral
properties we use are listed in Table~\ref{ta:viscosity}.
Note that the prescription for the mantle viscosity may include uncertainty.
The formula is based on the theoretical rate equation for creep law of rocks.
In this formula, the most important parameter to study 
thermal evolution of super-Earths is the activation volume ($V^*$),
since $V^*$ determines the dependence of the viscosity on pressure
and the pressure in the mantle can be increased by orders of magnitude 
as the planetary mass increases.
The activation volume is related to atomic volume, but the exact values 
under extremely high pressure is not well determined. 
In the nominal case, we use $V^*=10\times 10^{-6} \mathrm{m^3mol^{-1}}$,
but we also test a smaller value of $V^*=3\times 10^{-6} \mathrm{m^3mol^{-1}}$.
In section 3.4, we will discuss how the conclusion in the present paper
depends on a formula for the viscosity. 

\subsection{Release of gravitational energy by thermal contraction}

Although the thermal contraction 
is negligible for physical radius, the 
gravitational energy released by the thermal contraction 
cannot be neglected
(it is about 50kJ/kg for 100K change of the core in an 
Earth-mass planet). 
In our model, the released energy is regarded as 
increase in the specific heat of the core \citep{Yukutake2000}:
\begin{equation}\label{eq:thermo_cont}
\Delta C_p=\frac{\alpha P}{\rho}.
\end{equation}
The gravitational energy released by thermal
contraction is more effective in deeper regions. 
For an Earth-mass planet, 
$\Delta C_p/C_p$ is as large as 50\% at the CMB
in our calculation.

\subsection{Initial conditions}
\begin{figure}[htbp]
\begin{flushleft}
\includegraphics[scale=1.0,clip]{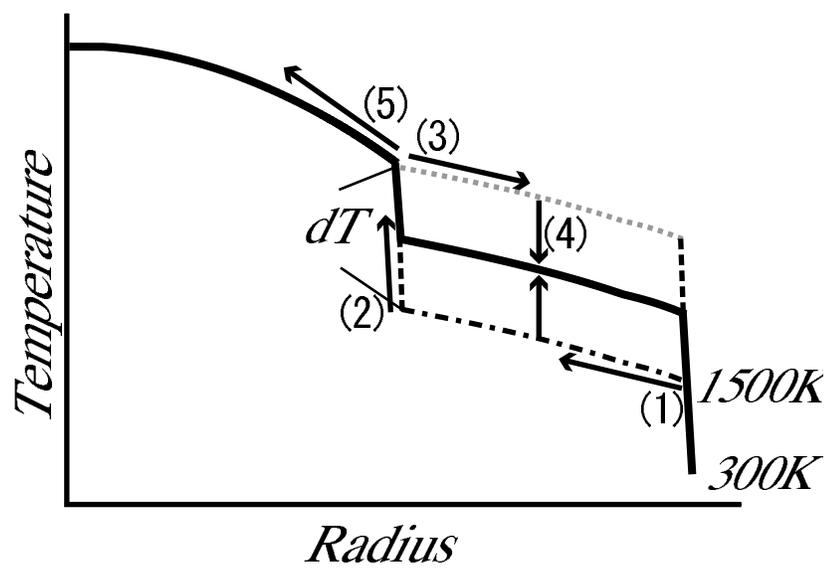}
\end{flushleft}
\caption{The schematic diagram of the procedure to obtain
initial temperature distributions. For more detail, see text.}
\label{initial_condition}
\end{figure}

Initial temperature distribution in the mantle is
determined by the procedure following \citet{Yukutake2000},
as illustrated in Fig. \ref{initial_condition}:
(1) An adiabat is drawn from the bottom of the
surface boundary layer with 1500 K down to the top of
the boundary layer at the CMB (the obtained tentative
temperature is denoted by $T_2$), assuming
efficient thermal convection,
(2) the initial
temperature at the bottom of the CMB is 
assumed to be $T_{\rm CMB} = T_2 + \Delta T_{\rm CMB}$, where
$\Delta T_{\rm CMB}$ is determined by step 6,
(3) the adiabat is drawn from the CMB 
with temperature $T_{\rm CMB}$ to the surface,
(4) the initial temperature distribution in the mantle is
given by the average of the two adiabats obtained by
steps 1 and 3, 
(5) core temperature is determined by the
procedure given in section 2.2 with 
$T_{\rm CMB}$,
and (6) the amount of $\Delta T_{\rm CMB}$ ($\sim 1000$K) 
is determined by the requirement that
the predicted surface heat flux and inner core radius
for an Earth-mass planet are comparable to 
the observed values for the present Earth.
We use this value for all the cases with various planetary masses.
Note that the temperature distribution is quickly relaxed to
an equilibrium distribution, as long as
we use the initial conditions created by the above procedures.

\subsection{Simulation parameters}

We summarize parameters for the ``nominal'' case:
\begin{itemize}
\item Boundary conditions 
   \begin{itemize}
   \item surface temperature: $T_{\rm surf}=300$K
   \item a mass ratio between mantle and core: $\zeta_{\rm m/c}=7:3$
   \item the CMB is a barrier for convection, while convection
         penetrates the upper/lower mantle boundary
   \end{itemize}
\item Initial conditions 
   \begin{itemize}
   \item impurity fraction: $x_{\rm 0S} = 0.1$ 
   \end{itemize}
\item Rheological conditions
   \begin{itemize}
   \item activation volume: $V^*=10\times 10^{-6} \mathrm{m^3mol^{-1}}$
   \end{itemize}
\end{itemize}

We first investigate planetary mass dependence for
planets with the above nominal parameters.
For Earth-mass planets, we adopt $\Delta T_{\rm CMB}=1000$K in most runs,
because the nominal case with $\Delta T_{\rm CMB}=1000$K reproduces the present Earth.
We also systematically study dependences of the results on
$\Delta T_{\rm CMB}$, because $\Delta T_{\rm CMB}$ is not well determined.
In some runs, the upper/lower mantle boundary is
treated as a barrier for convection.
We also carry out calculations with different 
values of $T_{\rm surf}$, $\zeta_{\rm m/c}$, and $x_{\rm 0S}$
to reproduce the results that are consistent
with the current magnetic fields of Mercury, Venus, and Mars
in Appendix B
(we do not systematically survey the dependences on these
parameters).

\subsection{Definition of lifetime of planetary intrinsic magnetic field}

To drive dynamo action, liquid metallic core must be
in active convection state.
Following \citet{Stevenson1983},
we adopt the threshold heat flux in the core 
for generation of dynamo action (conducted heat flux along the core adiabatic thermal structure) as,
\begin{equation}
F_{\rm crit}=k_c\left(\frac{\partial T_{\rm CMB}}{\partial r}\right)_{S}=k_c \frac{\rho g\gamma_G}{K_s}T_{{\rm CMB}}.
\label{eq:F_crit}
\end{equation}
We define the lifetime of magnetic field as
a period during which the core heat flux exceeds the threshold value.

\section{Numerical Results}
\subsection{Thermal evolution of the Earth}

\begin{figure}[htbp]
\begin{flushleft}
(a)\hspace{5.5cm}(b)\\
\includegraphics[scale=0.3, angle=0]{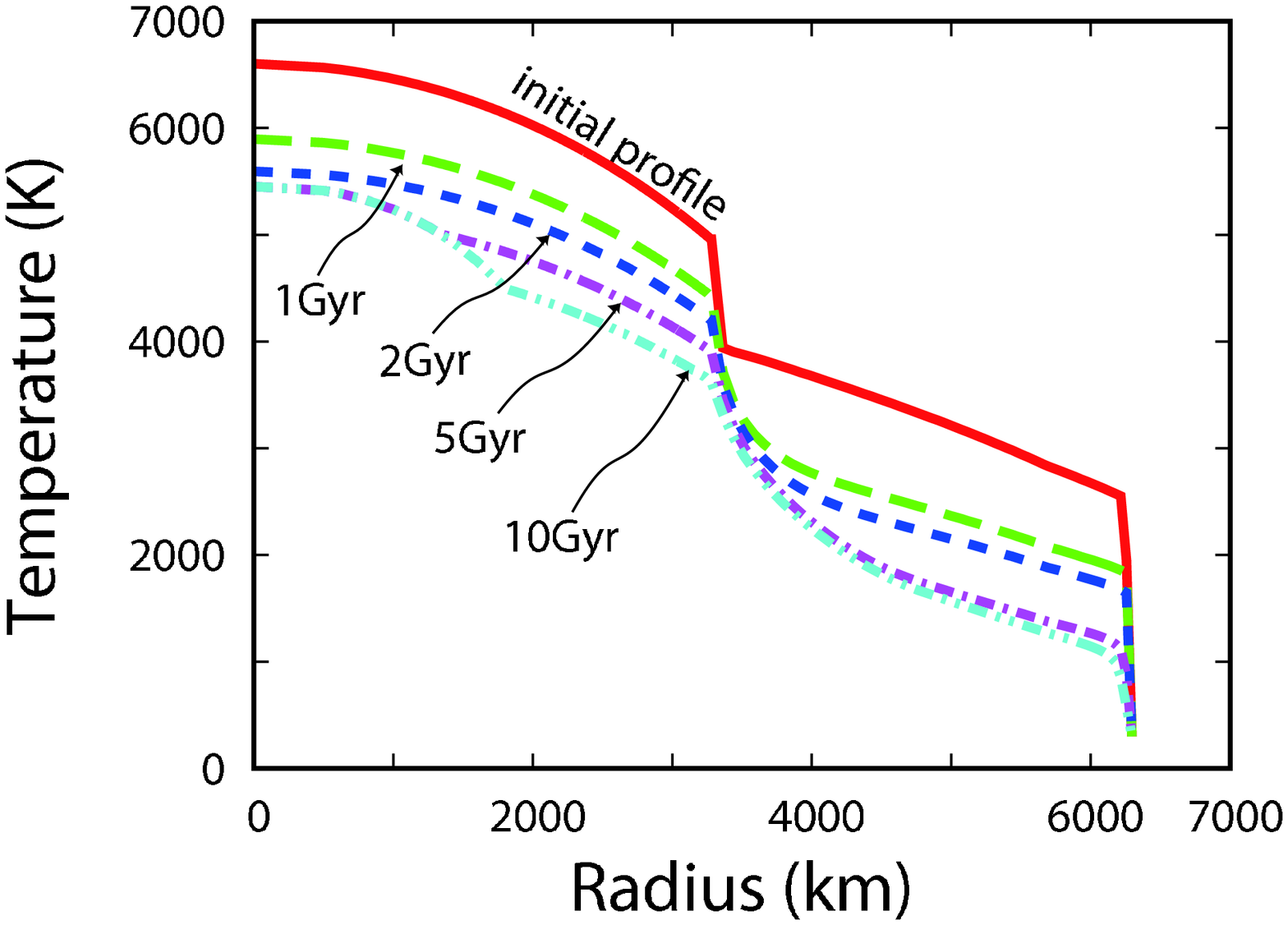}
\includegraphics[scale=0.3, angle=0]{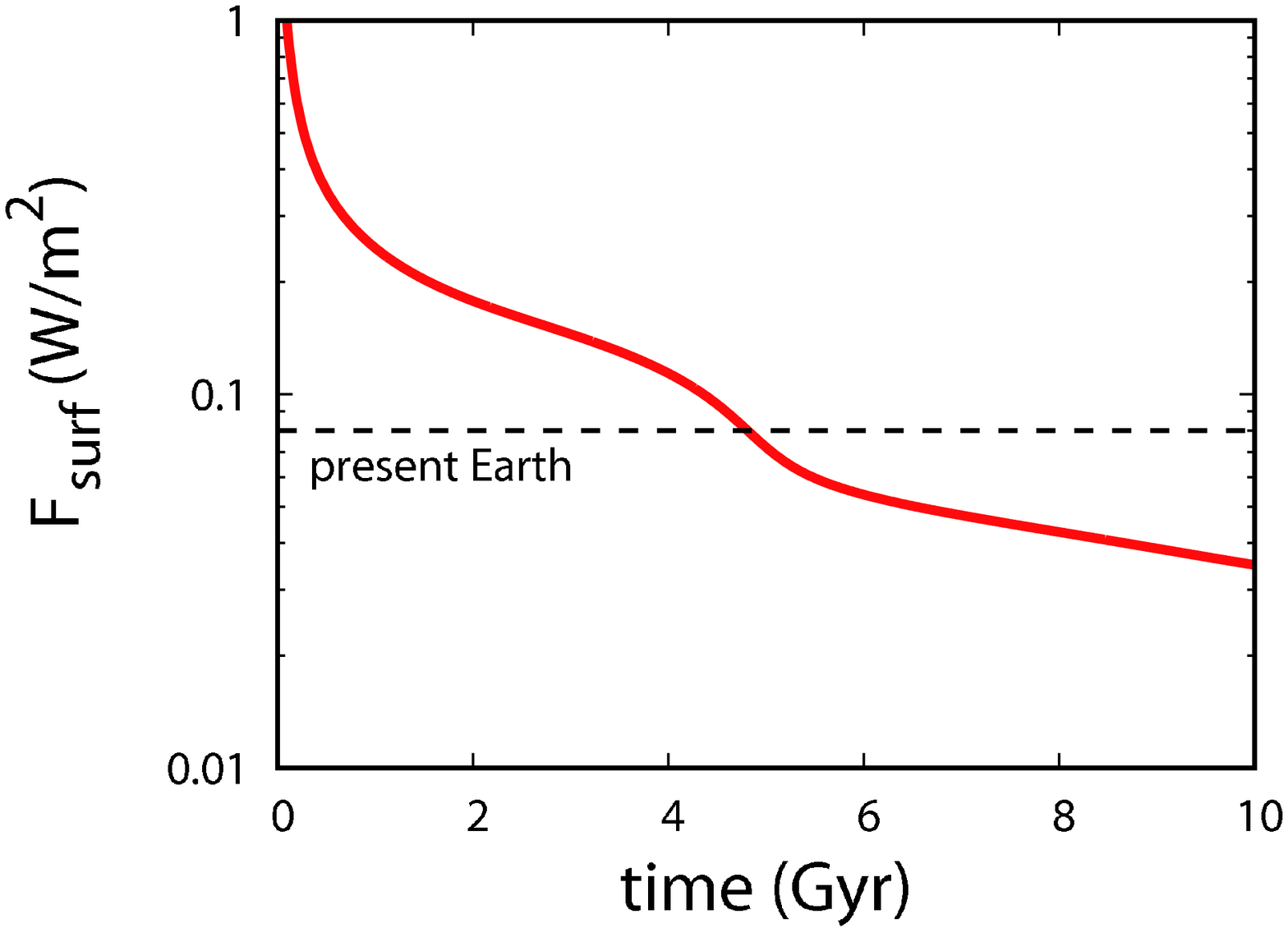}\\
(c)\hspace{5.5cm}(d)\\
\includegraphics[scale=0.3, angle=0]{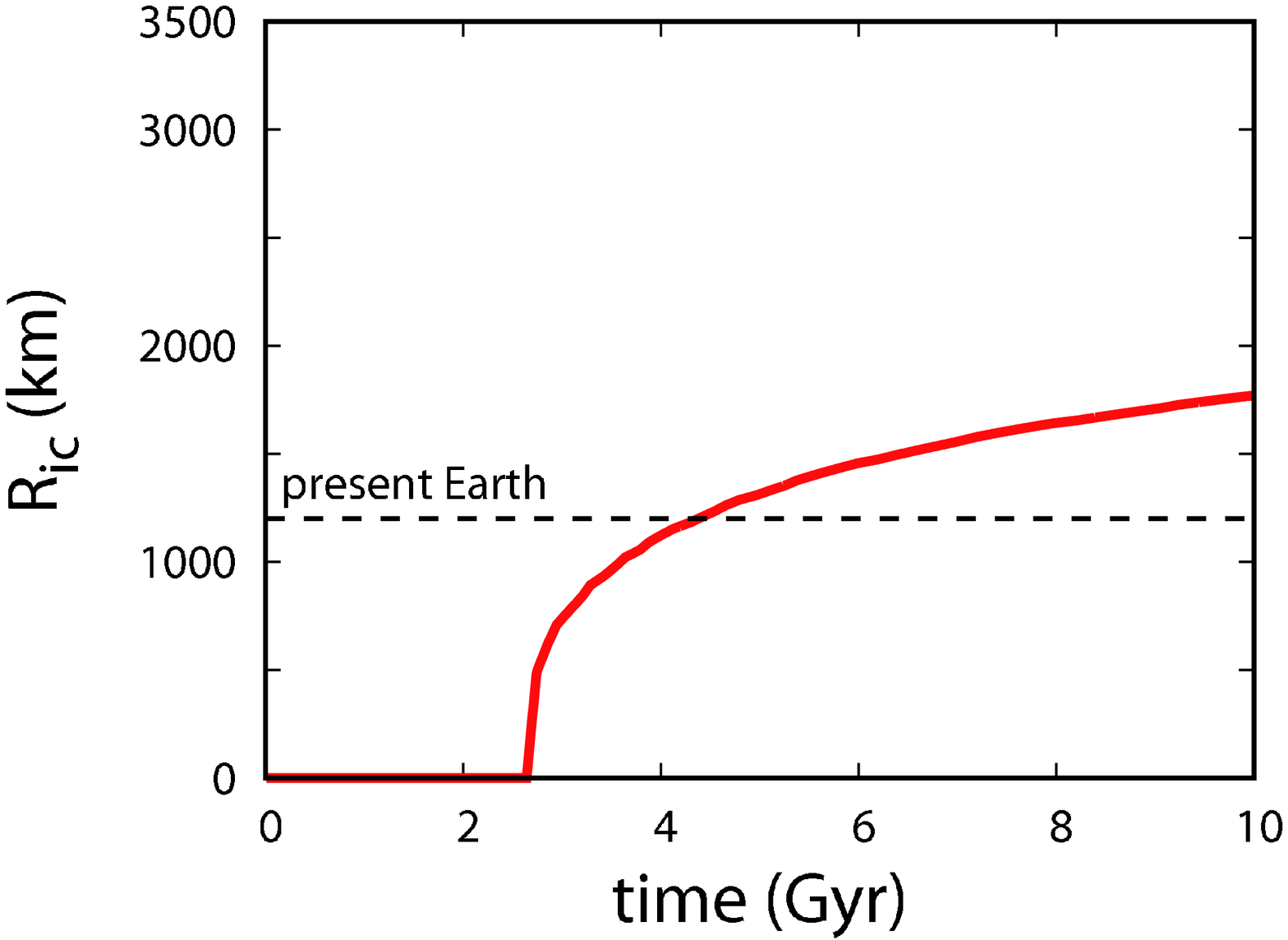}
\includegraphics[scale=0.3, angle=0]{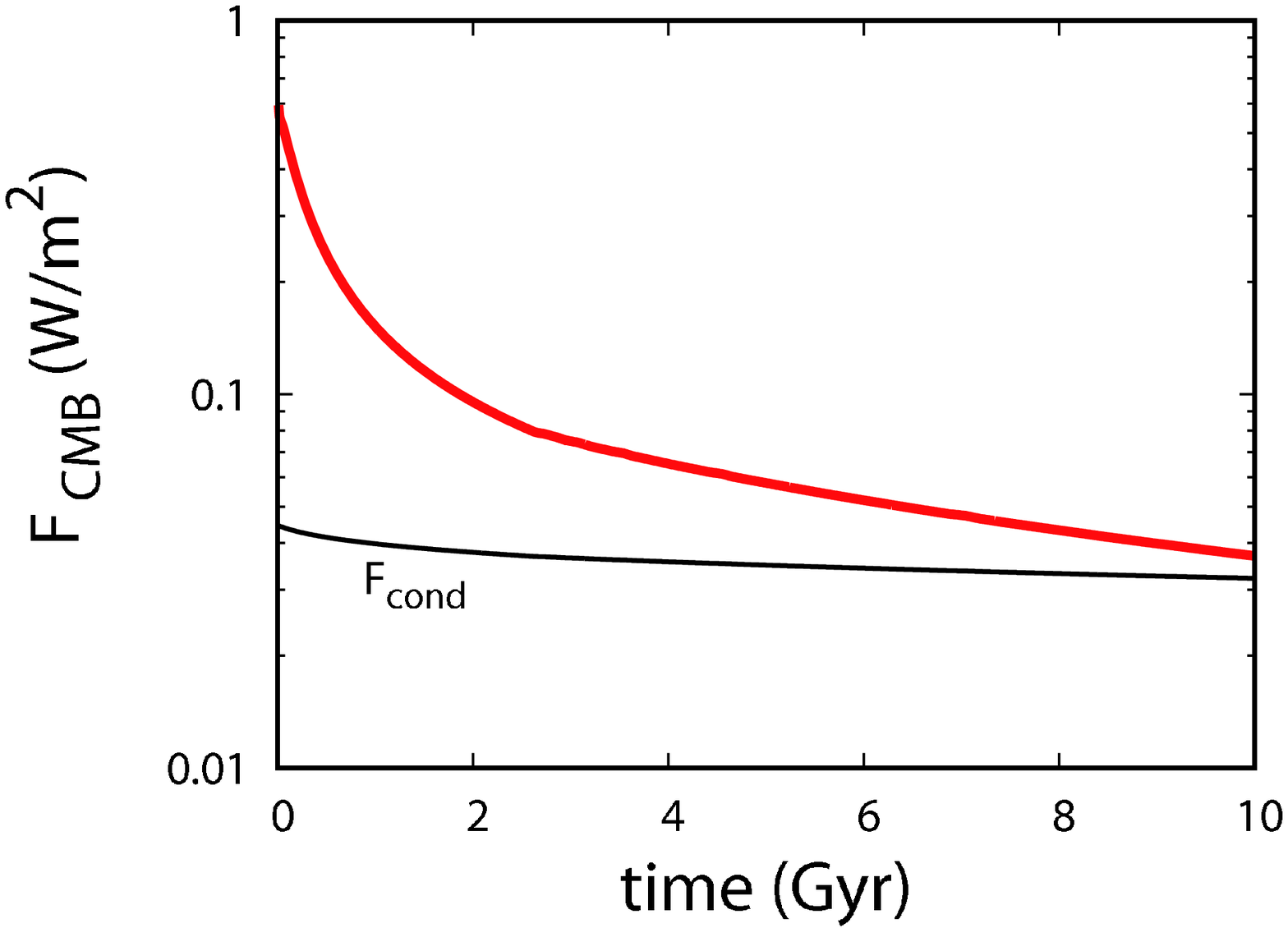}
\end{flushleft}
\caption{Time evolution of (a) the temperature profile, (b) the surface heat flux, (c) the inner core radius 
and (d) the heat flux through CMB for an Earth-mass planet with 
$\Delta T_{CMB} = 1000K$. 
(a)Time evolution of internal temperature distribution
in the case of $\Delta T_{CMB}=1000K$ with 1 $M_{\oplus}$. The planet radius of 6385km and 
core radius of 3375km.
The surface heat flux declines to $\sim 0.08$ ${\rm Wm^{-2}}$ and the inner core grows up to 1200 km at 4.5 Gyr after, which is nearly equal to the present observed value of the inner core radius of the Earth.
The core heat flux monotonically decreases with time 
but its time derivative discontinuously changes at the initiation of the inner core at around 2.2 Gyr. 
The solid curve represents the threshold flux ($F_{\rm cond}$)
to maintain dynamo activity in the outer core (Eq.~[\ref{eq:F_crit}]). 
}
\label{fig:Eartha1whole}
\end{figure}

We now show the evolution of temperature distribution calculated by the procedures in section 2.
Figure \ref{fig:Eartha1whole}a shows the evolution of thermal structure of 
an Earth-mass ($M=1M_{\oplus}$) planet for the nominal case.
Cooling of the mantle slows down with time, 
since the decrease in temperature enhances
the mantle viscosity (eq.~[\ref{eq:viscosity}]) 
and hence depresses the efficiency of heat transfer in the mantle.
This implies that the initial mantle temperature distribution 
hardly affects the thermal evolution on timescales longer than Gyr,
as long as the initial temperature is high enough \citep{Stevenson1983}.
However, since the core works as a heat bath for the mantle, 
the initial $T_{\rm CMB}$ 
would affect the thermal evolution of the mantle.

Figures \ref{fig:Eartha1whole}b, c and d show 
the time evolution of the surface heat flux ($F_{\rm surf}$), 
the heat flux across the CMB ($F_{\rm CMB}$)
and the inner core radius ($R_{\rm ic}$), respectively, 
in the nominal case. We adopt $\Delta T_{\rm CMB} = 1000$K. 
The inner core emerges at 2.2 Gyr. Its radius reaches 1200km at 4.5Gyr 
that agrees with the observed value of the present Earth,
as expected.
With higher/lower values of $T_{\rm CMB}$, 
the inner core radius at 4.5Gyr is smaller/larger.
The growth rate decreases with time, because
not only geometric effect but also the increase in
impurity concentration in the outer core depresses the melting
point of the outer core. 
After the emergence of the inner core,
the heat flux reduction becomes more slowly
because the inner core growth releases
gravitational energy and latent heat that work 
as internal heat sources. 
The heat flux remains larger than the critical value
given by Eq.~(\ref{eq:F_crit}), which is expressed by
the dot-dashed line in Fig.~\ref{fig:Eartha1whole}d,
during first 12 Gyr and the lifetime of dynamo
activity is expected to be 12 Gyr for this nominal case.

Some paleomagnetism data suggests that the magnetic field
of the Earth is enhanced to be the present level at $\sim 2$ Gyr
\citep[e.g.,][]{Hale1987}.
It might be due to formation of the inner core because nucleation of the inner core provides additional heat source.
Paleomagnetism data may include a large uncertainty.
If more detailed data is provided, it will constrain
the condition of generation of intrinsic magnetic field.

\begin{figure}[htbp]
\begin{flushleft}
(a)\hspace{5.5cm}(b)\\
\includegraphics[scale=0.3, angle=0]{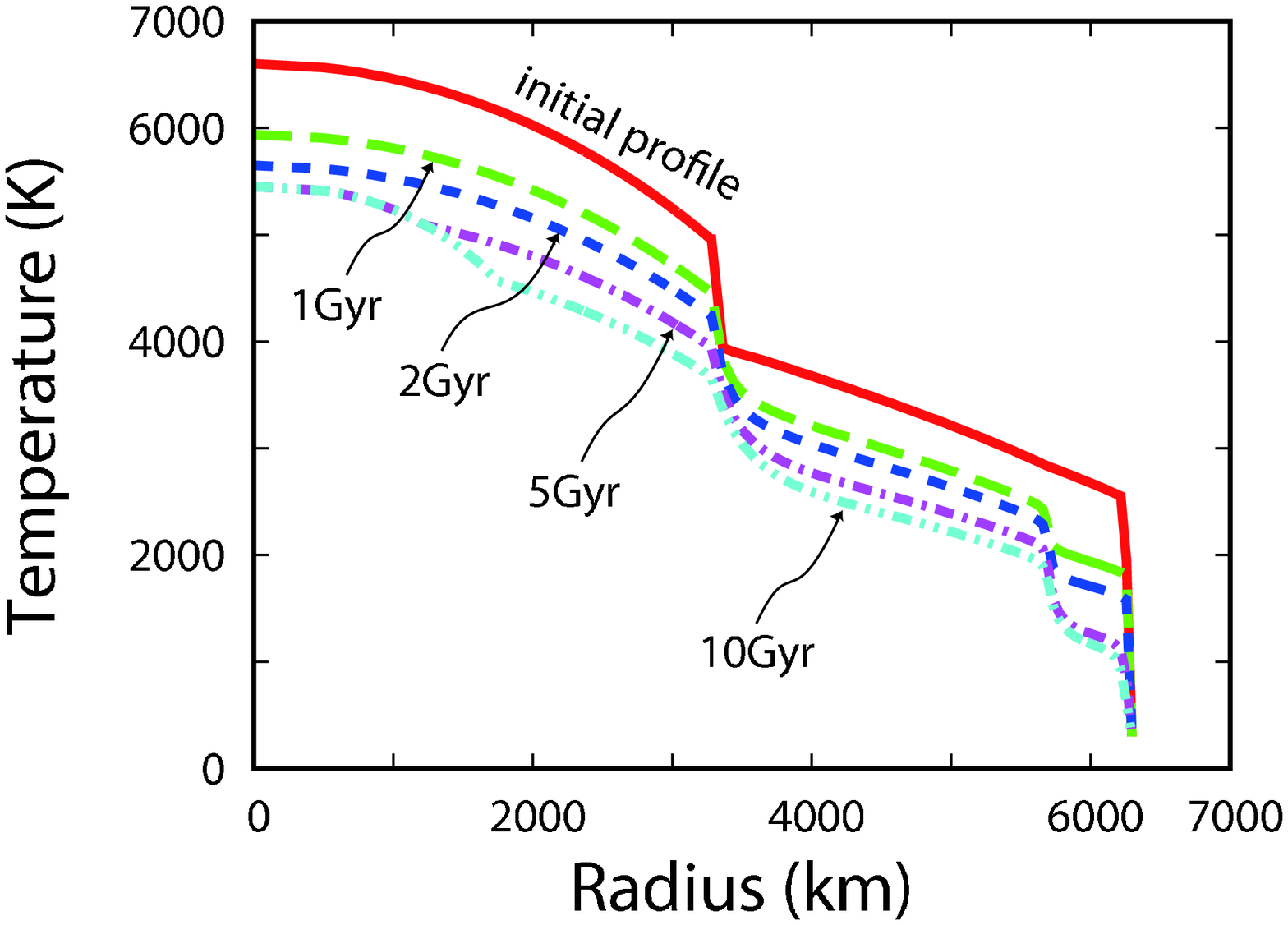}
\includegraphics[scale=0.3, angle=0]{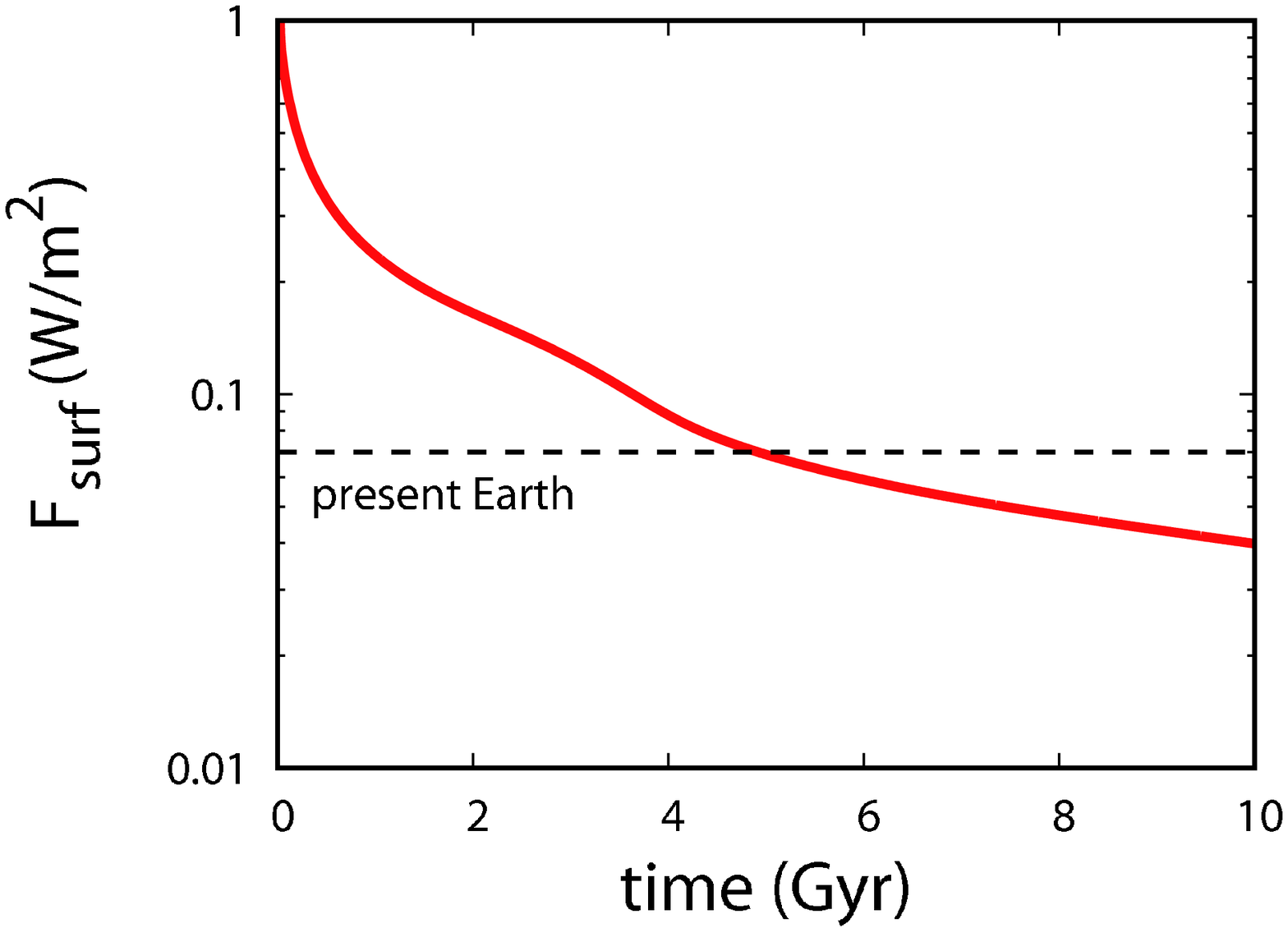}\\
(c)\hspace{5.5cm}(d)\\
\includegraphics[scale=0.3, angle=0]{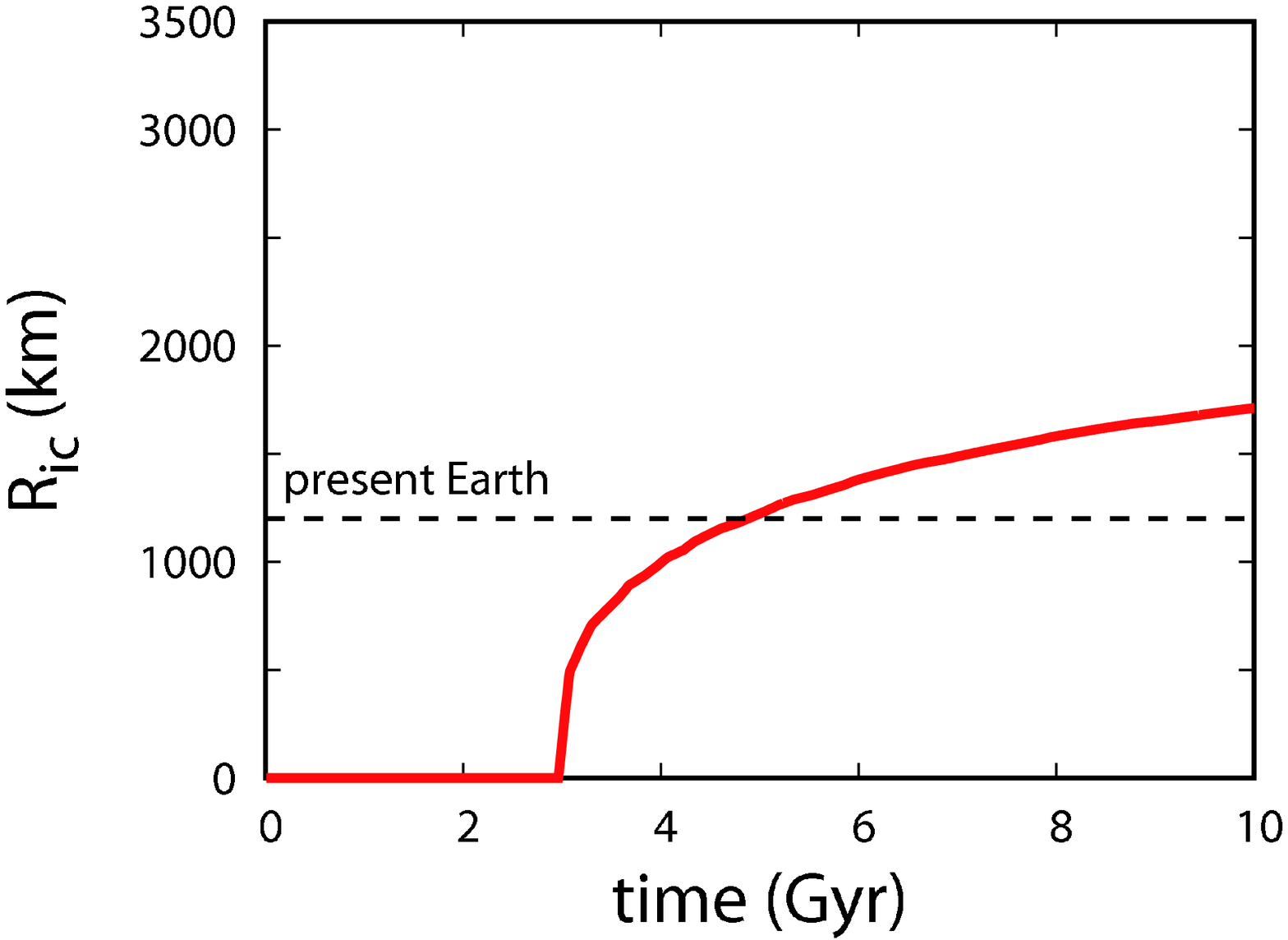}
\includegraphics[scale=0.3, angle=0]{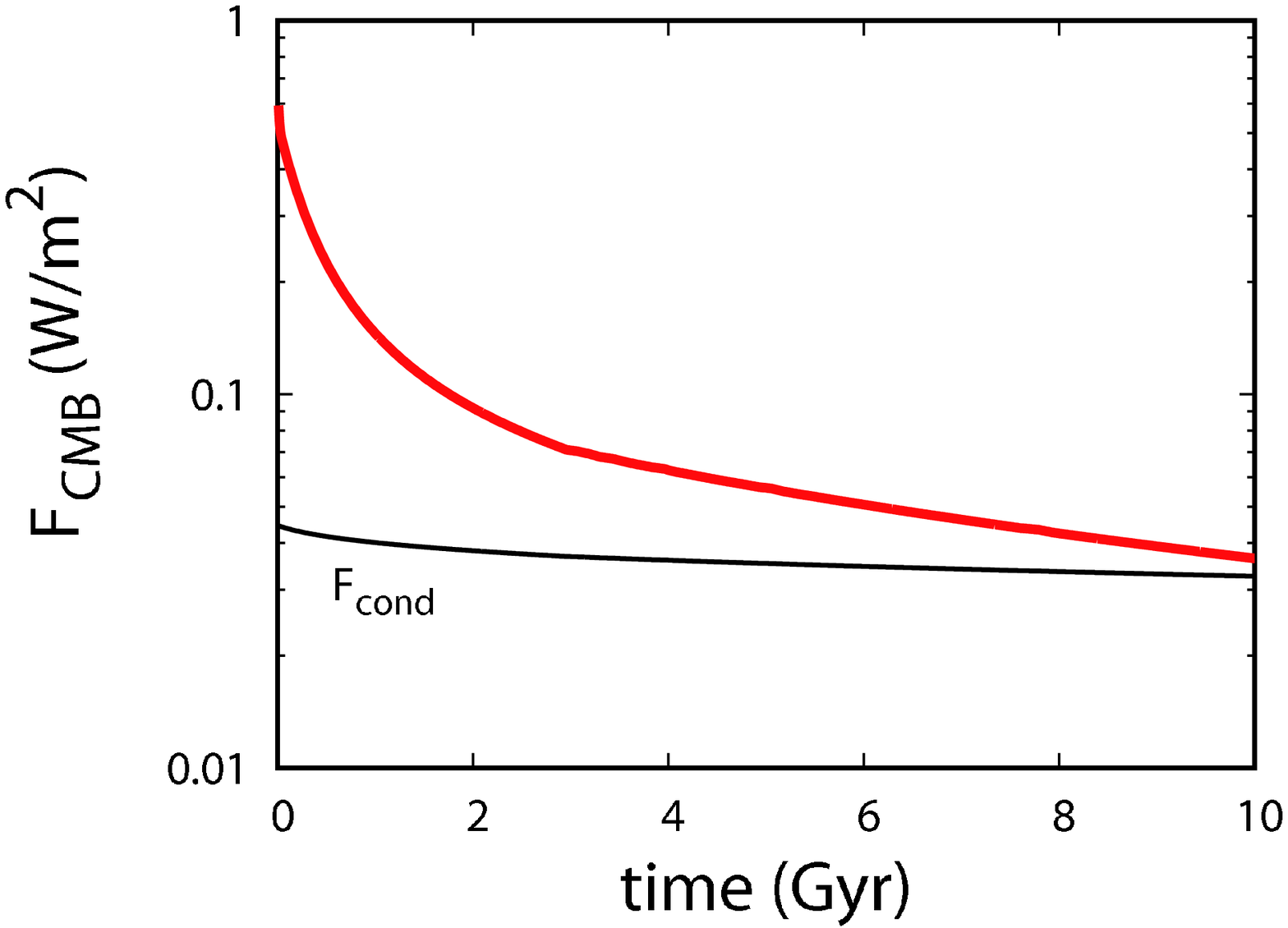}
\end{flushleft}
\caption{Same figures as Figs. \ref{fig:Eartha1whole} except for two-layered mantle convection model with 
 $\Delta T_{\rm CMB}=$1000K. 
}
\label{fig:Eartha0dT1000}
\end{figure}

\begin{figure}[htbp]
\begin{flushleft}
(a)\hspace{5.5cm}(b)\\
\includegraphics[scale=0.3, angle=0]{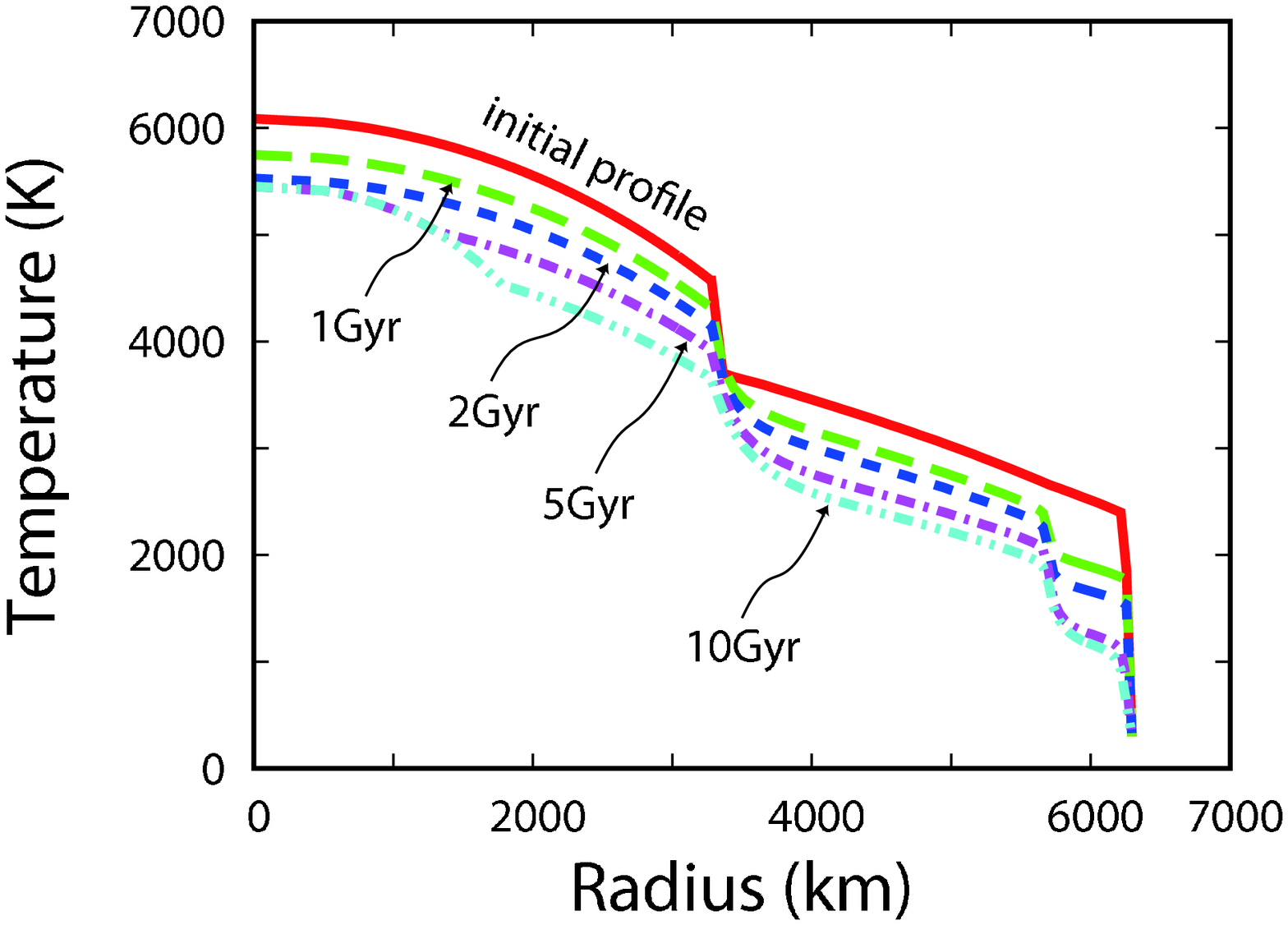}
\includegraphics[scale=0.3, angle=0]{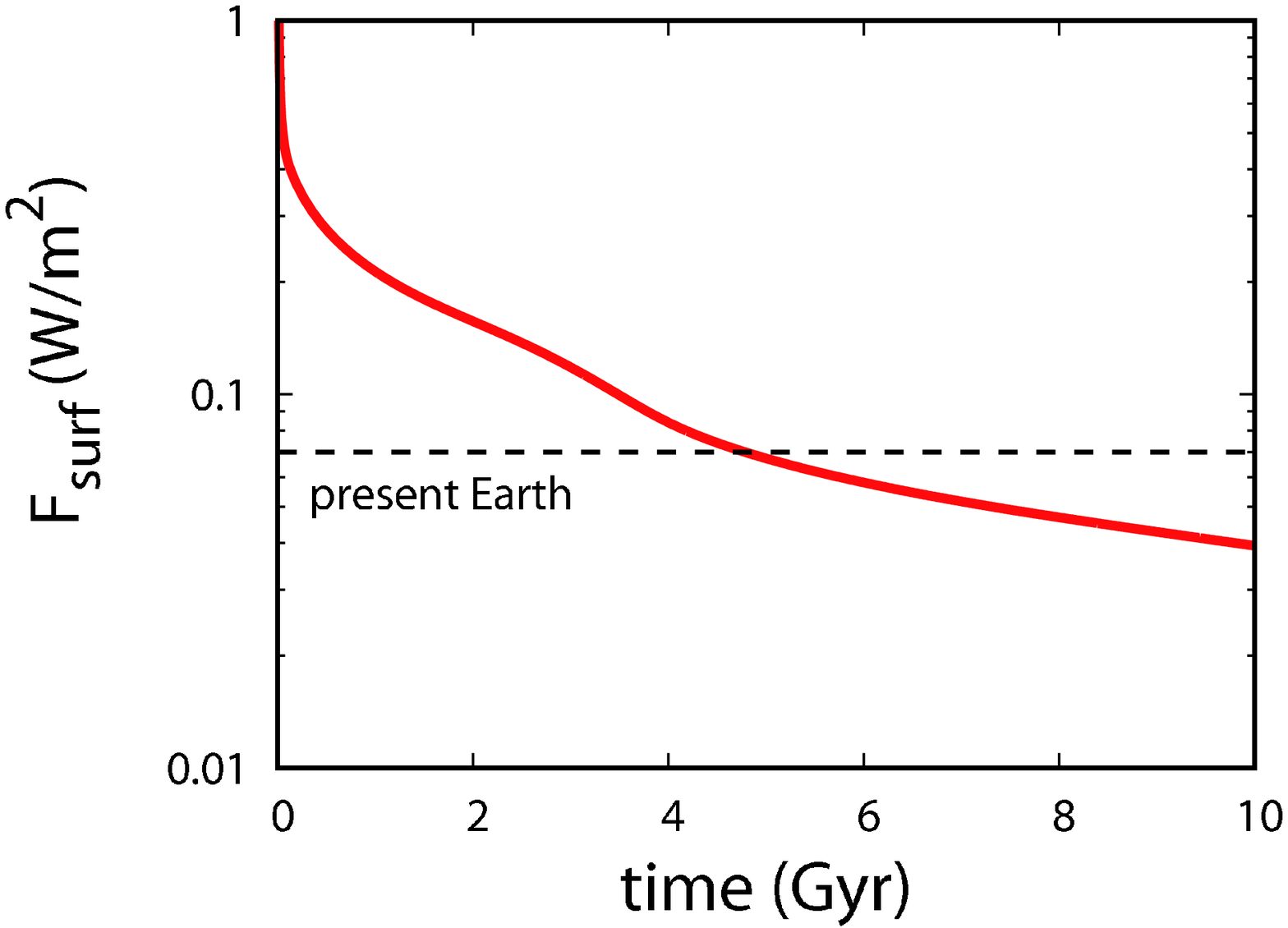}\\
(c)\hspace{5.5cm}(d)\\
\includegraphics[scale=0.3, angle=0]{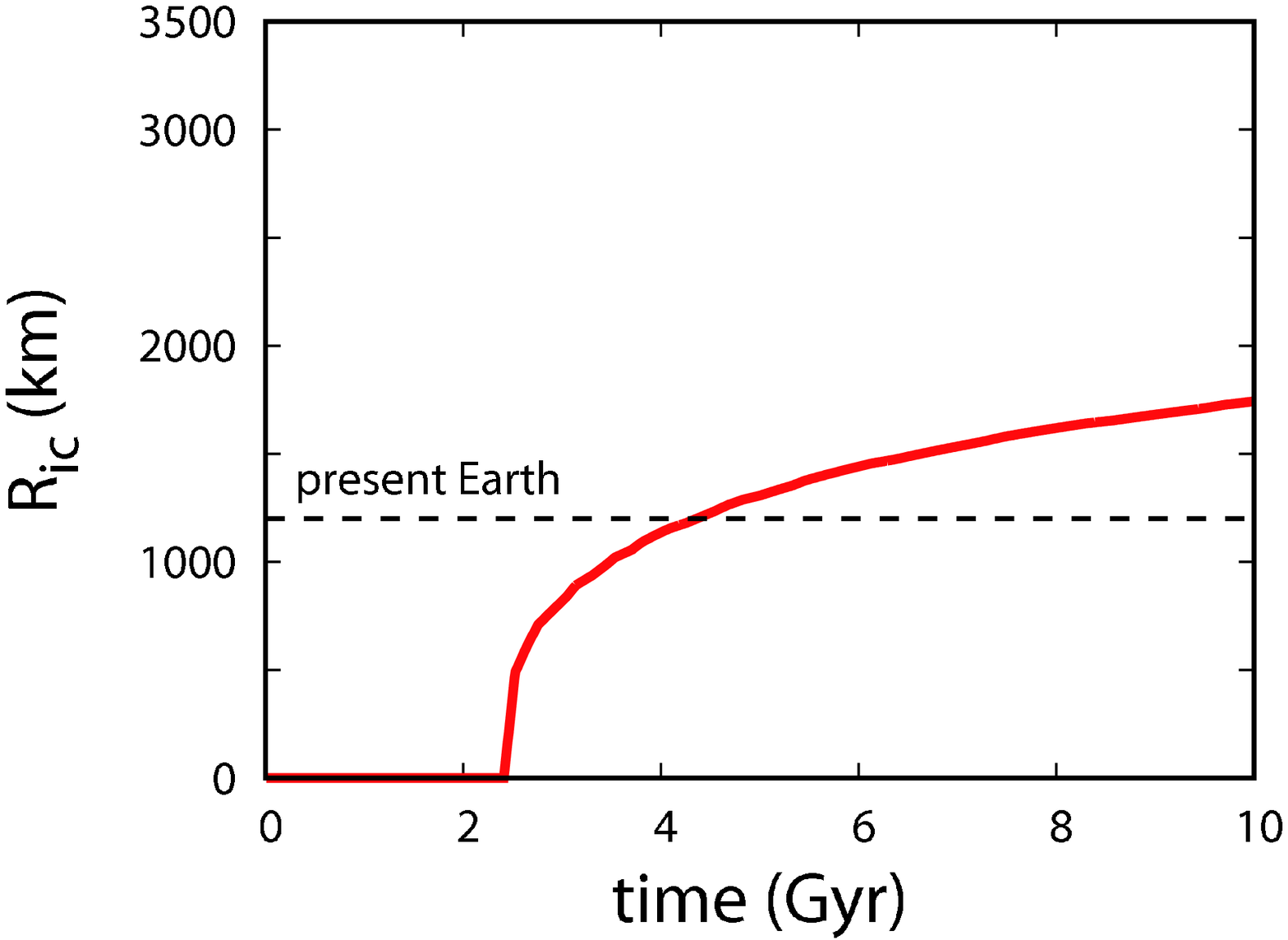}
\includegraphics[scale=0.3, angle=0]{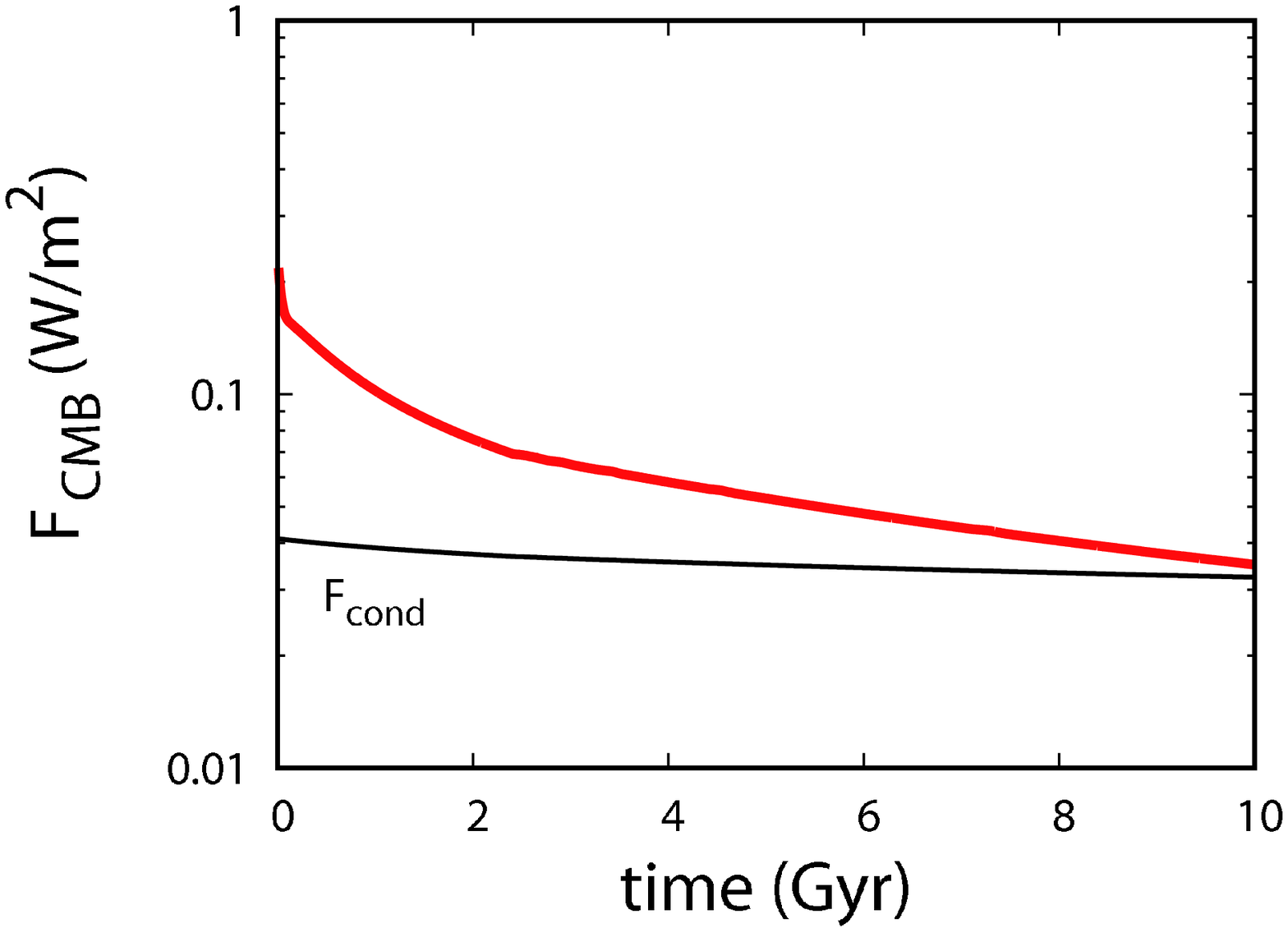}
\end{flushleft}
\caption{Same figures as Figs. \ref{fig:Eartha1whole} except that two-layered mantle convection model with 
$\Delta T_{CMB} = 800K$ is considered instead of one-layered model with $\Delta T_{\rm CMB}=$1000K. 
}
\label{fig:Eartha0whole}
\end{figure}

Figures \ref{fig:Eartha0dT1000} and \ref{fig:Eartha0whole} show the results of the two-layer convection.
In this calculation, we set 
the upper/lower mantle boundary as a barrier for convection. 
The other boundary conditions and the model parameters are 
the same as those in the nominal case. 
The mixing length is shorter in the entire regions of
mantle and cooling is slower than in the one-layer case.
As is shown in Figures \ref{fig:Eartha0dT1000},
if we adopt initial $\Delta T_{\rm CMB} = 1000$K
as in the case of one-layer convection, 
the inner core can not grow to 1200km because of the low heat transfer
efficiency of the layered convection and core temperature is somewhat higher
than that obtained by the one-layer convection. 
We also carried out a calculation with 
initial $\Delta T_{\rm CMB} = 800K$.
Figure~\ref{fig:Eartha0whole}a shows the evolution of thermal profile.
A thermal boundary layer 
at upper/lower mantle boundary is clearly established.
Because the lower initial $\Delta T_{\rm CMB}$ is compensated with
the inefficient two-layer convection,
evolution of heat flux through CMB ($F_{\rm CMB}$) 
and lifetime of magnetic field (11Gyr in this case)
are similar to those in the one-layer convection case.

If the two-layer convection is assumed only in the Archean and Hadean
($t < 2$Gyr),
with the same initial $\Delta T_{\rm CMB}$ (1000K)
the surface heat flux at 4.5Gyr is 
$F_{\rm surf} \sim 0.12 {\rm Wm^{-2}}$, which
is somewhat higher than the observed value ($\sim 0.09{\rm Wm^{-2}}$),
since the thermal energy beneath upper/lower mantle boundary have been stored until 2Gyr and then supplied to upper mantle after 2Gyr.
However, the evolution of $F_{\rm CMB}$ is not so different between
one- and two-layer convection.
The lifetime of magnetic field is about 12 Gyrs.

Thus, 
if we tune the initial $\Delta T_{\rm CMB}$
with the present observed values of $F_{\rm surf}$ and $R_{\rm ic}$, 
the expected lifetime of magnetic field is not
affected by the mode (one-layer or two-layer) of mantle convection.

\subsection{Thermal evolution of Mercury, Venus, and Mars}

The existence of the magnetic field for Mercury
and early decay of the magnetic field for Venus and Mars 
were addressed by \citet{Stevenson1983}, using the box model
with different parameter values such as surface temperature and a mantle-core mass ratio from those in 
the nominal case.
The validity of these parameter values is
discussed in Appendix B.
Adopting the same parameter values as \citet{Stevenson1983}, 
we have performed simulations for Mercury, Venus, and Mars
with our model.
As discussed in Appendix B, our model produces the
results that are not inconsistent with the
magnetic activity of Mercury, Venus, and Mars.

\subsection{Thermal evolution of super-Earths}

\begin{figure}[htbp]
\begin{flushleft}
(a)\hspace{5.5cm}(b)\\
\includegraphics[scale=0.3, angle=0]{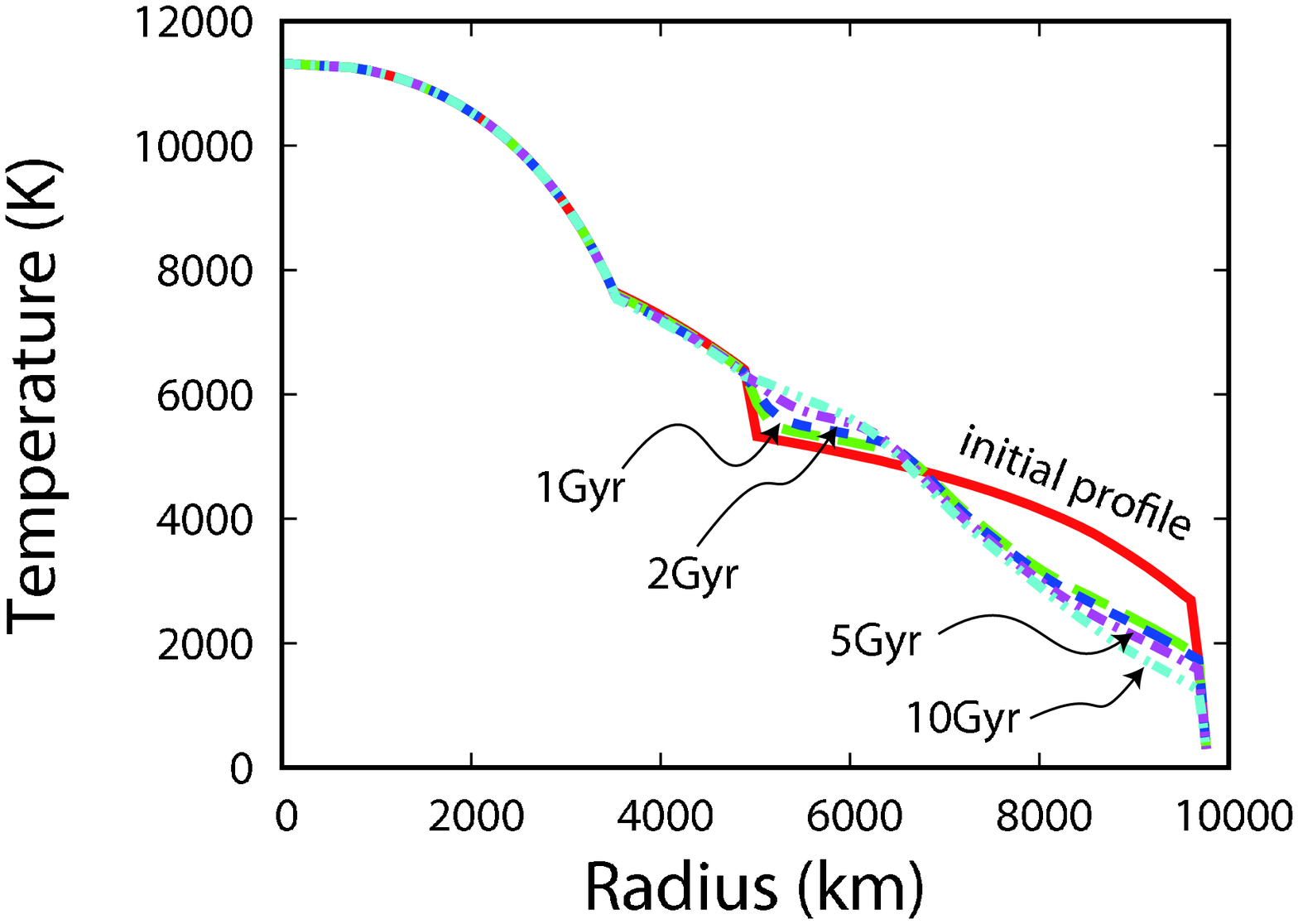}
\includegraphics[scale=0.3, angle=0]{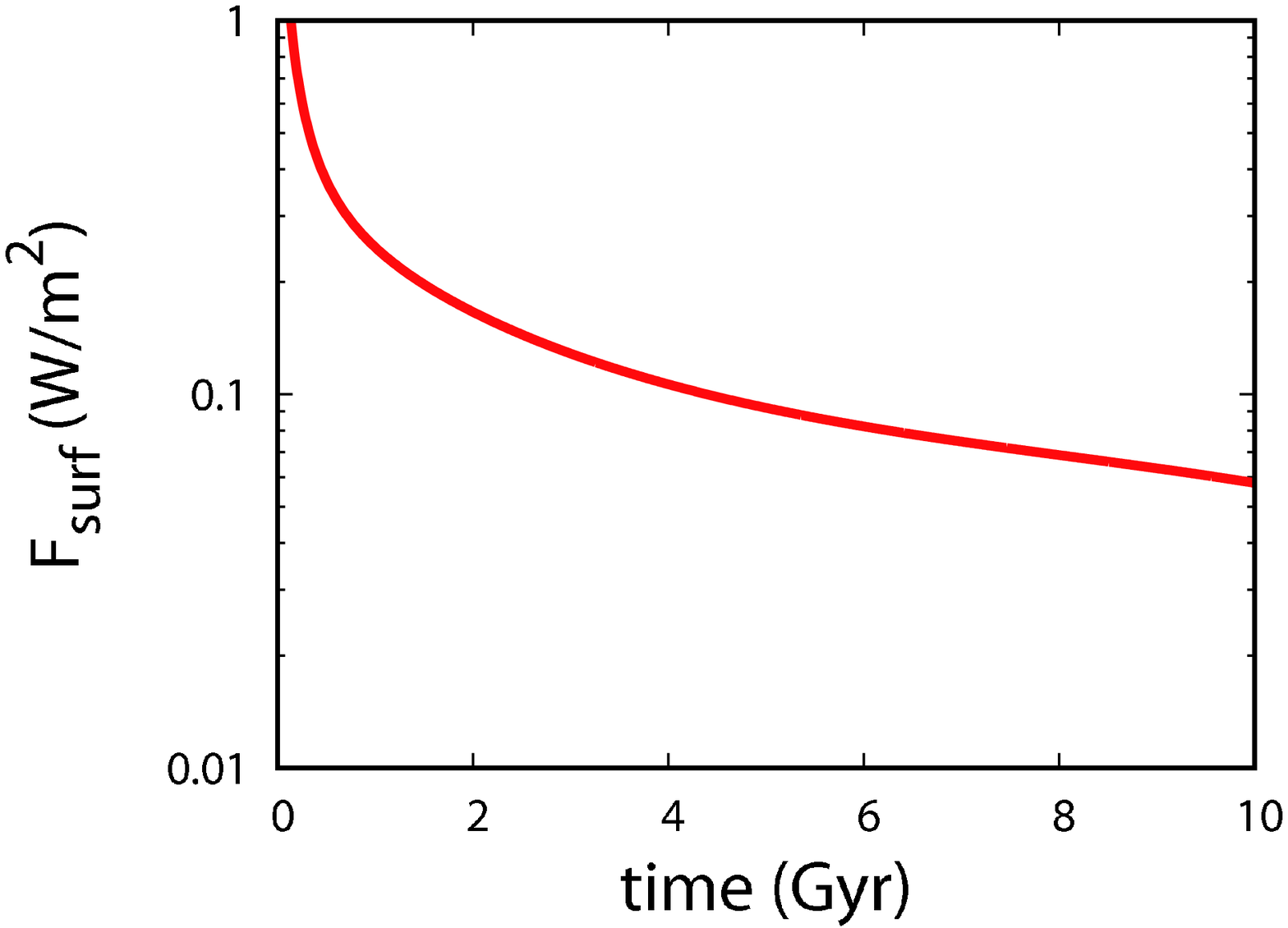}\\
(c)\hspace{5.5cm}(d)\\
\includegraphics[scale=0.3, angle=0]{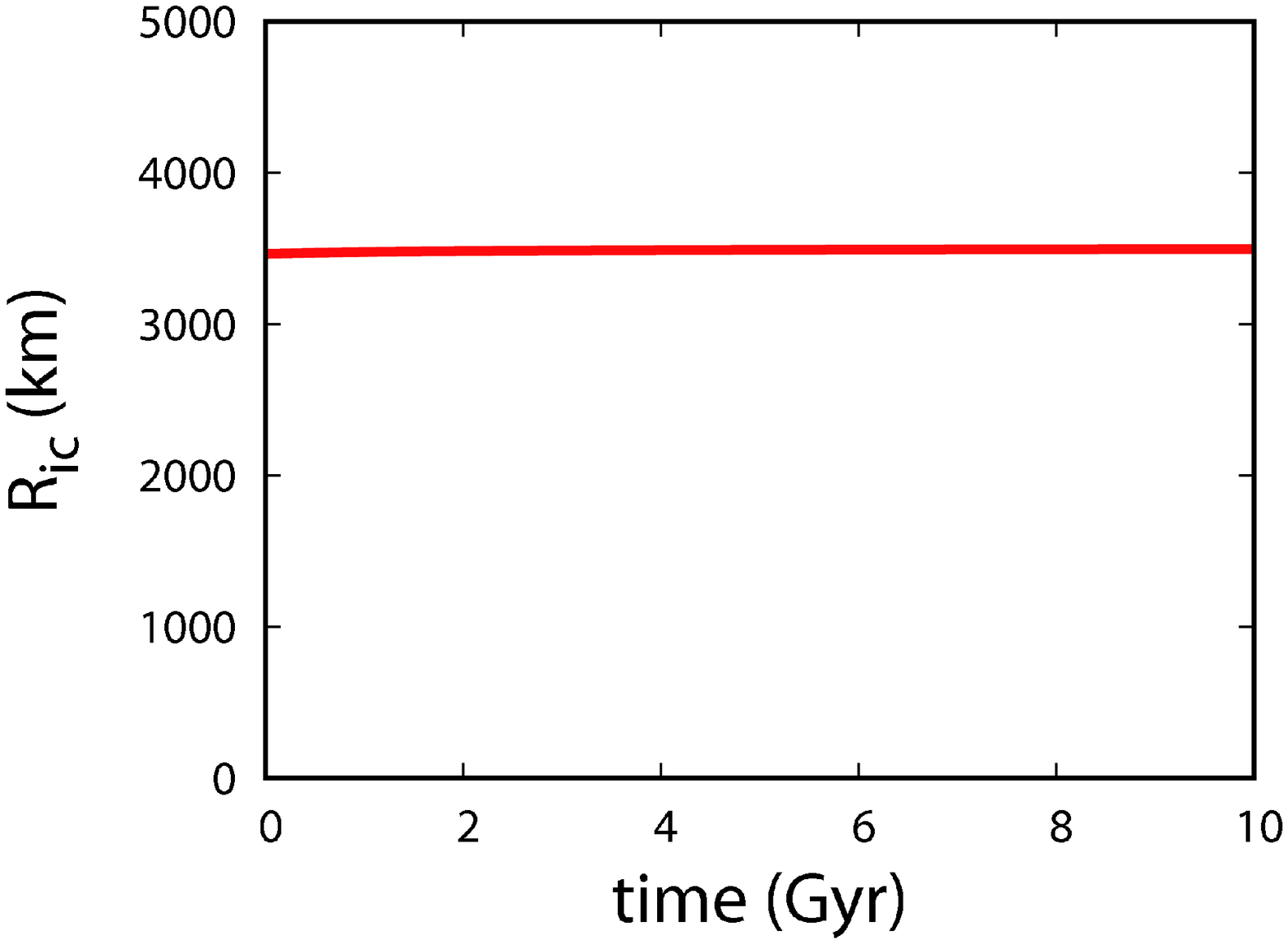}
\includegraphics[scale=0.3, angle=0]{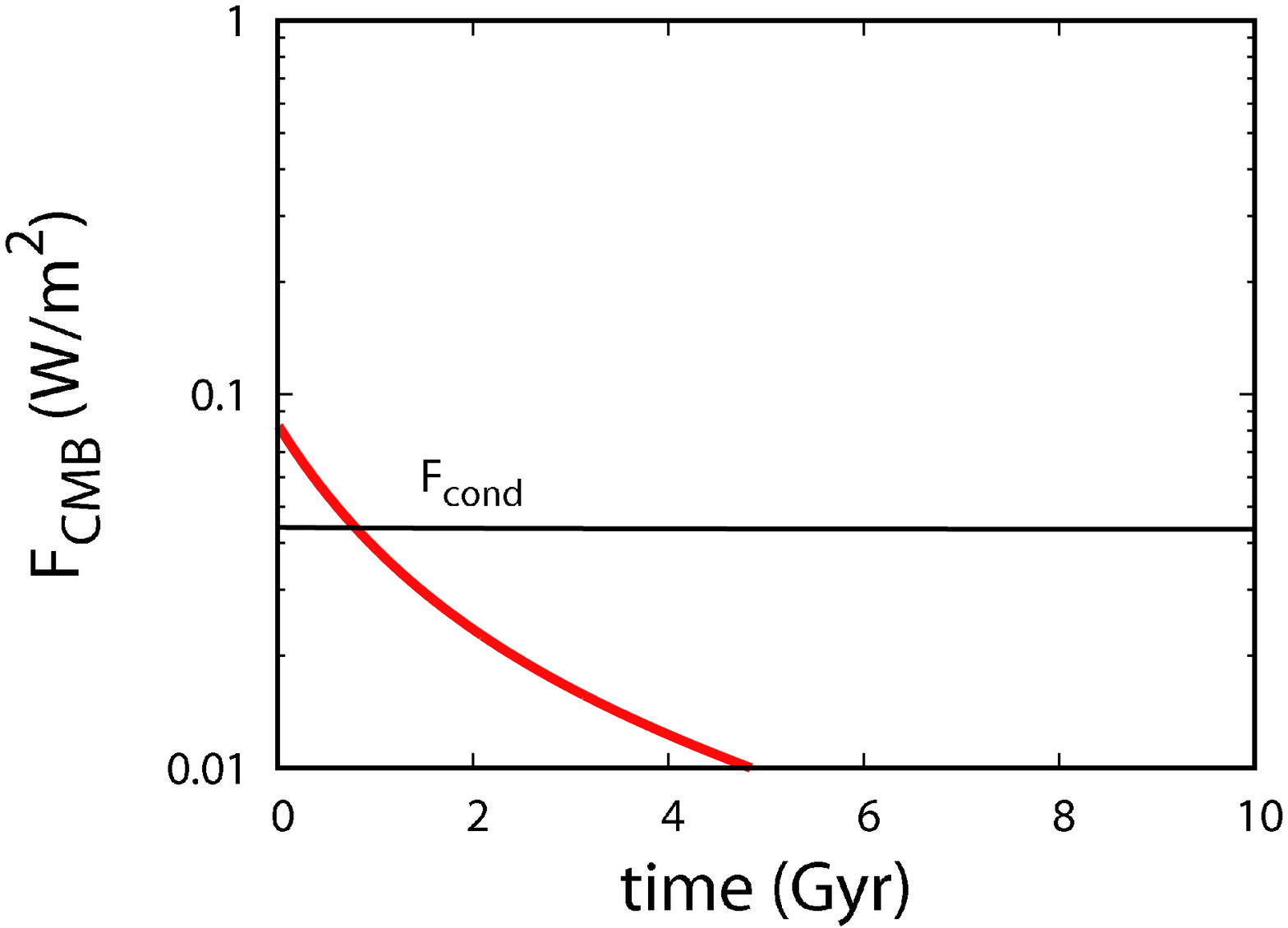}
\end{flushleft}
\caption{Same figures as Figs. \ref{fig:Eartha1whole} except for $M_p=5M_{\oplus}$. 
}
\label{fig:5Me}
\end{figure}

For super-Earths, we use the nominal parameters
(the surface temperature is 300K and
the mantle/core mass ratio is 7:3),
assuming that their orbits are nearly circular and in habitable zones.
We also assume one-layer convection throughout the mantle.
Detailed study on the effects of phase transitions is left to
future works. 
Figure \ref{fig:5Me}a shows evolution of temperature distribution
for a planet with mass $M_p = 5 M_{\oplus}$.
Compared with the case of $M_p=1M_{\oplus}$ in Figure \ref{fig:Eartha1whole}a, 
a thicker thermal boundary layer is established on the CMB 
within first few Gyrs, 
since the viscosity of the bottom of the mantle is higher. 
The increase in the viscosity due to higher pressure
dominates the decrease due to higher temperature
(Eq.~[\ref{eq:viscosity}]).
Thus, $F_{\rm CMB}$ is lower than the case of $M_p = 1M_{\oplus}$
(Figs.~\ref{fig:5Me}b and \ref{fig:Eartha1whole}b).
On the other hand, the effective heat capacity 
rapidly increases with $M_p$.
Figures~\ref{core_energy} in Appendix C show that
for fixed $T_{\rm CMB}$, thermal energy $E_{\rm th} \propto M_p^2$,
while the core surface area $S_{\rm core}$ 
increases with $M_p$ only weakly ($S_{\rm core} \propto M_p$). 
Therefore, the core for higher $M_p$ cools much more slowly.
It is shown that $R_{\rm ic}$ does not grow at all for 10Gyr.
On the other hand, $F_{\rm surf}$ is not so different from that in the case of $M_p= 1M_{\oplus}$.
The heat bath of surface heat flow is radiogenic elements in the mantle and that is proportional to $M_p$.
To balance heat generation and cooling, $F_{\rm surf}$ should be proportional to $M_p^{1/3}$, provided $R_p \propto M_p^{1/3}$.
Thereby $F_{\rm surf}$ changes by a factor of only $5^{1/3}\sim 1.7$.

\begin{figure}[htbp]
\begin{flushleft}
(a)\\
\includegraphics[scale=0.3, angle=0]{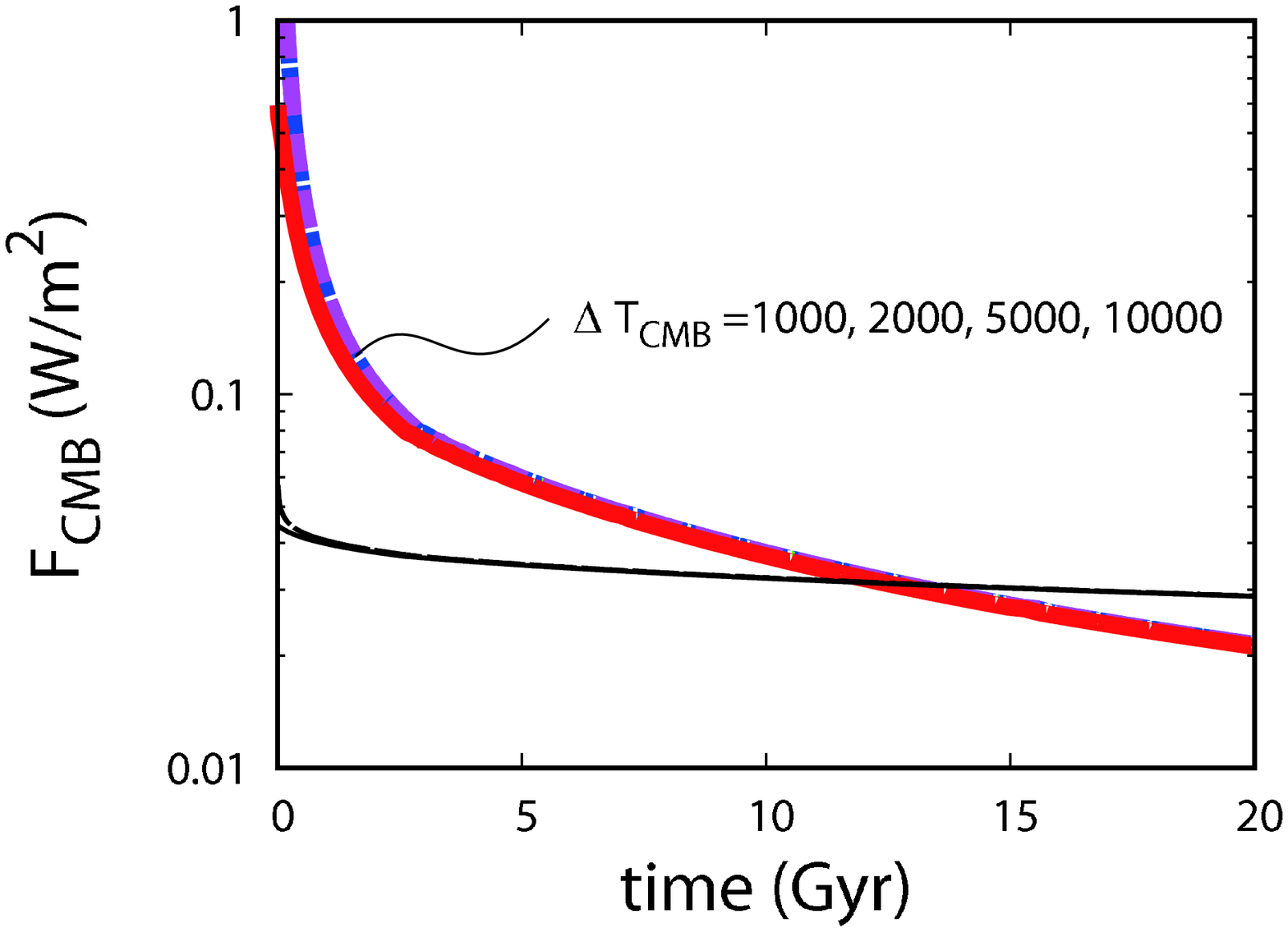} 
\includegraphics[scale=0.3, angle=0]{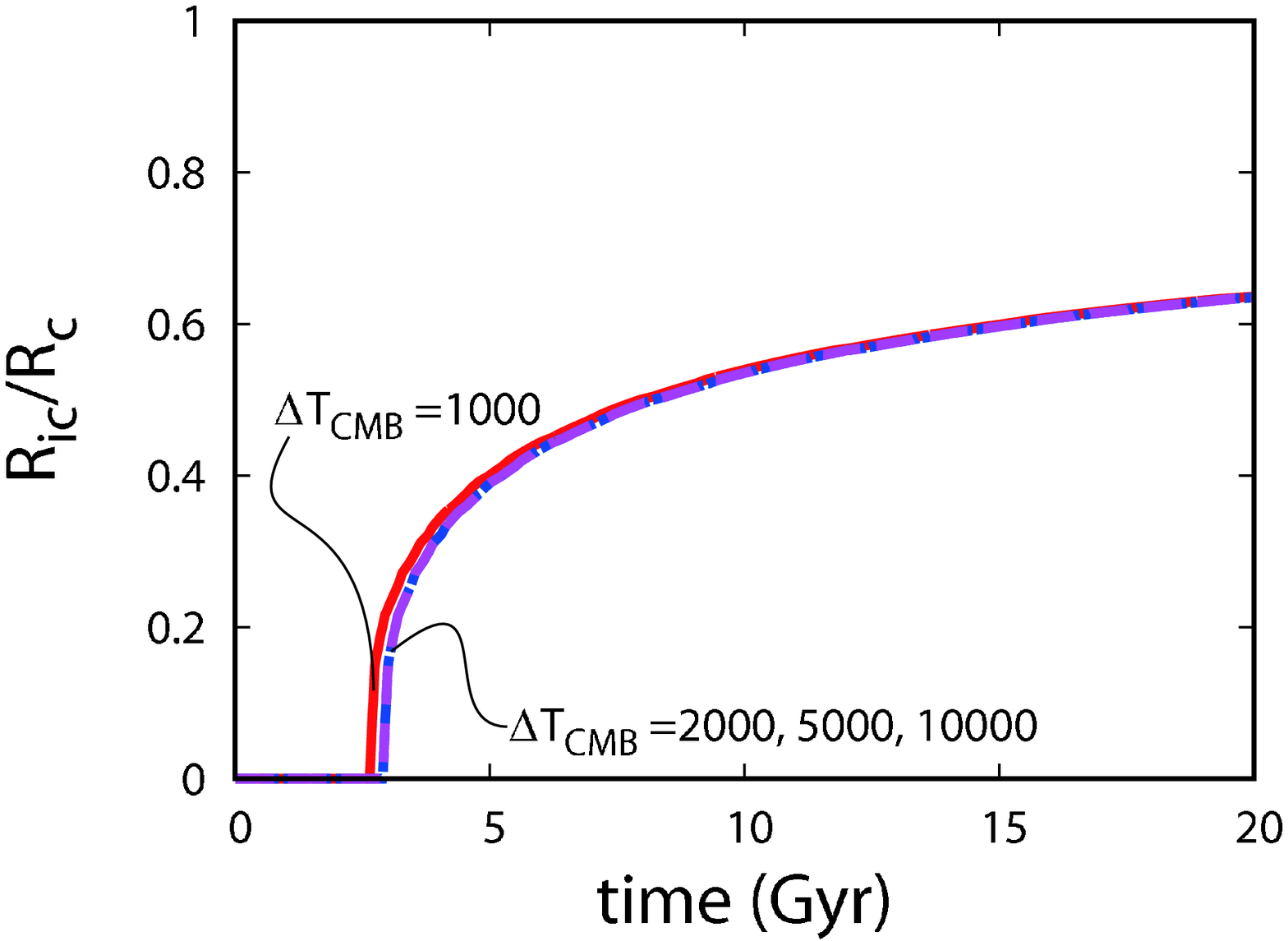}\\
(b)\\
\includegraphics[scale=0.3, angle=0]{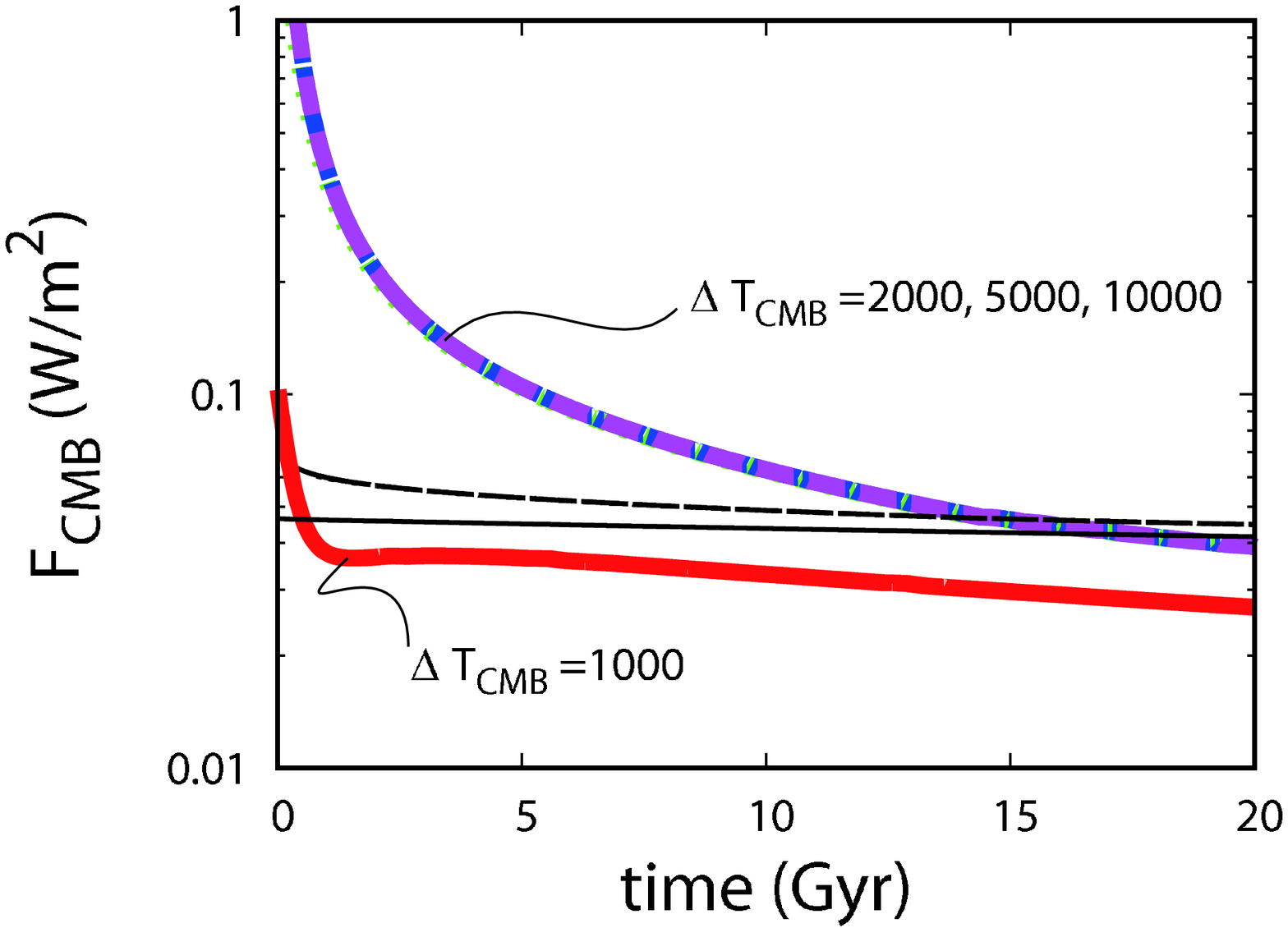} 
\includegraphics[scale=0.3, angle=0]{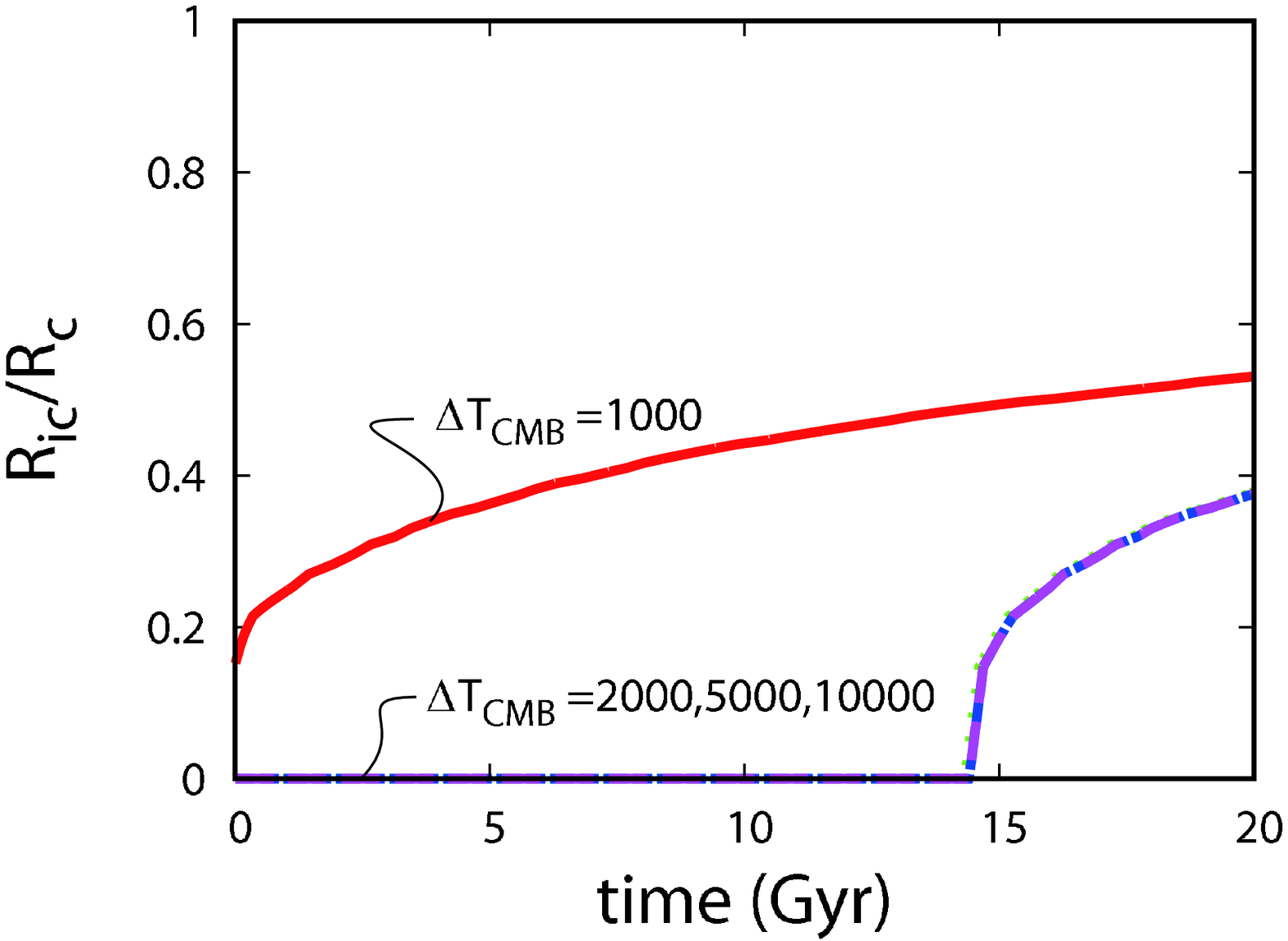}\\
(c)\\
\includegraphics[scale=0.3, angle=0]{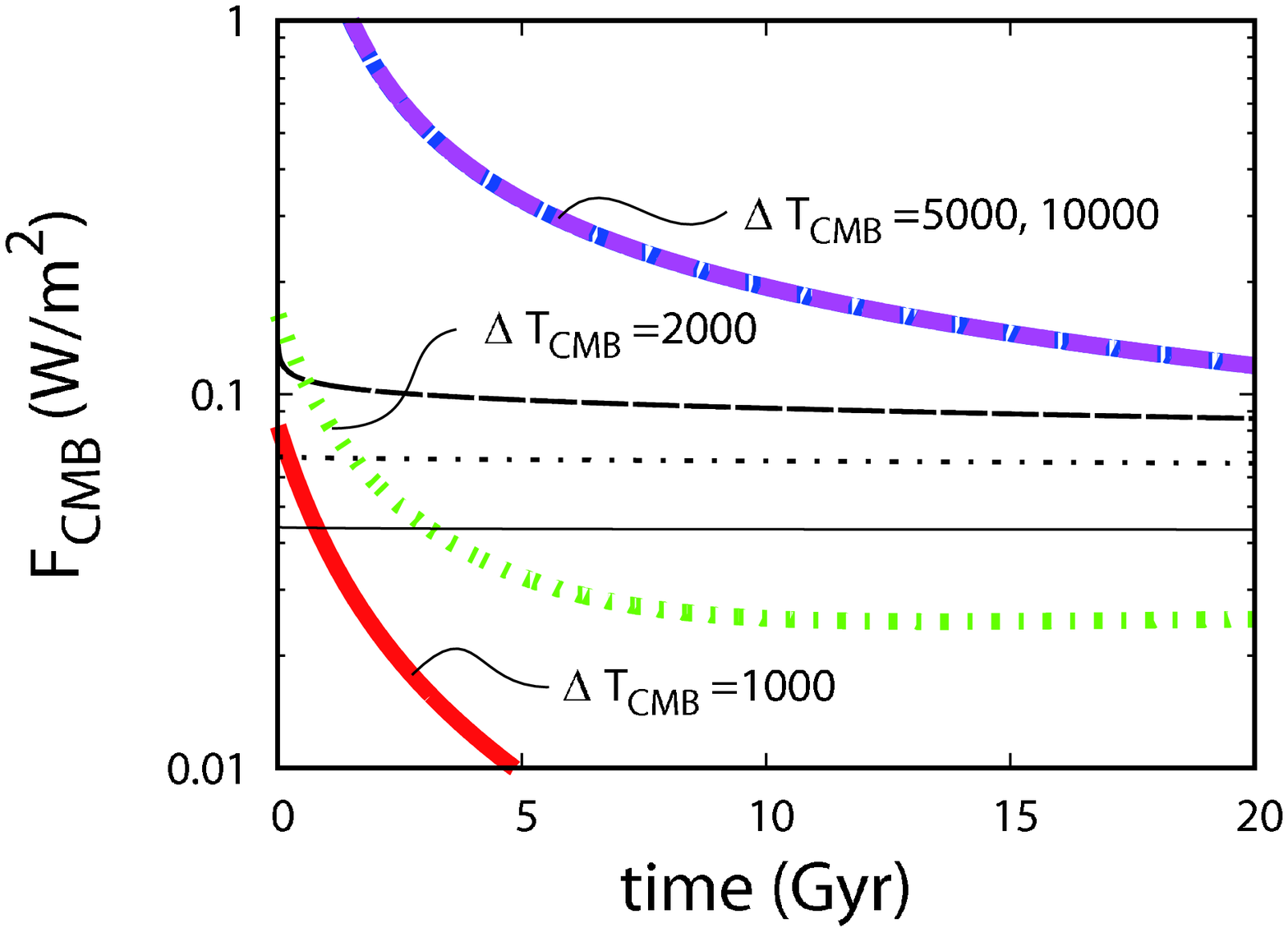} 
\includegraphics[scale=0.3, angle=0]{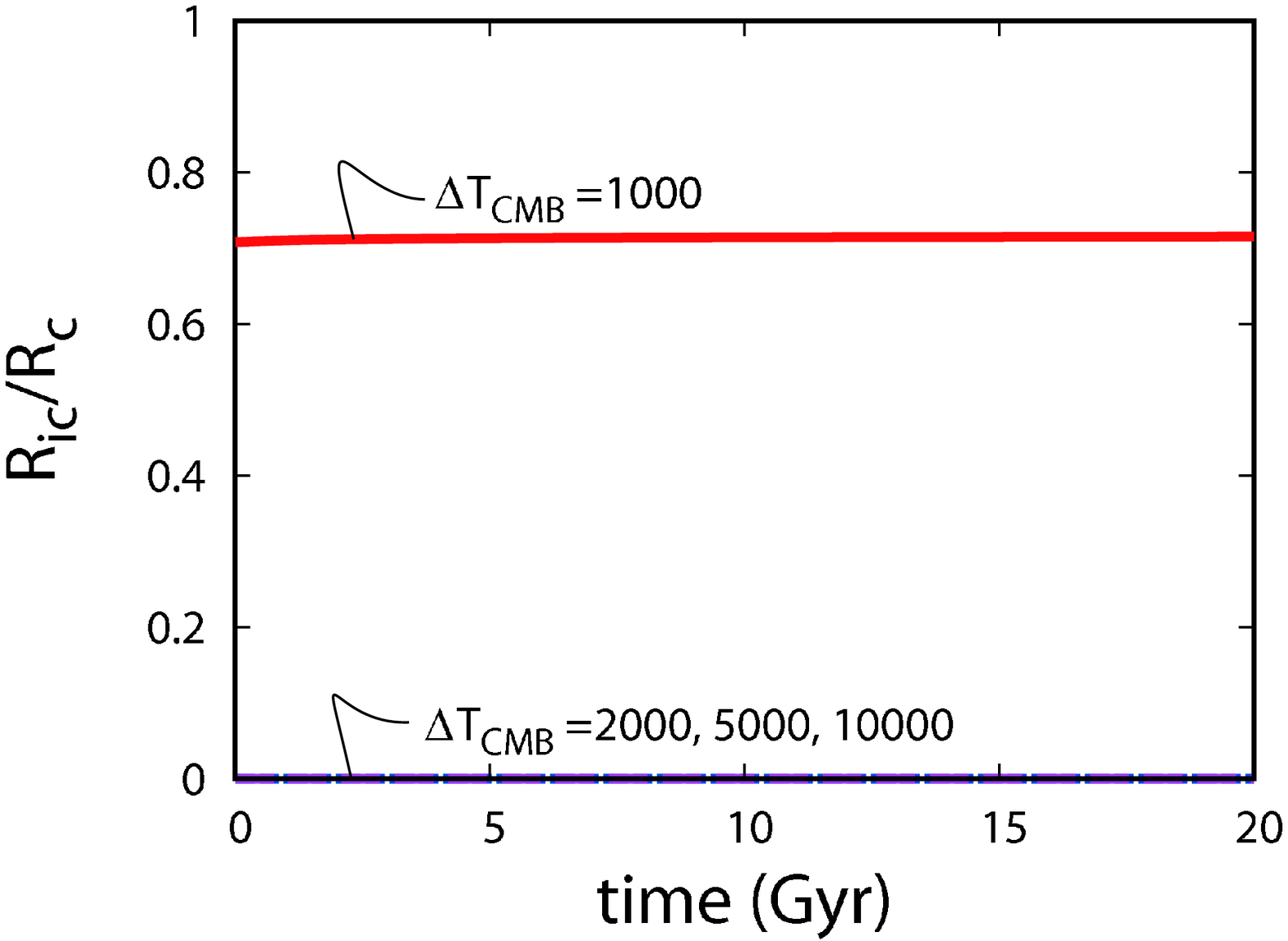}\\
(d)\\
\includegraphics[scale=0.3, angle=0]{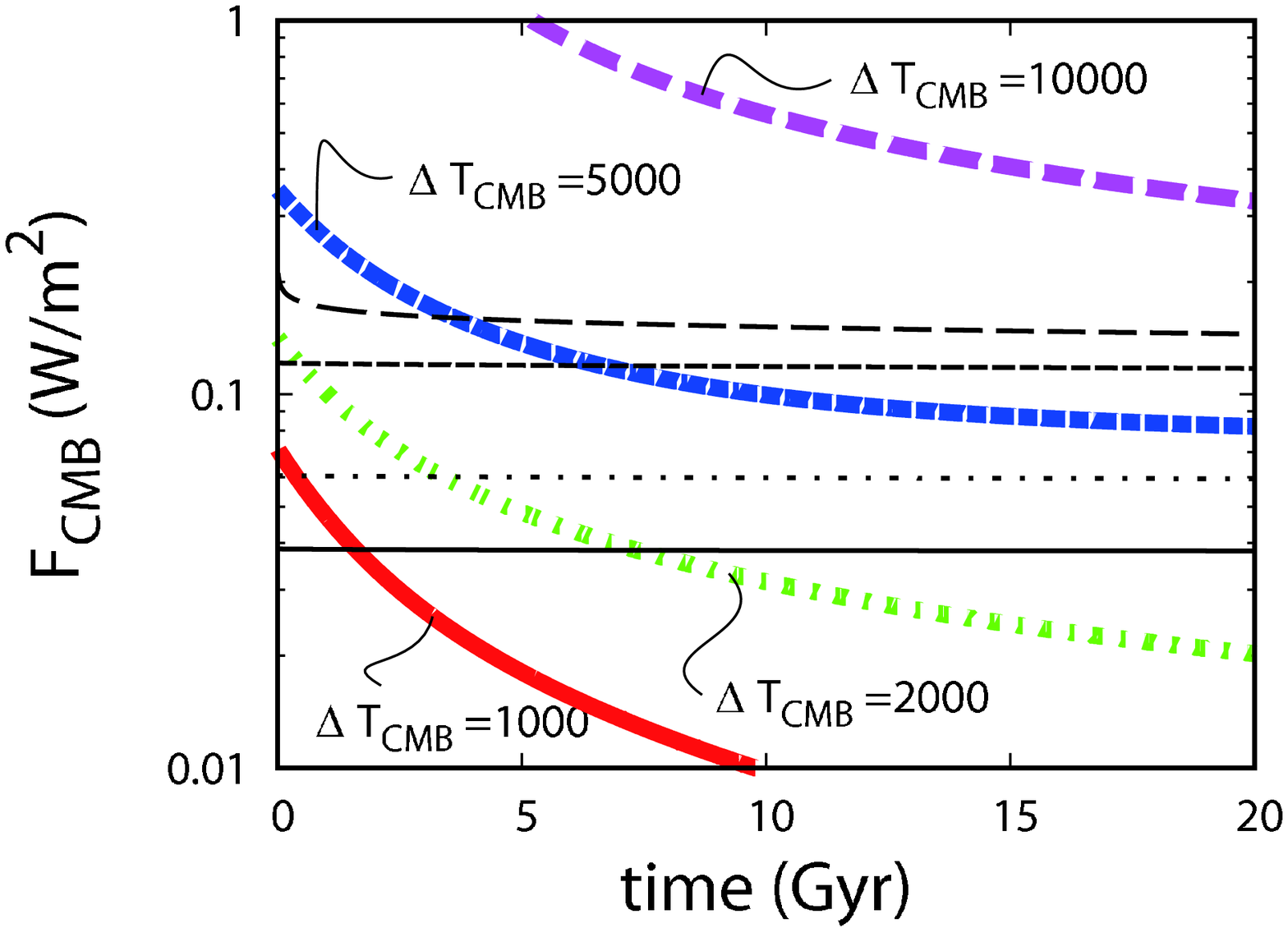}
\includegraphics[scale=0.3, angle=0]{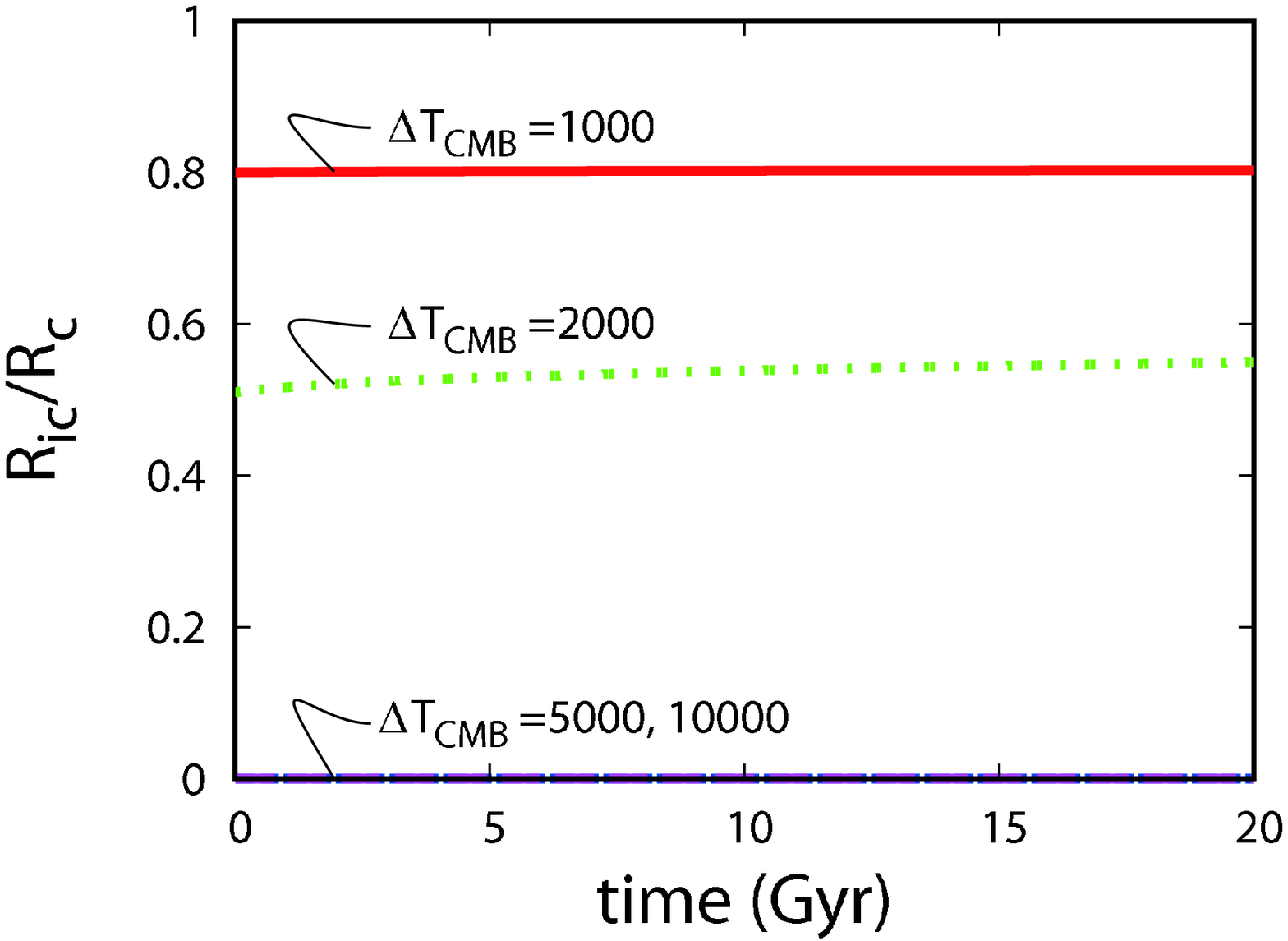}
\end{flushleft}
\caption{\scriptsize Evolution of the core heat flux (left column) and inner core radius (right column) for 
$M = $ (a)1, (b) 2, (c) 5, (d) and 10 $M_{\oplus}$ with various initial $\Delta T_{\rm CMB}$.
In the left column, $F_{\rm crit}$ for each $\Delta T_{\rm CMB}$ is expressed by thinner line with the same type.
Some lines with different initial $\Delta T_{\rm CMB}$ are overlapped by each other. 
For these parameters, the initial conditions do not affect the evolution since self-regulation of mantle heat transfer works
due to temperature dependence of the mantle viscosity.
Inner cores never nucleate in the cases of $\Delta T_{\rm CMB}=2000, 5000,$ and $10000$K for $M_p=5M_{\oplus}$ and $\Delta T_{\rm CMB}=5000$ and 10000K for $M_p=10M_{\oplus}$.
}
\label{fc_12510Me}
\end{figure}

As discussed in the above, 
thermal evolution of super-Earths differs from
that of Earth-mass planets in many aspects.
Here, we focus on evolution of heat flux through CMB, $F_{\rm CMB}$,
and inner core radius, $R_{\rm ic}$,
in order to study magnetic activity of super-Earths.
Figures~\ref{fc_12510Me} show 
the evolution of $F_{\rm CMB}$ (left column)
and $R_{\rm ic}$ (right column) 
with various initial $\Delta T_{\rm CMB}$ for the case of
(a) $M_p = 1 M_{\oplus}$, 
(b) $2 M_{\oplus}$,
(c) $5 M_{\oplus}$,
and (d) $10 M_{\oplus}$.
Solid, dot, dashed, and long-dashed lines 
represent the results with
initial $\Delta T_{\rm CMB} = 1000$K, 2000K, 5000K and 10000K,
respectively.
In all cases, $V^* = 10 \times 10^{-6} {\rm m^3/mol}$. 

The results in the left column show
that $F_{\rm CMB}$ is generally higher for 
higher initial $\Delta T_{\rm CMB}$.
The dependence is more pronounced for relatively large $M_p$ cases.
For $M_p = 1~M_{\oplus}$, the dependence is very weak.
We found the dependence is also very weak for $M_p < 1~M_{\oplus}$.
For $M_p \la 1~M_{\oplus}$, the temperature dependence of the
mantle viscosity (Eq.~[\ref{eq:viscosity}]) dominates over the
pressure dependence.
Then, $F_{\rm CMB}$ is high when the core temperature is high,
and $F_{\rm CMB}$ declines as the core cools.
Thus, the heat flux is self-regulated to be
quickly relaxed independent of the initial values.
On the other hand, as will be shown later,
when $\Delta T_{\rm CMB} \la 1000(M_p/M_{\oplus})$,
the pressure dependence is more effective.
Then, the self-regulation does not work and 
the dependence of $F_{\rm CMB}$ on initial 
$\Delta T_{\rm CMB}$ is retained for more than 20 Gyrs.
The threshold flux for driving dynamo action is
marked by an black lines in each case.
The decline of the threshold value is due to 
decrease of core surface temperature (Eq.~[\ref{eq:F_crit}]).
The duration for $F_{\rm CMB}>F_{\rm crit}$ determines lifetime of magnetic field generation.

\citet{Papuc2008} obtained $F_{\rm CMB} \propto M_p^{2/3}$,
whereas our results shows $F_{\rm CMB}\propto M_p$ provided that $\Delta T_{\rm CMB}$ is sufficiently high.
The difference may come from the assignment 
of specific heat of the core.
\citet{Papuc2008} assumed constant specific heat of the core, $C_p=1000{\rm JkgK^{-1}}$
for all sized planets.
As we discussed in section 2.6, however, 
thermal contraction results in increase in the effective $C_p$ and
the effect is more pronounced for larger $M_p$.
In our calculations that include this effect, the core tends to cool
less efficiently and the dependency of $F_{\rm CMB}$ on $M_p$ is stronger
than that obtained by \citet{Papuc2008}.

The right column shows the growth of inner solid cores
for $\Delta T_{\rm CMB}=1000, 2000, 5000$ and 10000K.
In the case of $M_p = 1~M_{\oplus}$,
an inner core is nucleated at 2-3 Gyrs, almost independent
of initial $\Delta T_{\rm CMB}$, since the core cooling is
self-regulated.
For $M_p = 2 M_{\oplus}$, the inner core growth depends
on $\Delta T_{\rm CMB}$ for $\Delta T_{\rm CMB} \ga 2000$K.
For such high $\Delta T_{\rm CMB}$, 
since the core has larger thermal energy initially and
the heat flux is not self-regulated, 
it takes more time for the core temperature
to become below the nucleation temperature.
For $M_p \ga 5M_{\oplus}$, core hardly cools on 20 Gyrs,
the inner core does not grow from the initial state.
In these cases the inner core size is determined by a relationship between adiabatic curve and melting curve of iron.
The inner core of super-Earths ($M_p > M_{\oplus}$) have 
never nucleated for $\Delta T_{\rm CMB} = 10000$K. 
For massive planets, the increase in the viscosity due to higher pressure
is overcome only by very high initial temperature.
The high $\Delta T_{\rm CMB}$ also delays nucleation of the inner solid core.
As a result, 
there is trade off between heat flux and inner core growth through the relation between melting point of iron core
and temperature- and pressure-dependency of mantle viscosity.

\begin{figure}[htbp]
\begin{flushleft}
(a)\hspace{5.5cm}(b)\\
\includegraphics[scale=0.3, angle=0]{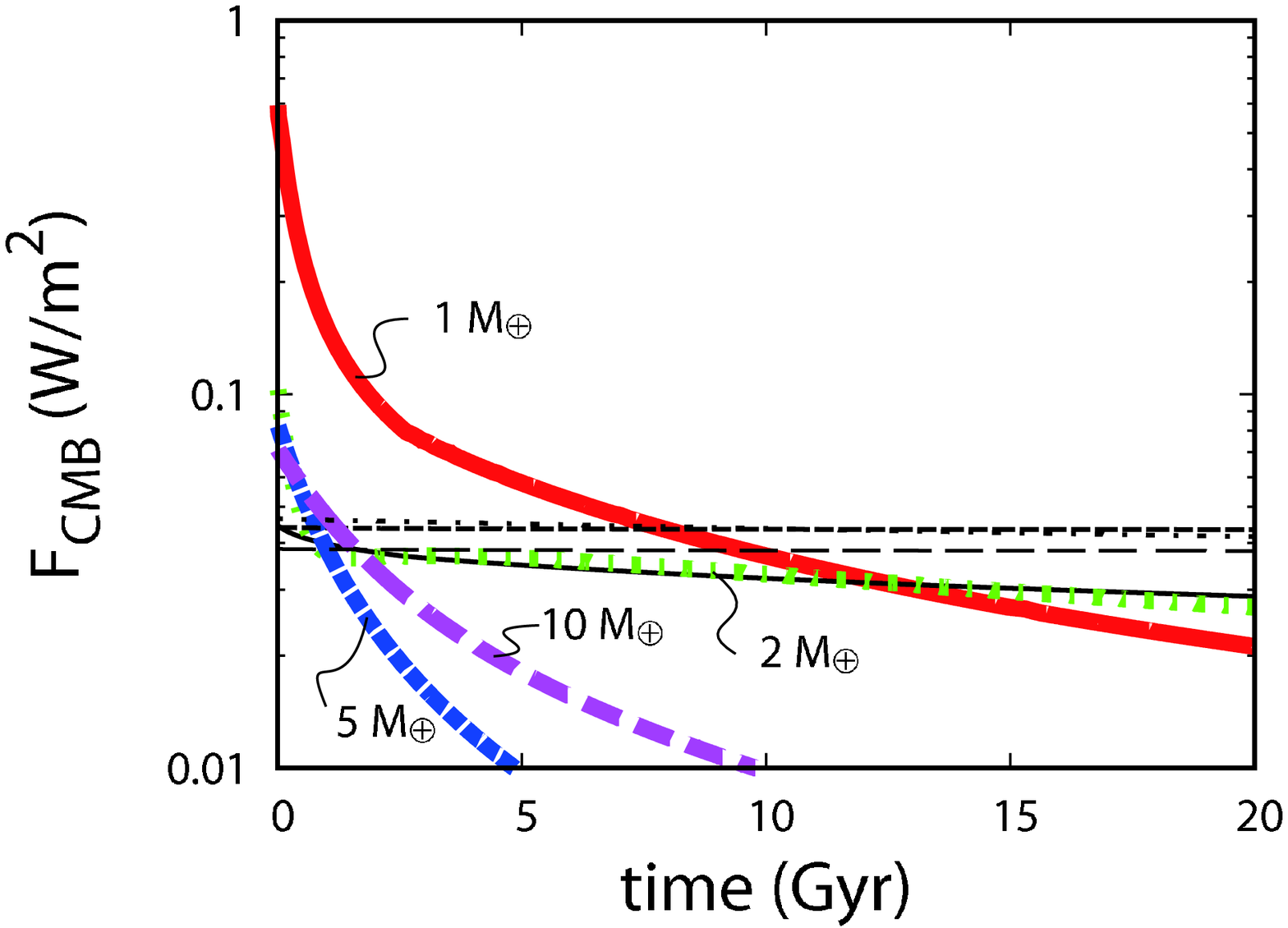} 
\includegraphics[scale=0.3, angle=0]{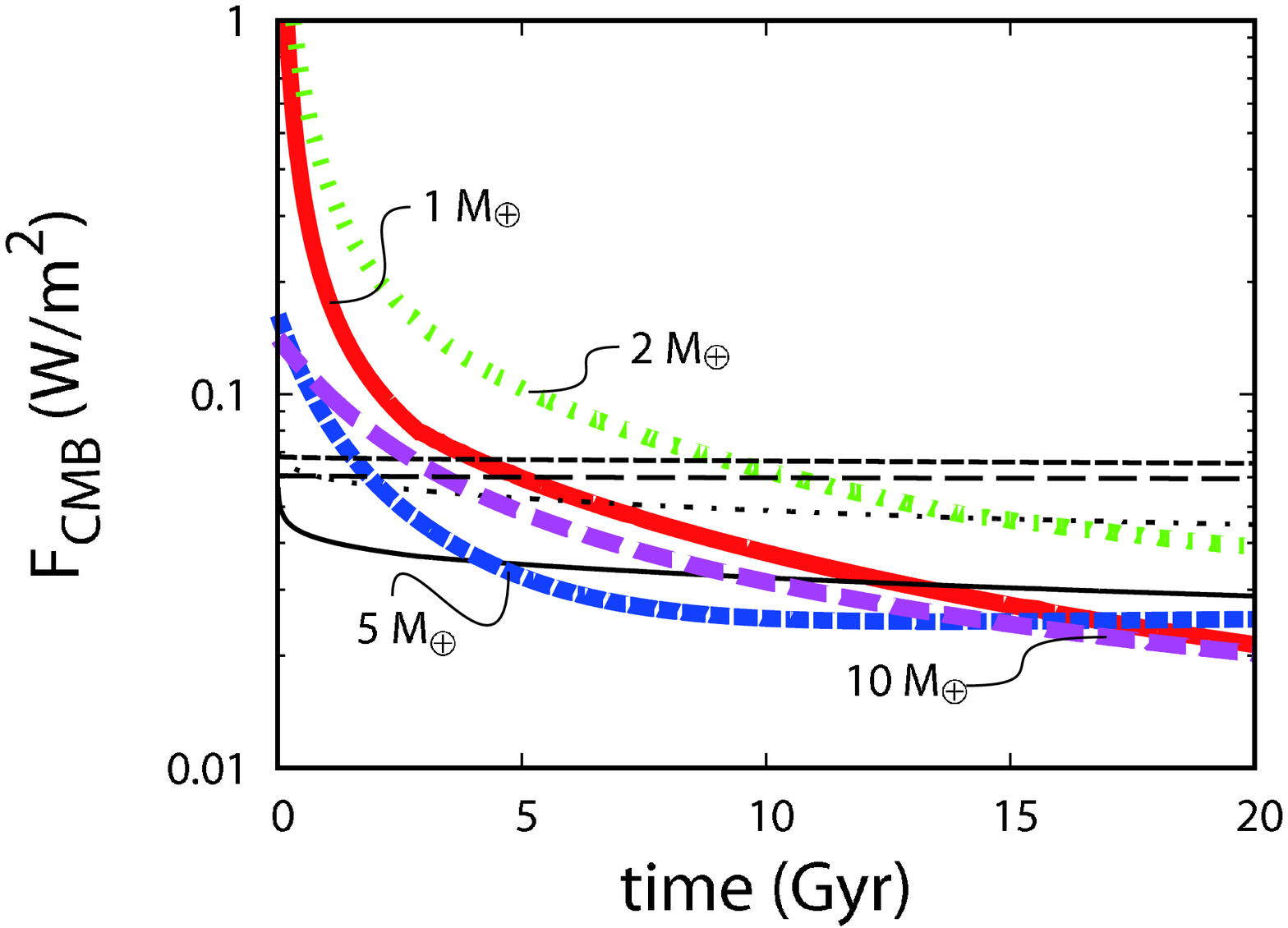} \\
(c)\hspace{5.5cm}(d)\\
\includegraphics[scale=0.3, angle=0]{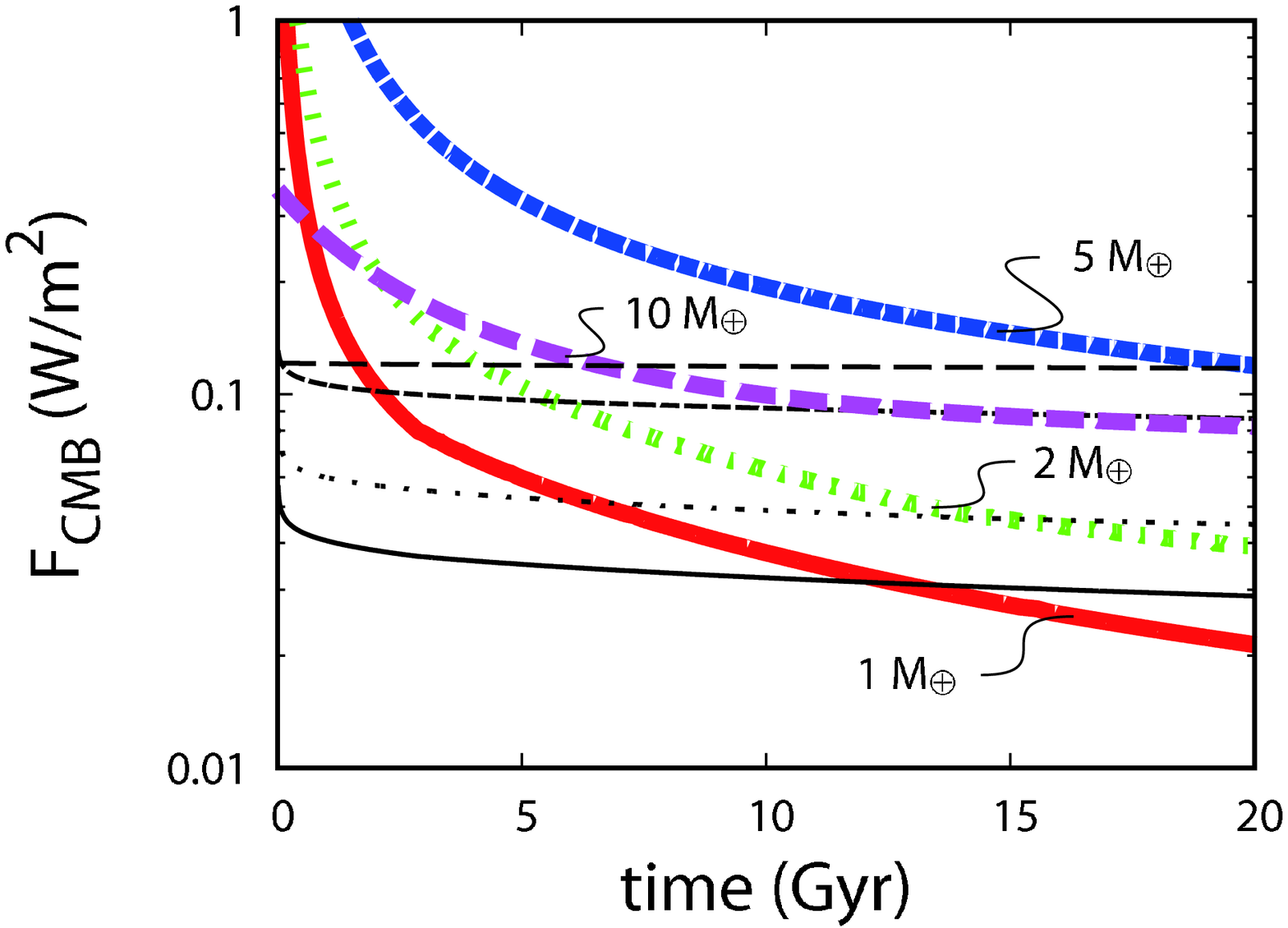} 
\includegraphics[scale=0.3, angle=0]{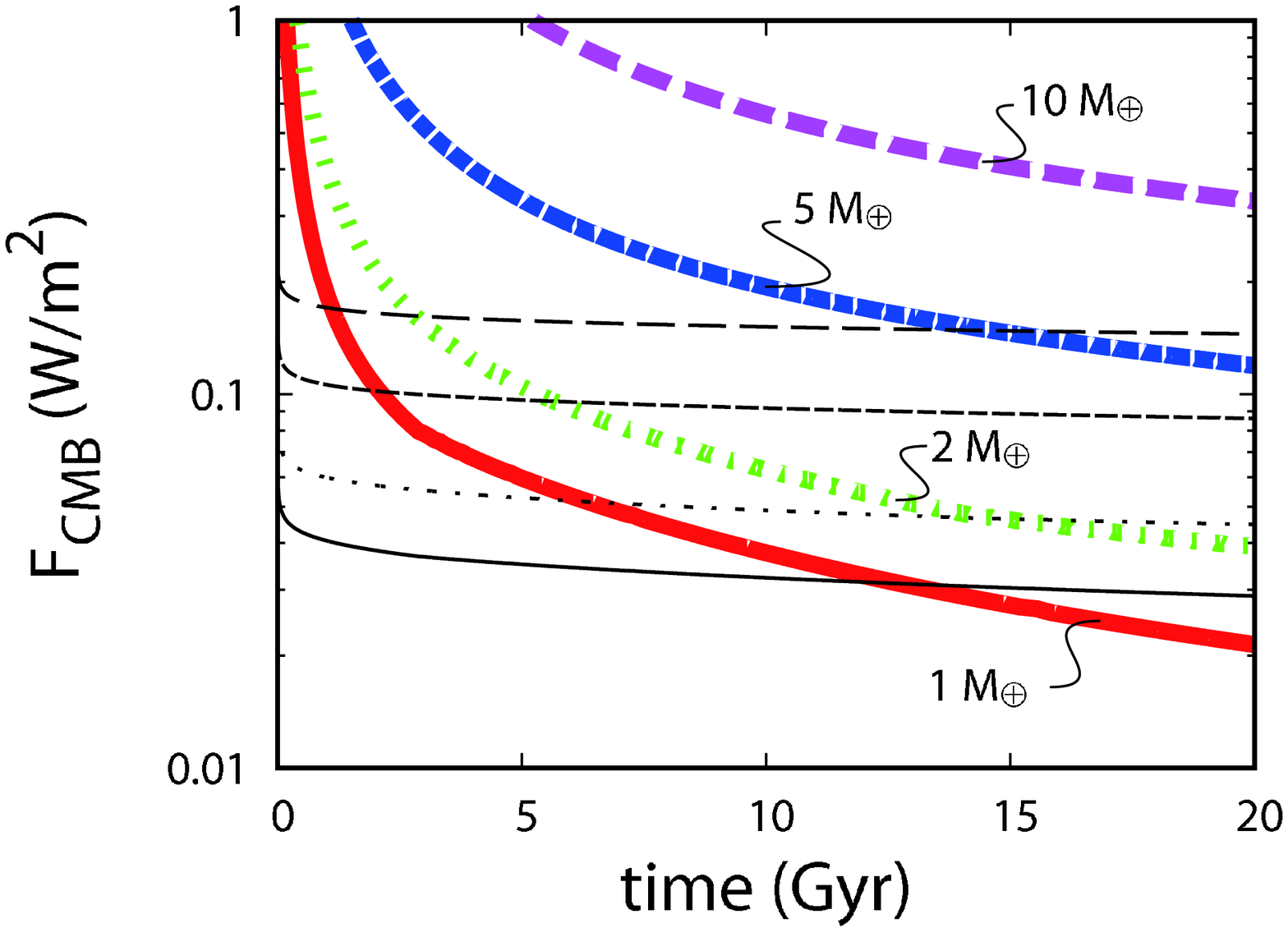}
\end{flushleft}
\caption{Evolution of the core heat flux for 
$\Delta T_{\rm CMB}$ =  (a)1000K, (b) 2000K, (c) 5000K, (d) 10000K with $M_{p} = 1, 2, 5, 10 M_{\oplus}$, respectively.}
\label{fig:fluxcomptemp}
\end{figure}

Figures~\ref{fig:fluxcomptemp} show dependence of
evolution of $F_{\rm CMB}$ on $M_p$ for
fixed values of $\Delta T_{\rm CMB}$.
For $\Delta T_{\rm CMB} = 1000$K,
we have already mentioned that
$F_{\rm CMB}$ is rather lower for $M_p = 5 M_{\oplus}$
than for $M_p = M_{\oplus}$ 
(Figs.~\ref{fig:5Me}b and \ref{fig:Eartha1whole}b),
because the increase in the viscosity due to higher pressure
dominates the decrease due to higher temperature 
for $M_p = 5 M_{\oplus}$.
This trend is clearly shown in Fig.~\ref{fig:fluxcomptemp}a.

However, this is not always the case.
If the core temperature is high enough (in other words,
$\Delta T_{\rm CMB}$ is high enough), or if pressure
is low enough ($M_p$ is small enough), the viscosity
should decrease with increase in $M_p$ due to the 
temperature effect.
For $\Delta T_{\rm CMB} = 10000$K (Fig.~\ref{fig:fluxcomptemp}d),
$F_{\rm CMB}$ is approximately proportional to $M_p$.
Even for $\Delta T_{\rm CMB} = 1000~\mbox{K}$,
$F_{\rm CMB}$ increases with $M_p$ for 
low mass regime ($M_p < 1 M_{\oplus}$).
Thus, $F_{\rm CMB}$ has a peak at some value of $M_p$ for
a given value of $\Delta T_{\rm CMB}$.
Figures~\ref{fig:fluxcomptemp}b and c show that
the critical planet mass ($M_{p,c}$) 
at which $F_{\rm CMB}$ takes the maximum value
is $2 M_{\oplus}$ for $\Delta T_{\rm CMB} = 2000$K 
and $5 M_{\oplus}$ for 5000K.
We empirically found that
\begin{equation}
M_{p,c} \simeq \frac{\Delta T_{\rm CMB}}{1000{\rm K}} M_{\oplus}.
\end{equation}

\begin{figure}[htbp]
\begin{flushleft}
(a)
\includegraphics[scale=0.3, angle=0]{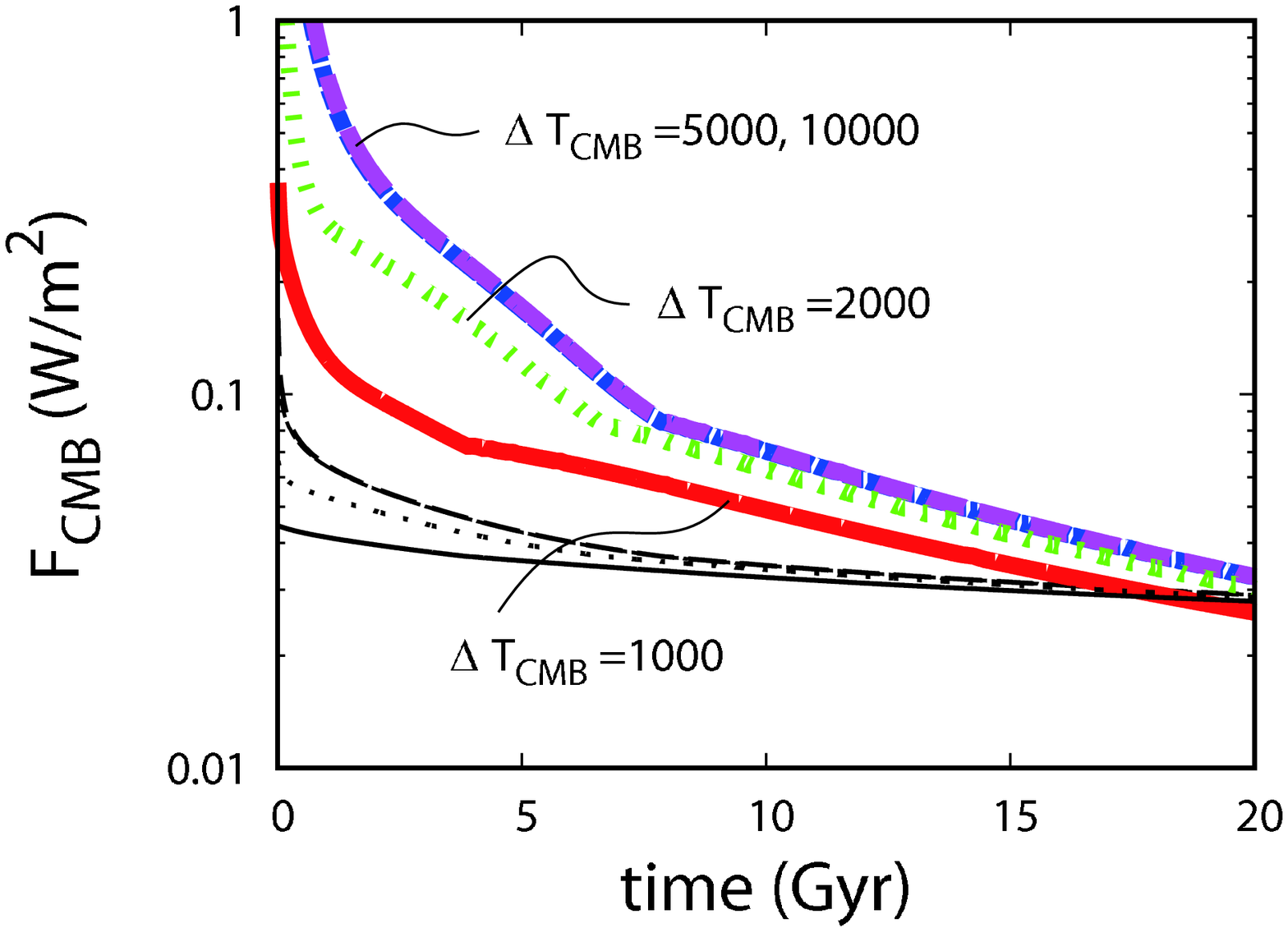} 
\includegraphics[scale=0.3, angle=0]{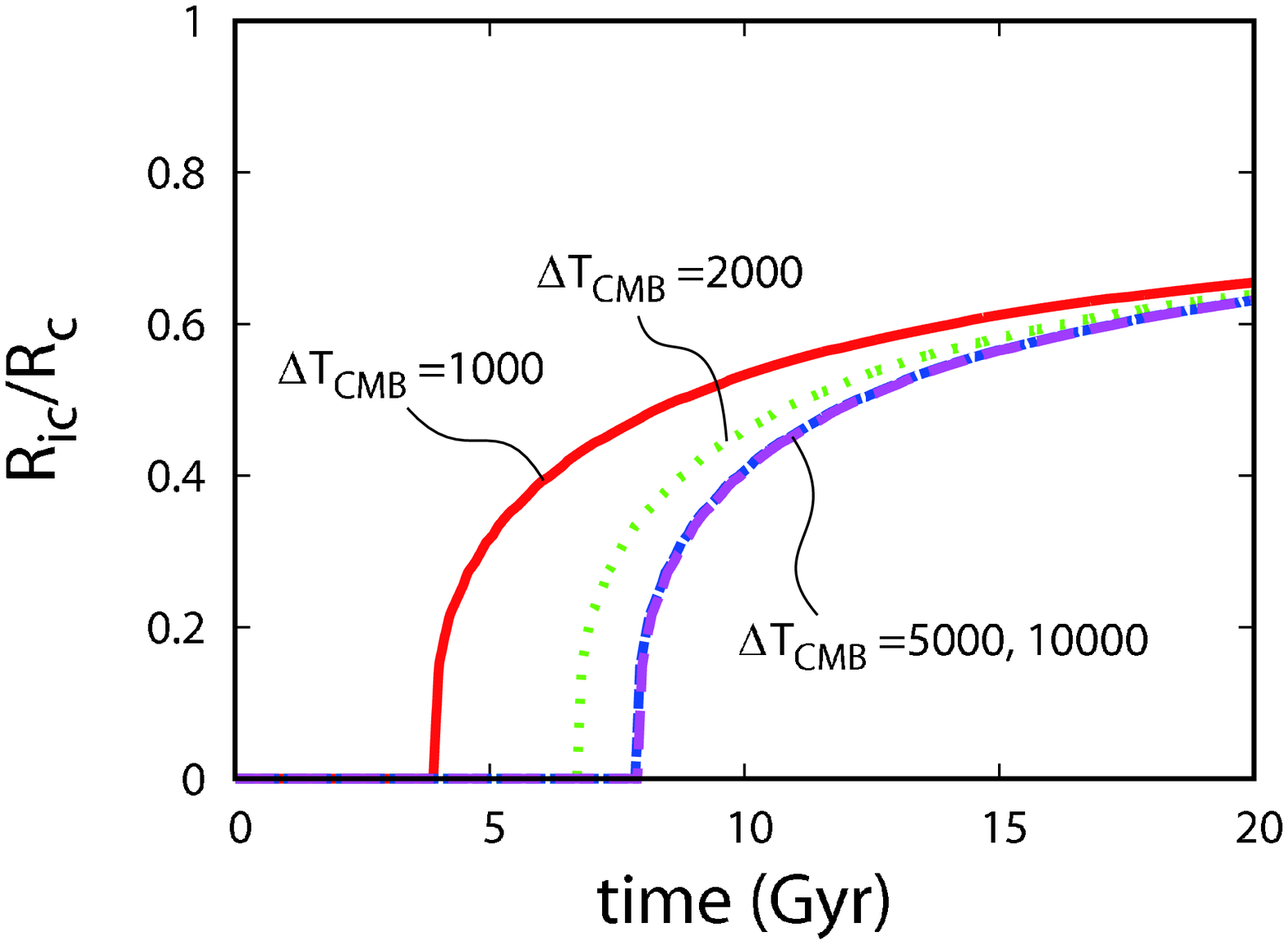}\\
(b)
\includegraphics[scale=0.3, angle=0]{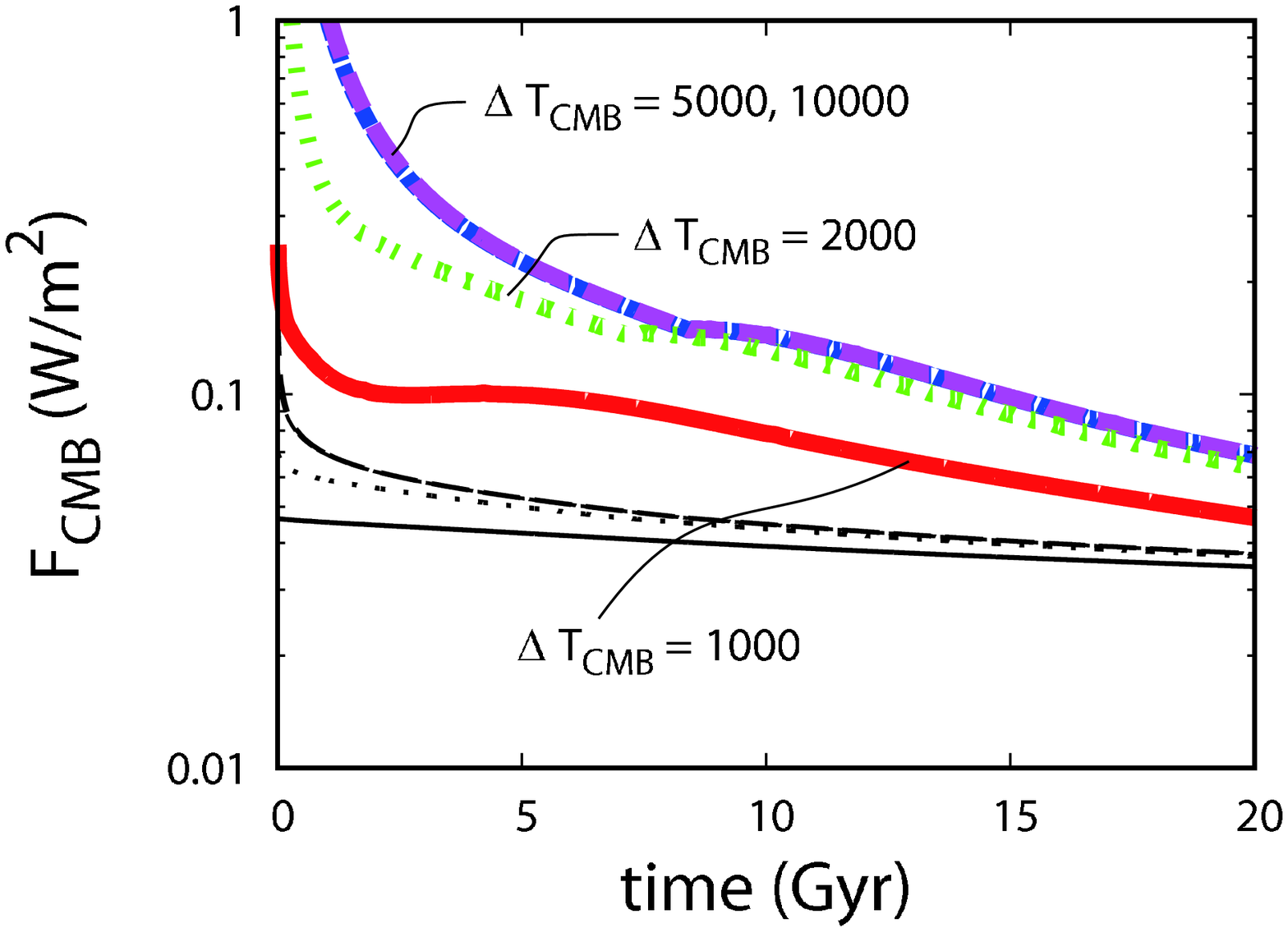} 
\includegraphics[scale=0.3, angle=0]{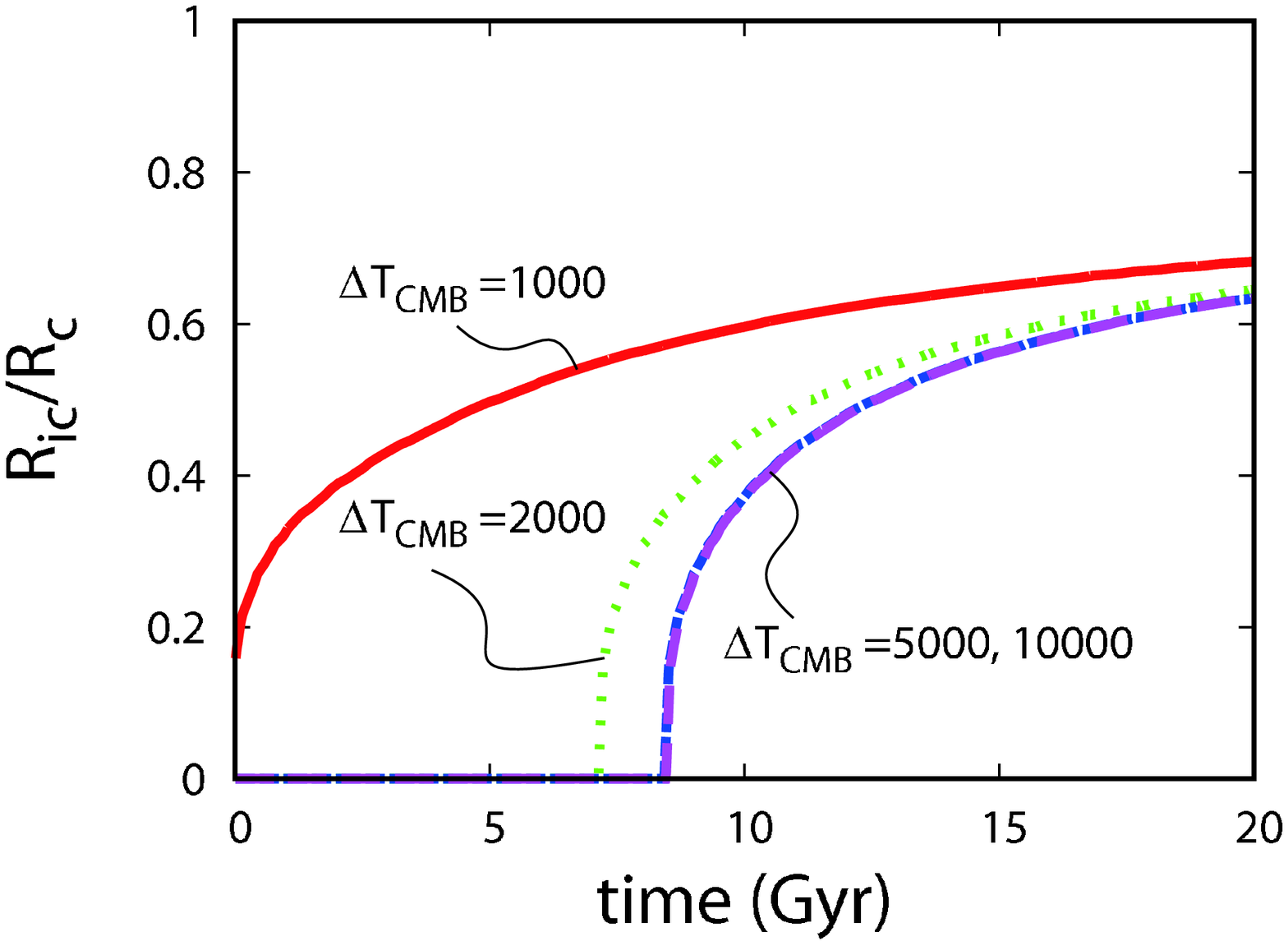}\\
(c)
\includegraphics[scale=0.3, angle=0]{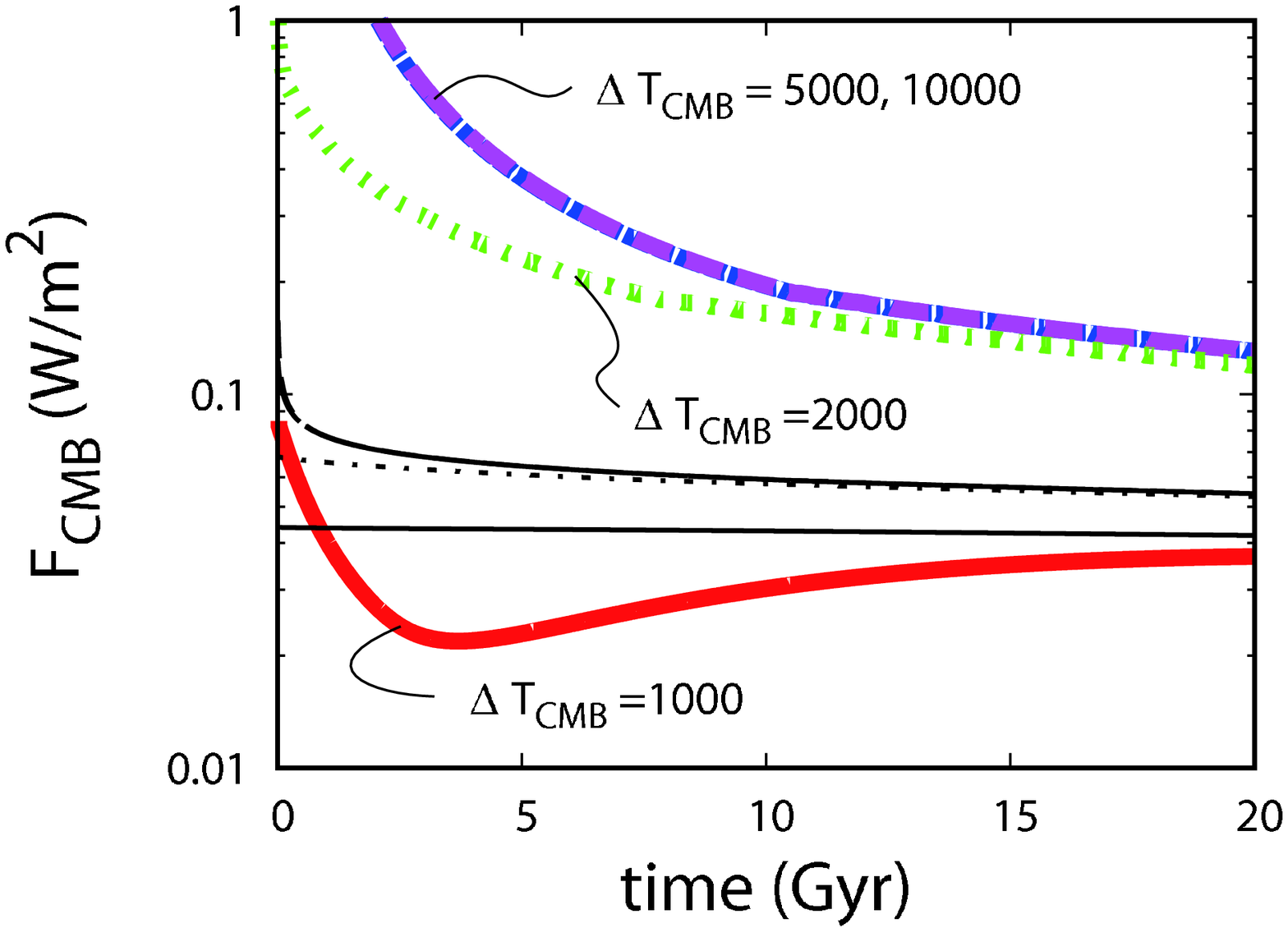} 
\includegraphics[scale=0.3, angle=0]{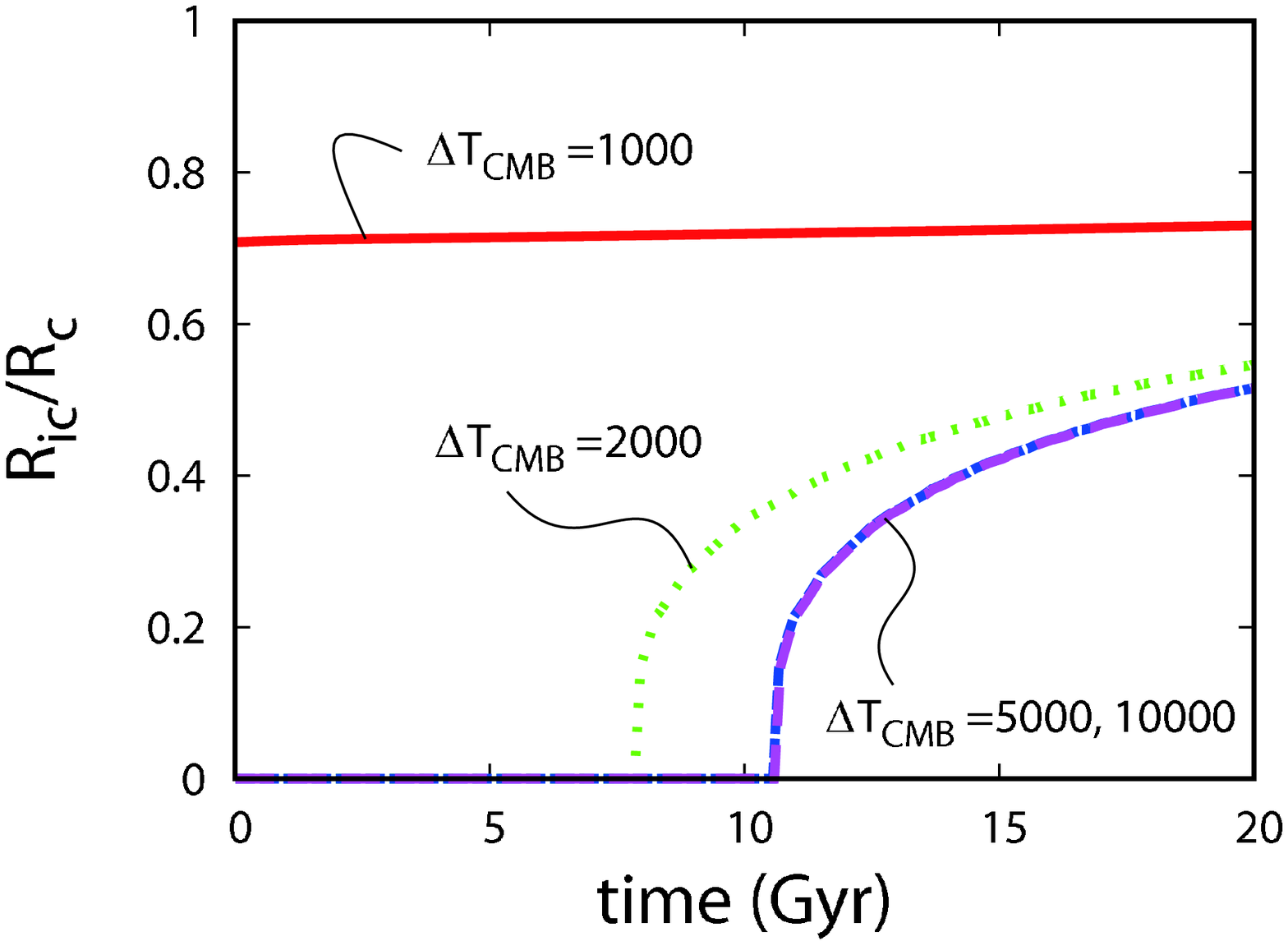}\\
(d)
\includegraphics[scale=0.3, angle=0]{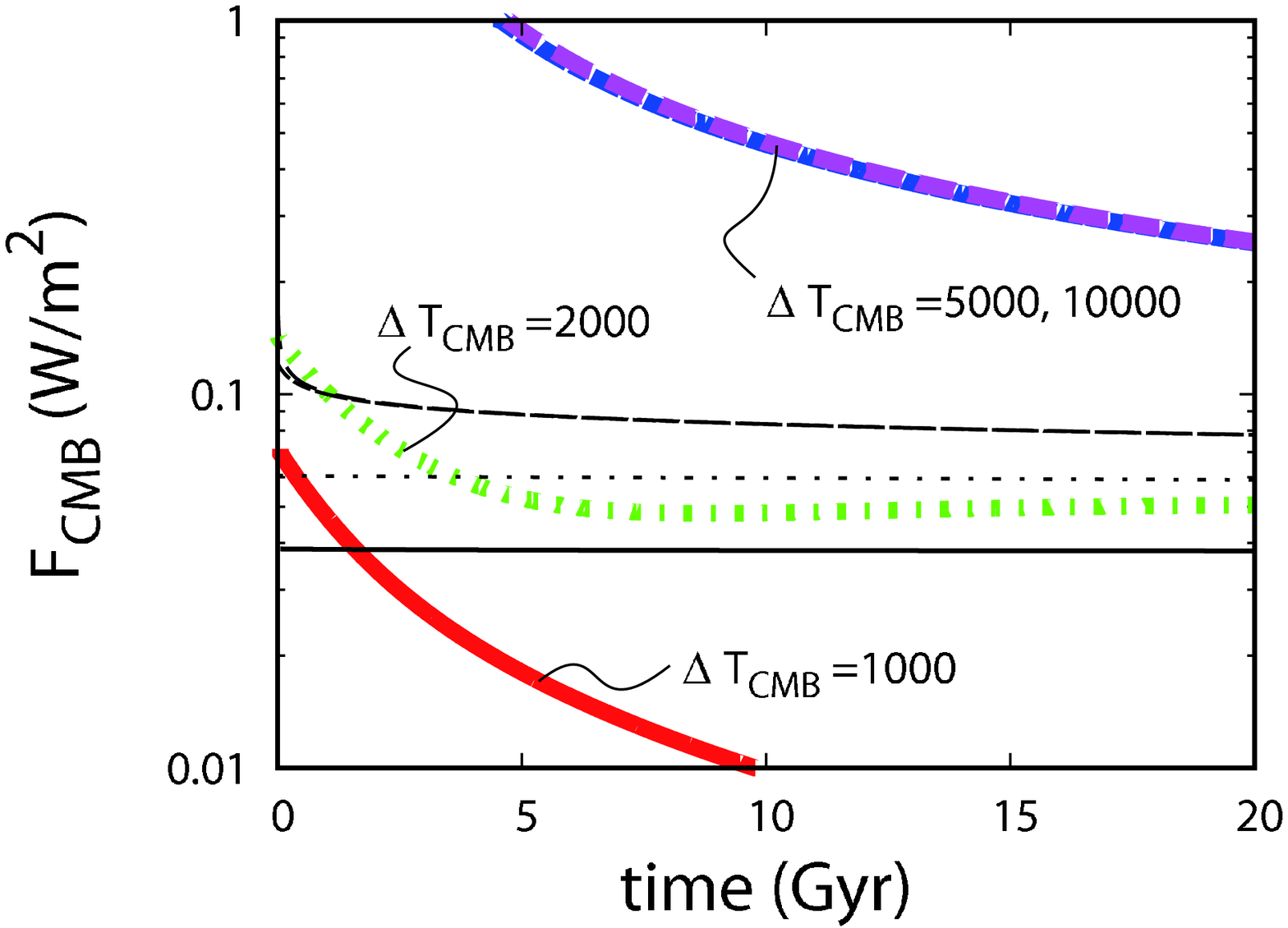}
\includegraphics[scale=0.3, angle=0]{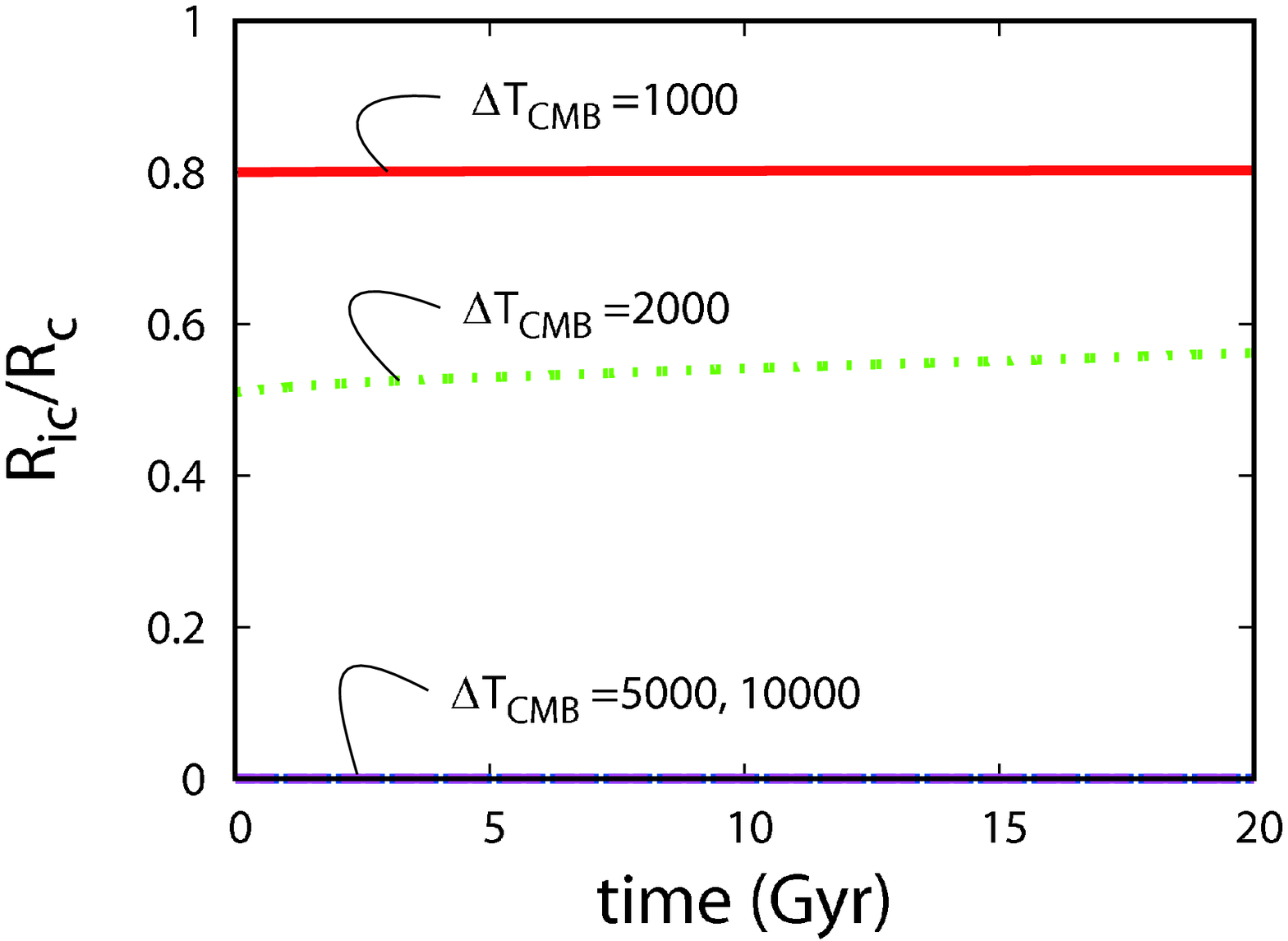}
\end{flushleft}
\caption{Same as Fig.~\ref{fc_12510Me}except $V^* = 3\times 10^{-6}{\rm m^3mol^{-1}}$}
\label{fig:actv3_fc_12510Me}
\end{figure}

Since the mantle viscosity depends on the
activation volume, $V^*$ (Eq.~[15]), and  
the values of $V^*$ may have uncertainty at high pressure,
we also performed calculations with a smaller value of $V^*$.
Figures~\ref{fig:actv3_fc_12510Me} show the results with 
$V^* = 3\times10^{-6}{\rm m^3mol^{-1}}$.
Due to the weakened pressure effect, $M_{p,c}$ is 
increased by a factor of a few.
In Figures~\ref{fig:actv3_fc_12510Me},
the viscosity is artificially increased 
(Eq.~[15] by a factor of 6000)
in order to compensate the smaller value of $V^*$
and reproduce Earth's observed values.
Note that the artificial increase does not affect $M_{p,c}$.

The critical planetary mass $M_{p,c}$ is approximately
derived by the $M_p$-dependence of the mantle viscosity at CMB.
Since we empirically found that
$P_{\rm CMB} \sim (M_p/M_{\oplus})P_{\oplus {\rm CMB}}$ 
and $T_{\rm CMB} \sim 5\Delta T_{\rm CMB}$,
Eq.~(\ref{eq:viscosity}) is reduced to
\begin{equation}
\eta_{\rm CMB}(M_p, T_{\rm CMB}) \propto
\exp\left(\frac{E^*+\left(\frac{M_p}{M_{\oplus}}\right)P_{\oplus}V^*}{nRT_{\rm CMB}}\right).
\end{equation}
When the argument of exponential is larger than unity,
the viscosity rapidly increases with $M_p$ to depress $F_{\rm CMB}$,
because $F_{\rm CMB} \propto \eta^{-1/3}$.
For $M_p < M_{p,c}$, we found that $F_{\rm CMB} \propto M_p$.
When the argument exceeds some critical value ($C > 1$),
the viscosity enhancement eventually overwhelms the factors for
the positive $M_p$-dependence of $F_{\rm CMB}$. 
Thus, $M_{p,c}$ is given by the value of $M_p$ with which
the argument of exponential is $\simeq C$, 
\begin{eqnarray}
M_{p, c} 
\simeq \frac{5nCR\Delta T_{\rm CMB} - E^*}{V^*}\frac{M_{\oplus}}{P_{\oplus{\rm CMB}}} \nonumber\\
\simeq \frac{5nCR\Delta T_{\rm CMB}}{V^*}\frac{M_{\oplus}}{P_{\oplus{\rm CMB}}} \nonumber\\
\sim \frac{C \Delta T_{\rm CMB}}{10000K} 
\left(\frac{V^*}{10\times10^{-6}{\rm m^3mol^{-1}}} \right)^{-1}M_{\oplus},
\end{eqnarray}
which explains the dependences on $\Delta T_{\rm CMB}$ and $V^*$
that we found numerically.
(If we adopt $C \sim 10$, the numerical factor is also explained.)

\subsection{Lifetime of intrinsic magnetic fields}

\begin{figure}[htbp]
\begin{flushleft}
(a)\hspace{5.5cm}(b)\\
\includegraphics[scale=0.3, angle=0]{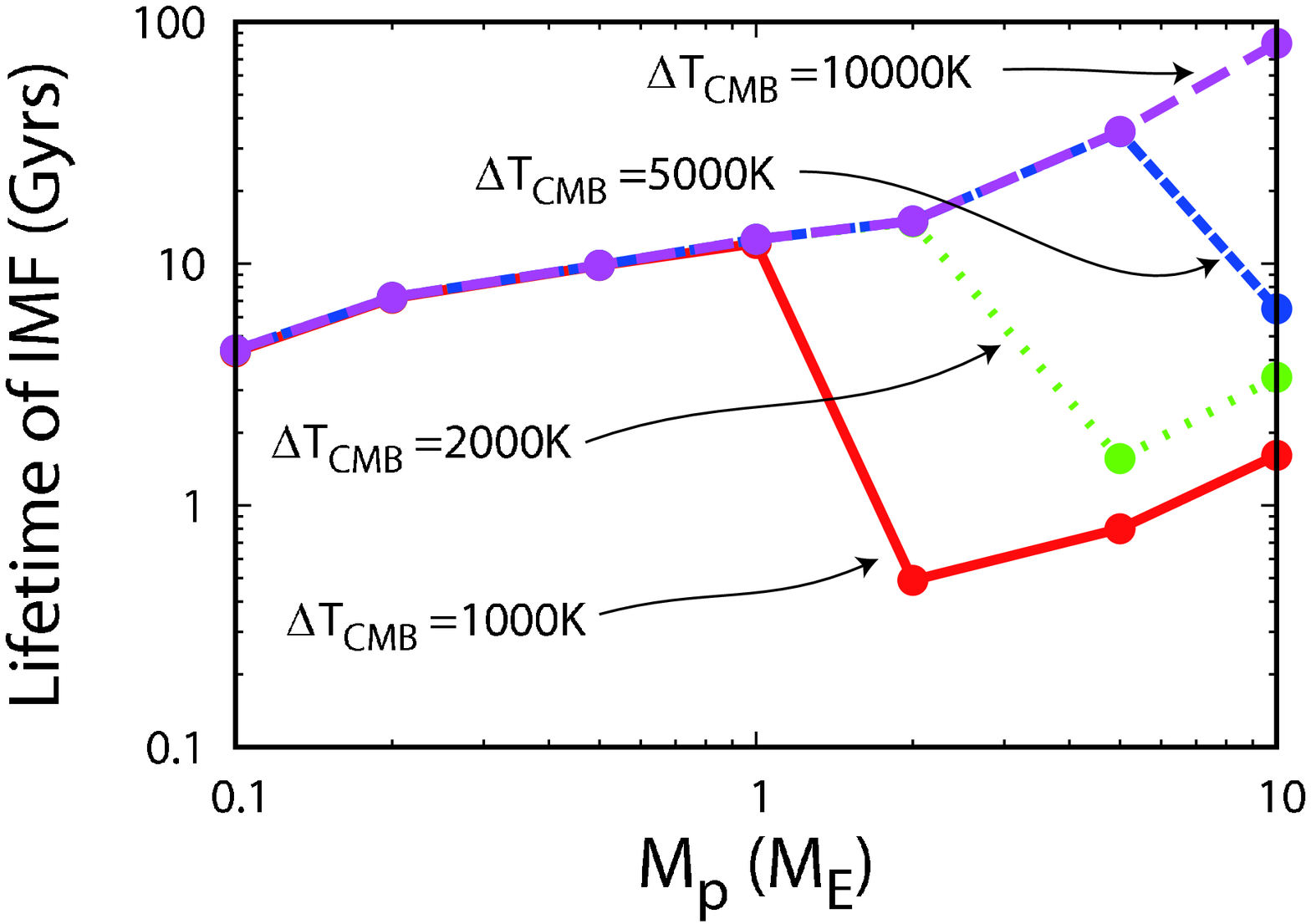} 
\includegraphics[scale=0.3, angle=0]{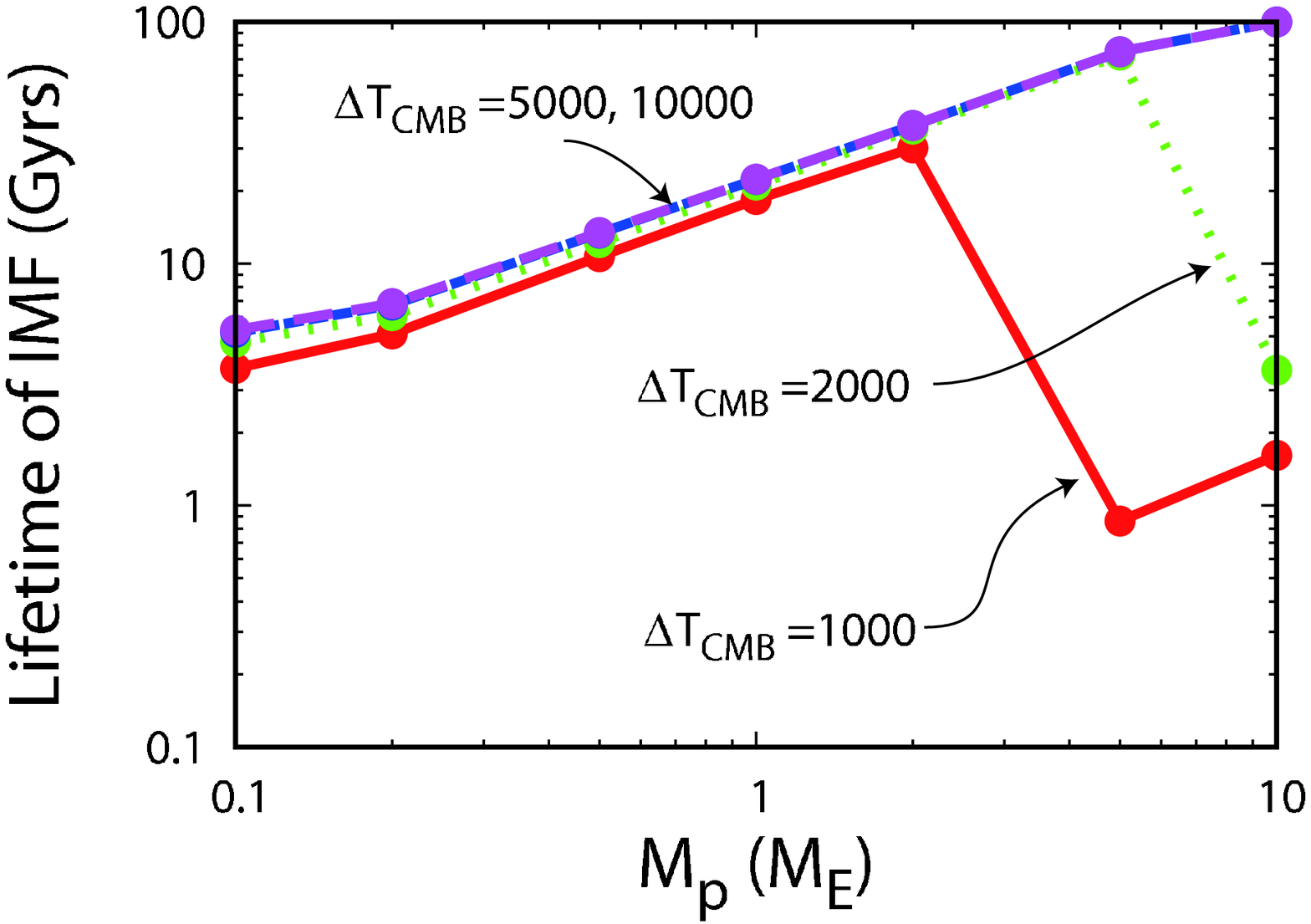} 
\end{flushleft}
\caption{Lifetime of magnetic fields as a function of planetary mass ($M_p$) with various $\Delta T_{\rm CMB}$ for (a) $V^*=10\times 10^{-6}{\rm m^3mol^{-1}}$ and (b) $V^*=3\times 10^{-6}{\rm m^3mol^{-1}}$.}
\label{ltcompmass}
\end{figure}

The lifetime of the intrinsic magnetic fields 
is calculated for $\Delta T_{\rm CMB}=1000, 2000, 5000$ and 10000 K
with a fixed value of 
$V^* = 10\times10^{-6}{\rm m^3mol^{-1}}$. 
The results are summarized in Fig.~\ref{ltcompmass}.
It is clearly shown that 
the lifetime declines for $M_p \ga M_{p,c}$ by
the increase in the mantle viscosity due to 
the pressure effect that we discussed in details in the
previous subsection.
The results for $V^* = 3\times10^{-6}{\rm m^3mol^{-1}}$
show a similar property. 

In Fig.~\ref{ltcompmass}, the lifetime weakly increases with $M_p$
for $M_p \la M_{p,c}$.
The dependence is explained as follows.
The lifetime is approximately given 
by $\tau_{\rm life} \sim E_{\rm th}/(S_{\rm CMB} F_{\rm CMB})$,
where $E_{\rm th}$ is thermal energy of the core
and $S_{\rm CMB}$ is surface area of the core.
According to our calculation,
$S_{\rm CMB} \propto M_p^{1/2}$ rather than $M_p^{2/3}$ 
due to self-compression.
Figures~7 and 13 show that
$F_{\rm CMB} \propto M_p$ and 
$E_{\rm th} \propto M_p^2$ for a fixed $T_{\rm CMB}$. 
As we mentioned in section 3.3, $T_{\rm CMB} \sim 5 \Delta T_{\rm CMB}$.
Thus, for a fixed $\Delta T_{\rm CMB}$,
it is predicted that
$\tau_{\rm life} \sim E_{\rm th}/(S_{\rm CMB} F_{\rm CMB})
\propto M_p^{2 - 1/2 - 1} = M_p^{1/2}$,
which is consistent with the numerical results in Fig.~\ref{ltcompmass}.

When $M_p > M_{p,c}$, 
the higher mantle viscosity due to the effect of higher pressure
significantly depresses heat transfer at the bottom of the mantle. 
The suppressed heat flux cannot 
maintain the vigorous core convection. 
As a result, the magnetic field lifetime is rather shorter for 
$M_p > M_{p,c}$.

Figure~\ref{ltcompmass} also shows 
that the lifetime for $M_p < M_{\rm p,c}$ does not depend
on $\Delta T_{\rm CMB}$ at all.
The initial temperature is high enough to 
overcome the pressure-dependency even for the case of
$\Delta T_{\rm CMB} = 1000$ K, resulting in the effective
self-regulation of $F_{\rm CMB}$.
As a result, the lifetime does not depend on the 
initial value of $\Delta T_{\rm CMB}$.

\subsection{Strength of magnetic fields}

The strength of magnetic fields is as important as their lifetime to
discuss habitability of the planets. 
Here we evaluate the strength of 
the magnetic fields, using the scaling law derived by \citet{Christensen2009}, 
For planets with sufficiently rapid spins,
\citet{Christensen2009} derived the magnetic strength at the core surface as
\begin{equation},
B_c \sim 0.5\mu_0^{1/2}\bar{\rho}_c^{1/6}F_{\rm conv}^{1/3},
\end{equation}
where $\bar{\rho}_c$ is average density of the core and $\mu_0$ is permeability.
If the magnetic moment is dipole-dominant
and the dipole moment is $\propto 1/r^3$ \citep{Gaidos2010}, 
the strength of magnetic dipole at the planetary surface is
$B_{\rm surf}=B_c (r_c/r_p)^3$.

\begin{figure}[htbp]
\begin{flushleft}
(a)\hspace{5.5cm}(b)\\
\includegraphics[scale=0.3, angle=0]{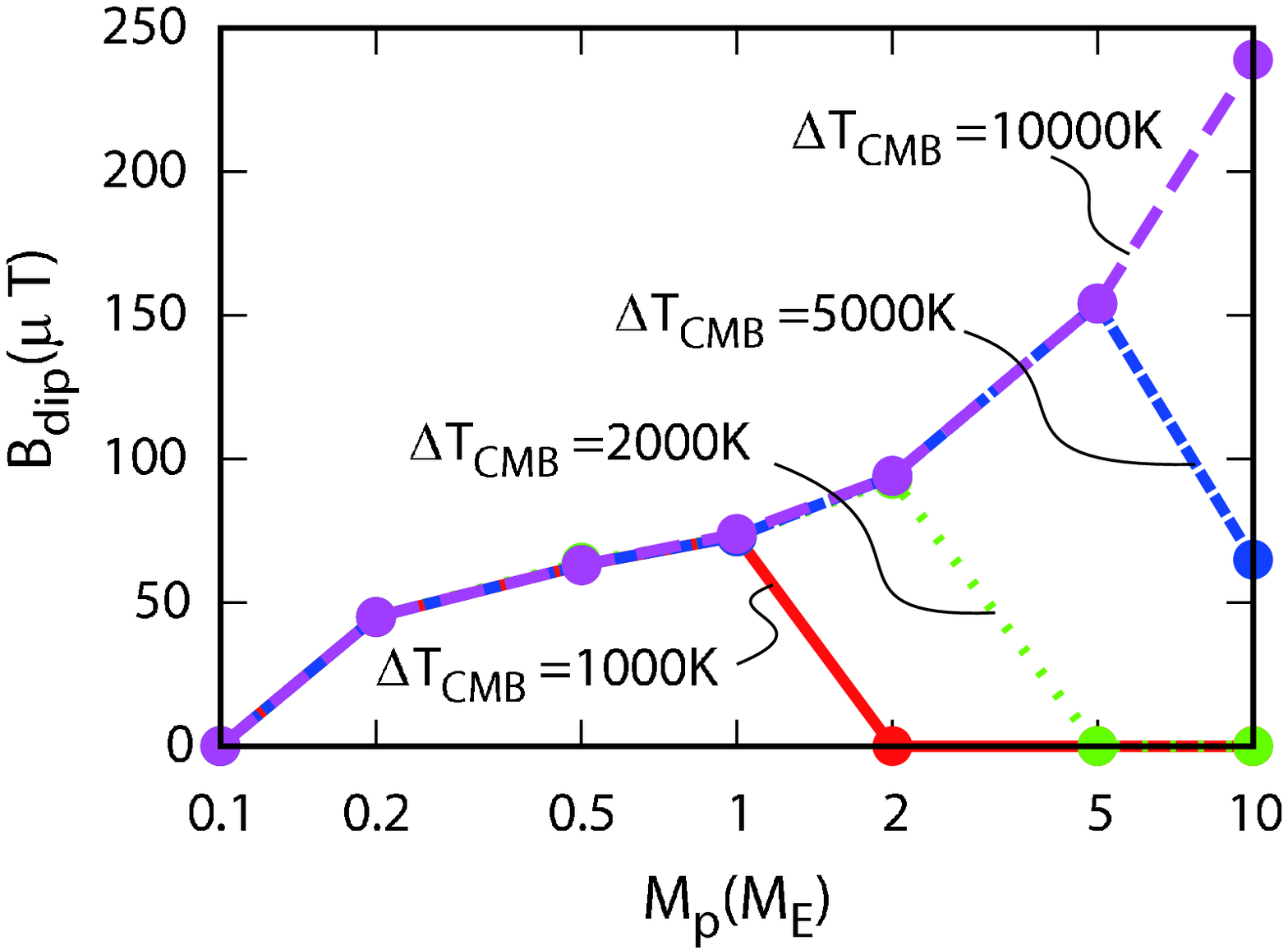} 
\includegraphics[scale=0.3, angle=0]{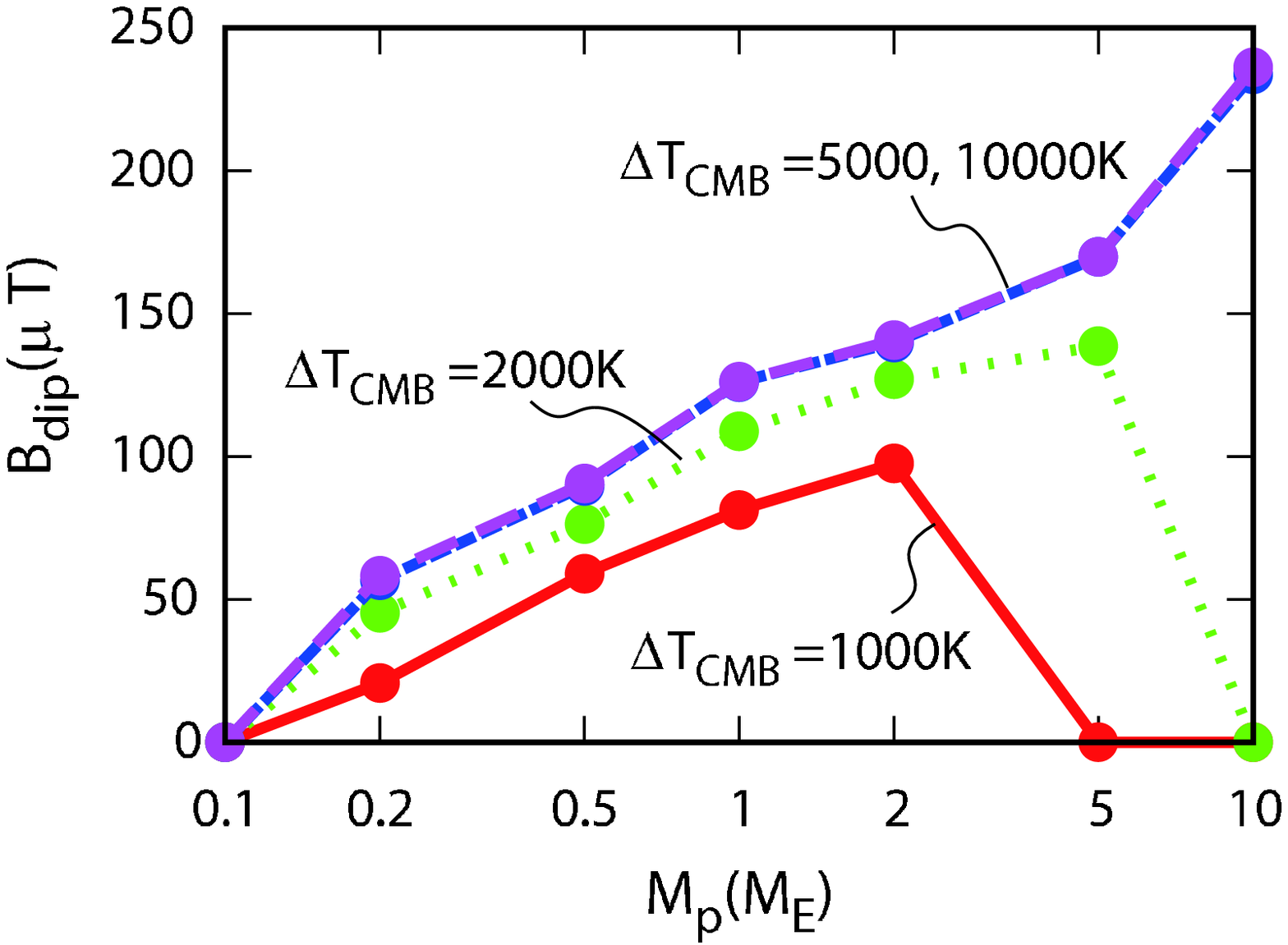} 
\end{flushleft}
\caption{Strength of magnetic fields after 5 Gyr as a function of planetary mass ($M_p$) with various $\Delta T_{\rm CMB}$ for (a) $V^*=10\times 10^{-6}{\rm m^3mol^{-1}}$ and (b) $V^*=3\times 10^{-6}{\rm m^3mol^{-1}}$.}
\label{strengthcompmass}
\end{figure}

With $F_{\rm conv} = F_{\rm CMB}-F_{\rm cond}$,
we calculated $B_{\rm surf}$ from our simulation results.
Figure~\ref{strengthcompmass} shows the calculated $B_{\rm surf}$ at
$t=5$Gyr for various $M_p$ and $\Delta T_{\rm CMB}$.
The strength monotonically increases if the initial $\Delta T_{\rm CMB}$ is sufficiently high.
The relatively weak dependence ($B\propto M_p^{1/3}$) comes
from the adopted the scaling law, 
$B\propto F_{\rm conv}^{1/3}$, and the numerically obtained relation,
$F_{\rm CMB}\propto M_p$.
If $\Delta T_{\rm CMB}$ is not high enough, the pressure effect is dominant and
the strength is significantly suppressed for $M_p \ga M_{p,c}$.

\section{Conclusion and Discussion}
We have developed a numerical model to simulate 
thermal evolution of various-mass terrestrial planets
in habitable zones.
The density distribution of the planetary interior
is calculated by Vinet EOS taking into account 
pressure dependence.
Using the interior structure model, we calculate
heat transfer through mantle,
using the astrophysical mixing length theory 
modified to mantle convection.
The modified mixing length theory is easily
applied to multi-layer convection that may be dominated
convection mode in super-Earths.
We have calibrated the modified mixing length theory with
the conventional parametrized convection model and
the boundary layer theory, in simple one-layer convection cases.

With nominal parameters of 
surface temperature $T_{\rm surf} = 300$K,
a mantle-core mass ration $\zeta_{m/c}=7:3$, 
initial core impurity $x_{\rm 0S}$ of 10 wt\%, 
and initial temperature gap at CMB $\Delta T_{\rm CMB}=1000$K,
our model for $M = 1 M_{\oplus}$
reproduces surface heat flow and inner core radius
of the present Earth.
With different parameter values suitable for Mercury,
Venus, and Mars, our model also reproduces the results
that are not inconsistent with present magnetic activity
of these planets.

With this model, we calculated thermal evolution of
terrestrial planets with mass $M_p = 0.1$--$10M_{\oplus}$
in habitable zones, using the nominal parameters, 
to study lifetime of intrinsic magnetic field that is
one of the important factors for the planets to be habitable. 
We found from the numerical calculations
that the lifetime is maximized at
\begin{equation}
M_{p, c} \sim \frac{\Delta T_{\rm CMB}}{1000K}
\left(\frac{V^*}{10\times10^{-6}{\rm m^3mol^{-1}}} \right)^{-1}M_{\oplus},
\label{eq:M_pc}
\end{equation}
where $V^*$ is activation volume of mantle material.
Planets with smaller masses
cool more rapidly, so that they cannot
maintain core heat flux to generate dynamo long enough.
For $M_p > M_{p,c}$, 
the rapid increase in the mantle viscosity caused by
high pressure significantly depresses
heat transfer throughout the mantle  and hence
that in the core.
As a result, dynamo cannot last long.
Although the temperature effect tends to decrease the mantle
viscosity as planetary mass becomes large,
the pressure effect to increase the viscosity overwhelms 
the temperature effect for $M_p > M_{p,c}$. 
With the numerically obtained empirical relation,
$T_{\rm CMB} \sim 5 \Delta T_{\rm CMB}$,
we can analytically derive Eq.~(\ref{eq:M_pc})
from the Arrhenius-type formula for the mantle viscosity 
that we adopt (Eq.~[15]).

We found that while the lifetime of magnetic fields
does not depend on $\Delta T_{\rm CMB}$ for $M_p < M_{p,c}$,
it sensitively depends on $\Delta T_{\rm CMB}$ for $M_p > M_{p,c}$
because $M_{p,c} \propto \Delta T_{\rm CMB}$ (Eq.~[\ref{eq:M_pc}]).
The initial $\Delta T_{\rm CMB}$, that is, 
the initial temperature profile of planetary interior, is one of the
most uncertain parameters, 
because it highly depends on
the processes of planetary formation and differentiation of 
the planetary interior. 
As is shown by SPH simulations, if a planet undergoes giant impacts,
its metallic core is heated as high as several tens thousands K for 
$M_p \sim 1~M_{\oplus}$ \citep{Canup2004}.
On the other hand,
if a planet accreted from small planetesimals without
giant impacts, 
the initial temperature profile is determined by
the balance between gravitational energy buried by planetesimals and thermal
transfer efficiency through rocky mantle. 
The process includes crystallization of magma ocean
and depends on the mechanical property of molten mantle \citep{Abe1986,Zahnle1988,Senshu2002}. 
Thus, to evaluate the lifetime of magnetic fields, 
in particular for super-Earths that are likely satisfy $M_p > M_{p,c}$,
detailed analysis for accretion and
early thermal evolution of terrestrial planets are needed.

It is also found that higher initial temperature profile delays 
the inner core nucleation.
For super-Earths, in order to maintain magnetic field 
more than 10 Gyr, 
the initial temperature has to be high enough to
overwhelm the pressure-dependence.
However, in that case, 
the temperature of the core center never reaches 
its condensation temperature
and the inner core cannot grow.
Some geo-dynamo simulations suggest that 
the presence of the inner core stabilizes the dipole moment
of geomagnetic field \citep{Sakuraba1999}. 
It is also suggested that because thermally
driven convection is not sufficient to drive dynamo action against the ohmic dissipation within the core of Earth \citep{Gubbins2003},
the compositional convection induced by light elements
released to the outer core by solidification of 
the inner core plays an essential role in
dynamo generation \citep{Stevenson1983, Gubbins2004}.
Since our results (Figures 6) show that inner core is not nucleated
and compositional convection
does not occur for $M_p \ga 5 M_{\oplus}$,
dipole magnetic fields of super-Earths might not be stable. 

The existence of magnetic field of extrasolar planets could be 
directly detected by the polarization observation of the photon 
from transiting planets or detection of H$_3^{+}$ trapped by
the magnetic fields in the future.
Another possibility of the detection of planetary magnetic field is, 
although it is indirect, observation of composition of 
planetary atmosphere or atmospheric tail. 
If the planet has intrinsic magnetic field, its atmosphere could 
keep H$_2$O molecules for long period. 
Venus may have lost H$_2$O molecules on a short time scale 
\citep{Bullock2001}. 
Thus if water series molecules, such as H$_2$O, H$_3$O, and HO, 
were detected in the planetary atmosphere, 
it would indicate the existence of intrinsic magnetic field,
although super-Earths might be able to sustain the H$_{{\rm X}}$O molecules in the atmosphere by their high gravity 
even without the protection by magnetic fields.

We need to elaborate our thermal evolution model,
by considering details of mantle convection mode that is affected by phase transition between 
$\gamma$-spinel to perovskite \citep{Christensen1985}
at upper/lower mantle.
We also should take into account further mineral transitions suggested by ab initio calculations \citep{Umemoto2006} that may
appear in super-Earths, because they may affect
internal density structure and the mantle convection mode.

The abrupt enhancement in the mantle viscosity due to the pressure
effect relies on the Arrhenius-type formula for the mantle viscosity we adopt here.
The critical mass beyond which the pressure effect dominates
is inversely proportional to activation volume (Eq.~[\ref{eq:M_pc}]).
Thus, detailed rheological properties affect habitability of super-Earths.
The values of the activation volume are not clear at such high pressure
as in deep mantle in super-Earths.
The mechanism to inhibit dynamo activity in super-Earths proposed by 
\citet{Gaidos2010} also depends on high pressure material
properties (melting and adiabatic curves), which also need to be confirmed.
These provide new motivations to high pressure experiments
and first principle simulations.
Super-Earths provide good links between astronomy and 
high-pressure material science.

\section*{Acknowledgment}
The authors thank to useful discussions with 
Diana Valencia, Masahiro Ikoma, and Hidenori Genda. This work is partly supported by Global COE program "From the Earth to Earths".

\section*{Appendix A. Comparison among 
Nu-Ra relationship model, thermal boundary layer model and mixing length theory model.}

In evaluation of thermal transfer of mantle convection,
we compare the modified mixing length theory
\citep{Sasaki1986,Abe1995} with the conventional 
parameterized convection model \citep[PCM; e.g.,][]{Sharpe1979}
and commonly used thermal boundary layer model \citep[BLT; e.g.,][]{Stevenson1983}.

The original mixing length theory \citep[MLT; e.g.,][]{Vitense1953,Spiegel1963} is 
often used in the thermal transfer 
within the stellar interior to simulate the stellar evolution.
\cite{Sasaki1986} modified the mixing length theory 
for very low Reynolds number convection 
in which the vertical flow is characterized by 
the Stokes velocity determined by a
balance between buoyant force and resident force of viscosity 
rather than by free fall velocity.
In the modified version,  
a distance ($D$) from the closest barrier 
such as the CMB or the top of the mantle layer
is adopted for the mixing length $\ell$, while
in the original theory, the density scale height is 
usually adopted for $\ell$.

The PCM uses the empirical $Nu$-$Ra$ relationship,
\begin{equation}
Nu = \zeta \left( \frac{Ra}{Ra_c} \right)^{1/3},
\label{eq:Nu_Ra}
\end{equation}
where Nusselt number represents the ratio between total heat flux and heat flux only due to conduction without convection,
\begin{equation}
Nu = \frac{F_{\rm total}}{F_{\rm cond}},
\end{equation}
\begin{equation}
F_{\rm cond}=k\frac{\Delta T}{d},
\label{eq:F_cond}
\end{equation}
and Rayleigh number is a dimensionless number representing the strength of convection,
\begin{equation}
Ra=\frac{\rho g \alpha \Delta T d^3}{\kappa \eta},
\end{equation}
where $g, \alpha, \Delta T$ and $d$ are gravitational acceleration, thermal expansion, temperature difference between top and bottom and thickness of convective region, respectively,
and $Ra_c$ is critical Rayleigh number ($\sim 650$)
for thermal convection.
Because when $Ra \sim Ra_c$, $Nu$ must be $\sim 1$,
$\zeta$ is $O(1)$.
Sotin et al. (1999) derived $\zeta \sim 1.5$-2.0 through
3D fluid dynamical simulation although the value of $\zeta$ is somewhat lower in high $Ra$ region.
We here adopt $\zeta = 1.7$.

From eqs.~(\ref{eq:Nu_Ra}) to (\ref{eq:F_cond}),
total heat flux through a fluid layer is represented by Rayleigh number as
\begin{equation}
F_{\rm total} = \zeta \left( \frac{Ra}{Ra_c} \right)^{1/3} F_{\rm cond}.
\end{equation}
This model is very simple, but $Ra$ is
``mean'' value of the whole mantle that is 
difficult to evaluate for real mantle in which
viscosity changes by order of magnitude throughout the mantle.
In particular, it may not have enough resolution to evaluate
$F_{\rm CMB}$ that we are concerned with in the present paper.

In the BLT, heat flux is evaluated in the boundary layer.
The thickness of boundary layer is estimated by an
assumption that the layer is marginally stable against
thermal instability.
Then, the local Rayleigh number of the thermal boundary layer 
($Ra_l$) is nearly equal to the critical Rayleigh number
for thermal instability, that is,
\begin{equation}
Ra_l = \left( \frac{\rho g\alpha} {\kappa \eta} \right)_l \delta^3 \Delta T_{\rm TB} \sim Ra_c,
\end{equation}
where $\delta$ is thickness of thermal boundary layer
and $\Delta T_{\rm TB}$ is temperature difference between the bottom
and the top of the boundary layer, 
and subscript ``$l$'' denotes the values in the
thermal boundary layer.
Thus, the heat flux through the layer is calculated as
\begin{equation}
F_{\rm total}=k\frac{\Delta T_{\rm TB}}{\delta},
\end{equation}
where $\delta$ is calculated as
\begin{equation}
\delta = \zeta' \left[ \left(\frac{\kappa \eta}{\rho g \alpha}\right)_l 
       \frac{Ra_{\rm c}}{\Delta T_{\rm TB}}\right]^{1/3},
\label{eq:delta}
\end{equation}
and the factor $\zeta'$ $\sim O(1)$ is determined as follows.
If $\kappa, \eta, \rho, g,$ and $\alpha$ are constant,
$\Delta T_{\rm TB} \sim \Delta T/2$, so that
\begin{equation}
F_{\rm total} = \frac{1}{2^{4/3} \zeta'} 
\left( \frac{Ra_l}{Ra_c} \right)^{1/3} F_{\rm cond}.
\end{equation}
To be consistent with 3D fluid dynamical simulation
by Sotin et al. (1999), we set $2^{4/3} \zeta' = 1/1.7$,
that is, $\zeta' = 0.23$.

Since the heat flux is expressed by quantities 
only in the thermal boundary layer (eqs. [27] and [28]), which is localized
in the mantle, the BLT has better resolution than the PCM,
in particular, for evaluation of $F_{\rm CMB}$.
However, since the values of viscosity change by 
order of magnitude even in the thin thermal boundary layer,
it is not clear which value has to be chosen as
a representative value of the viscosity in Eq.~[\ref{eq:delta}].
For the terrestrial planets in our Solar system, 
observational data can be used to constrain the uncertainty.

Since the modified MLT uses local values of physical quantities
(Eq.~[\ref{TT}]), 
it quantitatively agrees with the BLT for wide range of parameters
as shown below.
There is no uncertainty for choice of
a representative value of viscosity in the MLT, while
choice of the mixing length has uncertainty. 
For calculation of thermal evolution of the Earth, 
it is proposed that a distance ($D$) from the closest barrier 
such as the CMB or the top of the mantle layer
is appropriate for $\ell$
\citep[][and references therein]{Sasaki1986,Abe1995,Senshu2002,Kimura2009}.
Through comparison with the calibrated PCM and BLT,
we adopt $\ell = 0.82D$ as shown below.

To compare these models, we calculate the heat flux
in the case of radially constant $\eta$ with the individual calibrated models.
Internal heat generation due to radioactive elements is neglected.
Figure~10 shows the heat flux 
at the base of the mantle as a function of $Ra$,
obtained by each model.
The values are normalized by $F_{\rm cond}$, that is,
Nusselt number.
Although the MLT does not assume the relation of
$Nu \propto Ra^{1/3}$, it produces the relation.
To match the absolute values, we set $\ell = 0.82D$.
The maximum value of $\ell$ is proportional
to $d$ and heat flux is proportional to $\ell^4$
(Eq.~[\ref{eq:k_v}])
The sensitive dependence on $d$ is canceled out
to result in the rather weak dependence, $Nu \propto Ra^{1/3}
\propto d$, because we found that 
$[(\partial T/\partial r) - (\partial T/\partial r)_s]$
decreases with increase in $\ell$ (Eq.~[\ref{eq:k_v}]).
Analytical argument for it is found in \cite{Abe1995}.

We also examined a case in which the viscosity 
is strongly temperature-dependent,
\begin{equation}
\eta(T)=\eta_0 \exp[\log(\eta_1/\eta_0)(1-T)],
\end{equation}
where $\eta_0$ and $\eta_1$ are viscosity at the top ($T = 0$) and the bottom ($T = 1$) of convective region.
Figure~\ref{fig:nucompvvis} shows Nusselt number obtained by 
PCM, BLT and MLT as a function of $Ra$.
In the PCM, $Ra$ is a mean value for a whole mantle.
The representative viscosity is evaluated using average temperature of mantle, that is, $T=0.5$
if mantle is thermally equilibrated because the PCM assume constant heat flux throughout mantle.
In the BLT and MLT, the heat flux is evaluated by local quantities.
The BLT and MLT produce the same heat flux within 1\%
in all cases, 
while the results by the PCM deviate from those by
the BLT and the MLT for high $Ra$ or high $\eta_0/\eta_1$. 
These results show that MLT is as good as BLT to calculate
thermal evolution of terrestrial planets.
Since MLT is more easily to be applied for
super-Earths that may have barriers for convection in their mantle
(section 2.3), we adopt MLT.

\section*{Appendix B. On the magnetism of planets in Solar system}

In order to confirm the validity of our model,
we show that our model produces thermal evolution 
for individual terrestrial planets 
in the Solar system that is not inconsistent with
their current magnetic activity, with appropriate non-nominal
parameter values, in a similar way to  \citet{Stevenson1983}.
Currently, Earth and Mercury have self-generating magnetic 
fields induced by dynamo action,
while Venus and Mars do not (although some parts of the Martian crust 
have remnant magnetic field in the past \citep{Acuna1999}).

To apply our model to Mercury, Venus and Mars, we
need to use non-nominal parameter values:
\begin{itemize}
\item Mercury: $\zeta_{m/c} = 3:7$ (a significantly 
large metallic core) and $T_{\rm surf} = 440$K.
These are observed values.
We also tested smaller values of $x_{\rm 0S} = 0.01,0.05$ 
according to \citet{Stevenson1983}.
We also tested higher mantle viscosity than Eq.~(15) by multiplying viscosity increase factor
$\Delta \eta = 100$

\item Venus: $T_{\rm surf} = 737$K,
while the nominal values are used for  $\zeta_{m/c}$ and $x_{\rm 0S}$. 
We also tested higher mantle viscosity as well as in the case of Mercury.
Note that two-layer convection model is used for Venus,
because the spinel-perovskite transition also could work
as a barrier for Venusian mantle.
 
\item Mars: $T_{\rm surf} = 210$K.
$\zeta_{m/c}$ is nominal value and $x_{\rm 0S}=0.1, 0.15, 0.2$. 
The standard formula, Eq.~(15), is used for mantle viscosity.
\end{itemize}
 
$\Delta \eta$ is viscosity increase factor due to lack of water in the case of 
Mercury and Venus.
It is suggested by experiment that dry rock has factor of 100 higher viscosity than that of hydrated rocks.
Thereby, we multiply $\Delta \eta=100$ in the case of Venus and Mercury.

The lifetime of magnetic fields calculated by our model
is shown in Fig. \ref{fig:ltmfss}.
In order to be consistent with current Mercury, Venus and Mars,
the lifetime must be longer than 4.5Ga for Mercury 
and shorter than 4.5Ga for Venus and Mars.
Because Martian crust of age $\sim 4$Gyr retains
paleomagnetic field, the lifetime of Martian magnetic field
may be longer than 0.5 Gyr. 

Figures \ref{fig:ltmfss} show that for Mercury, the lifetime is
longer than 4.5Ga for relatively small values of $x_{\rm 0S}$
($\sim 0.01-0.05$)
except for extremely small $\Delta T_{\rm CMB}$
($< 200-300$K).
The relatively long lifetime is resulted by nucleation of inner core
due to lower solidification temperature corresponding to
small values of $x_{\rm 0S}$.
If the nominal value of $x_{\rm 0S}$ is used,
the lifetime is short.
The small value of $x_{\rm 0S}$ for Mercury was discussed
by \citet{Stevenson1983}.

The predicted lifetime of magnetic field for Venus is quite short
for relatively high mantle viscosity ($\Delta \eta > 100$).
Observation suggests that Venus is lack of H$_2$O.
That may be due to runaway greenhouse effect of H$_2$O itself and
consequent dissipation by UV dissociation and heating of the molecules.
Because melting temperature of the mantle viscosity
is lowered by H$_2$O, relatively high mantle viscosity
is more likely, although we do not know exact values of
Venus' mantle viscosity.

The predicted lifetime of magnetic field for 
for Mars is longer than 1 Gyr but shorter than 4.5Gyr, 
if initial $\Delta T_{\rm CMB}$ is $\sim 10-500$K.
If Mars has never undergone giant impacts that cause 
significant heating of metallic core, such low initial
$\Delta T_{\rm CMB}$ is likely.

Thus, with non-nominal parameters that reflect distance 
from the Sun and accretion history of individual planets,
our model can produce the results that are not inconsistent with
the current terrestrial planets in the Solar system. 
However, in order to clarify intrinsic physics in 
generation of magnetic field of extra solar terrestrial planets,
we focus on the results with
the nominal parameters ($T_{\rm surf} = 300$K,
$\zeta_{\rm m/c}=7:3$, and $x_{\rm 0S}=0.1$),
which correspond to the parameters of terrestrial planets
with the same compositions as the Earth in habitable zones.

\section*{Appendix C. Energy in the core} \label{coreenergy}
Figure \ref{core_energy} shows the thermal and gravitational energy,
released latent 
heat, and their sum as a function of $T_{\rm CMB}$ 
for the nominal cases with planetary mass $M=1, 2, 5$ and $10M_{\oplus}$, 
which are calculated
by the procedures in section 2.2.
We set that each value is zero at the temperature
at the initiation of inner core growth. 
As is shown in this figure, the loss of
thermal energy occupies about one third of the total energy loss of the
core for the case of $M=1M_{\oplus}$. Released latent heat corresponds to about one fifth of the total
energy loss, which depends on $T_{\rm CMB}$ because of the
nonlinear density-dependency of the melting temperature of metal. 
 
The gradient of the total energy in Fig.~\ref{core_energy} corresponds to
an effective specific heat of the core. 
The total heat 
capacity is twice larger than the specific heat of thermal
energy solely just after the inner core initiation
($T_{\rm CMB} \simeq 4100$K), while their values converge
as temperature decreases.
This is because impurity concentration increases with
the temperature decrease in the outer core. 
Inner core growth is moderated by the depression of melting
temperature of outer core due to the concentration 
of impurities into outer core. 
Released gravitational energy and latent
heat become smaller than thermal energy 
as the temperature decreases. 
Note that the gravitational energy released
by the thermal contraction of the core also works as resistance to cooling of
the core (see section \ref{sec:thermal evolution
of the core}). 

The ratio of gravitational energy, latent heat and thermal energy is varied with planetary mass.
Thermal energy is more dominant than other energies for more massive planet.
It means that the gravitational energy and latent heat are not main energy source to drive dynamo action within cores of
massive super-Earths.
This is mainly because of the change in slope of adiabatic curve within core.
The higher gravity causes steeper adiabatic thermal structure, and then core posses large amount of thermal energy inside it for the case of massive planets.
This is also the reason why the effective specific heat of core is increased as planetary mass increases.

\bibliography{apjmnemonic,bibliography}

\bibliographystyle{apj}


\clearpage	
\onecolumn
{
\renewcommand{\baselinestretch}{1}
\small\normalsize

\begin{table}
\caption[Physical properties of mantle and core components we adopted]
	{\label{ta:PP} Physical properties of mantle and core components we adopted\citep{Valencia2007radius}}
\begin{tabular}{cccccccc}
\hline
\hline
material & $\rho_0$ & $K_0$ & $K^{\prime}_0$&$\gamma_0$ & $q$ & $\theta_0$ &Refs.\\
&(kgm$^{-3}$)&(GPa)&&&&&\\
\hline
ol&3347&126.8&4.274&0.99&2.1&809&a\\
wd+rw&3644&174.5&4.274&1.20&2.0&908&a\\
pv+fmw&4152&223.6&4.274&1.48&1.4&1070&a\\
ppv+fmw&4270&233.6&4.524&1.68&2.2&1100&b\\
Fe&8300&164.8&5.33&1.36&0.91&998&c,d\\
FeS&5330&126&4.8&1.36&0.91&998&c,d\\
\hline
\multicolumn{8}{l}{}\\
\end{tabular}
\\a \citep{Stixrude2005}, b \citep{Tsuchiya+2004}, c \citep{Williams+1997}, d \citep{Uchida+2001}

\end{table}

\begin{table}
\caption[Parameters of radiogenic elements we adopted]
	{\label{ta:radiogenic heat} Parameters of radiogenic elements we adopted \citep{VanSchmus1995}}
\begin{tabular}{cccc}
\hline
\hline
element & $U$(ppb) & $H$($\mu$Wkg$^{-1}$) & $\lambda$(yr$^{-1}$)\\
\hline
K$^{40}$   & 28.0 & 29.17 & 5.54$\times10^{-10}$\\
Th$^{232}$ & 76.4 & 26.38 & 4.95$\times10^{-11}$\\
U$^{235}$  & 0.14 & 568.7 & 9.85$\times10^{-10}$\\
U$^{238}$  & 20.1 & 94.65 &1.551$\times10^{-10}$\\
\hline
\end{tabular}
\end{table}

\begin{table}
\caption[Parameters of the viscosity in upper and lower mantle
 \citep{Ranalli2001}]
	{\label{ta:viscosity} Parameter of the viscosity in upper mantle and lower mantle \citep{Ranalli2001}}
\begin{tabular}{cccccc}
\hline
\hline
     & $B$(Pa$^{-n}$s$^{-1}$) & n & E$^*$(10$^3$Jmol$^{-1}$) &
 V$^*$(10$^{-6}$m$^2$mol$^{-1}$) & $\dot{\epsilon}$(s$^{-1}$)\\

\hline
upper mantle & 3.5$\times10^{-15}$ & 3.0 & 430 & 10 & 10$^{-15}$\\
lower mantle & 7.4$\times10^{-17}$ & 3.5 & 500 & 10 & 10$^{-15}$\\
\hline
\end{tabular}
\end{table}

\begin{table}
\caption[Physical properties of upper mantle, lower mantle, and core]{\label{ta:physical property} Physical property of upper mantle, lower mantle, and core \citep{Yukutake2000}}
\begin{tabular}{cccc}
\hline
\hline
      & $k_c$(W mK$^{-1}$) & $C_p$(J kg$^{-1}$K$^{-1}$) & $\alpha$(K$^{-1}$) \\

\hline
upper mantle & 5 & 1250 & 3.6$\times 10^{-5}$\\
lower mantle & 10 & 1260 & 2.4$\times 10^{-5}$\\
outer core   & 40 & 840 & 1.4$\times 10^{-5}$\\
\hline
\end{tabular}
\end{table}

}

\clearpage

\begin{figure}[htbp]
\begin{flushleft}
\includegraphics[scale=0.5, angle=0]{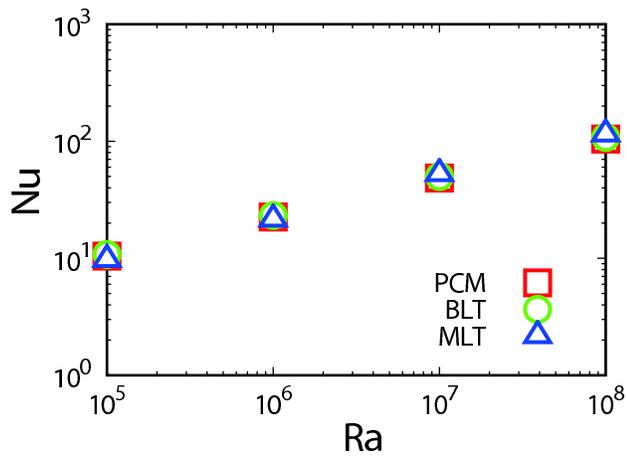} 
\end{flushleft}
\caption{Temporal averaged Nusselt numbers obtained by PCM, BLT and MLT models. Open triangles, circles and squares represent the results of PCM, BLT and MLT models, respectively.}
\label{nucomp}
\end{figure}

\begin{figure}[htbp]
\begin{flushleft}
(a)\hspace{4.5cm}(b)\\
\includegraphics[scale=0.3, angle=0]{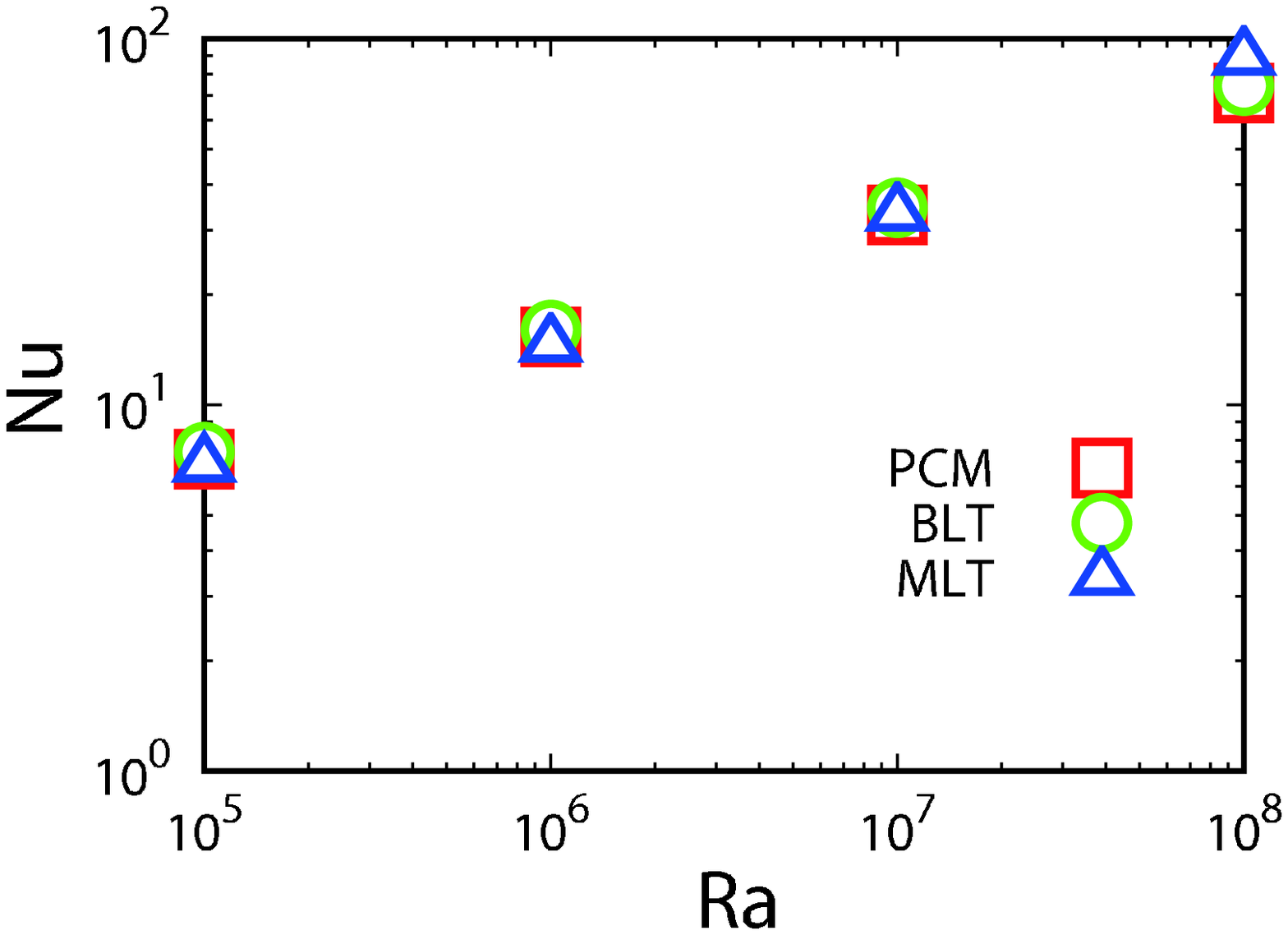} 
\includegraphics[scale=0.3, angle=0]{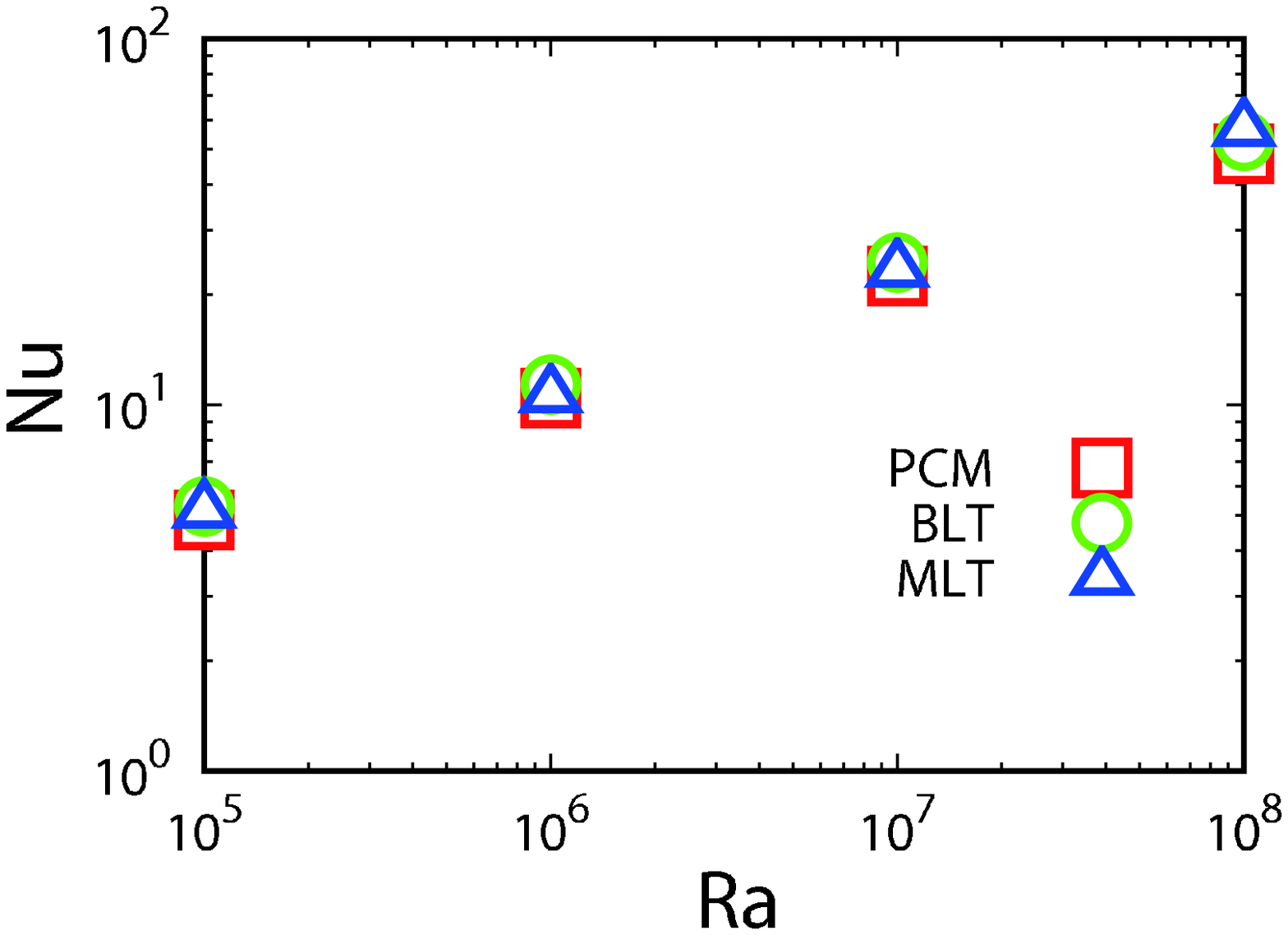} \\
(c)\hspace{4.5cm}(d)\\
\includegraphics[scale=0.3, angle=0]{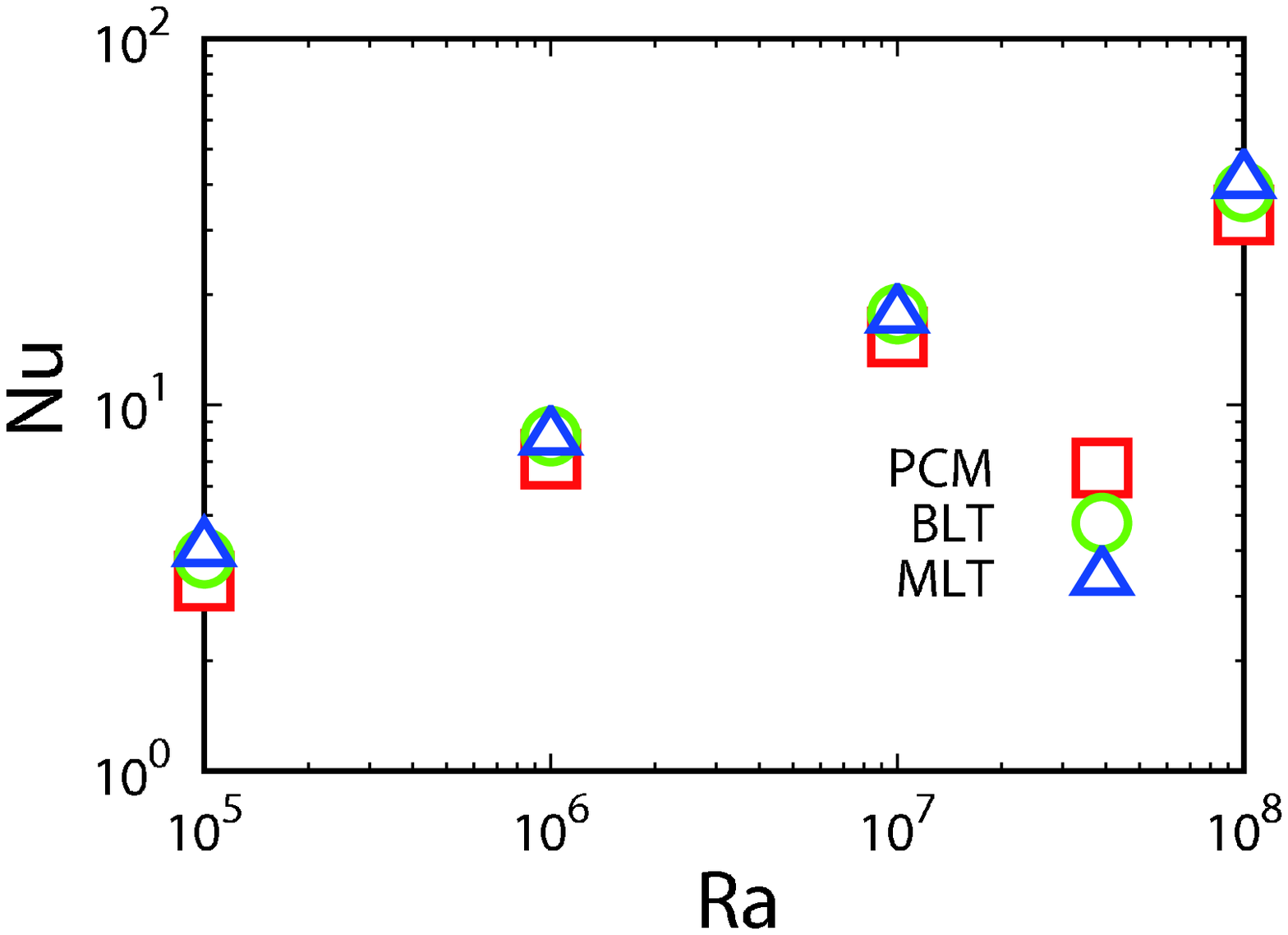} 
\includegraphics[scale=0.3, angle=0]{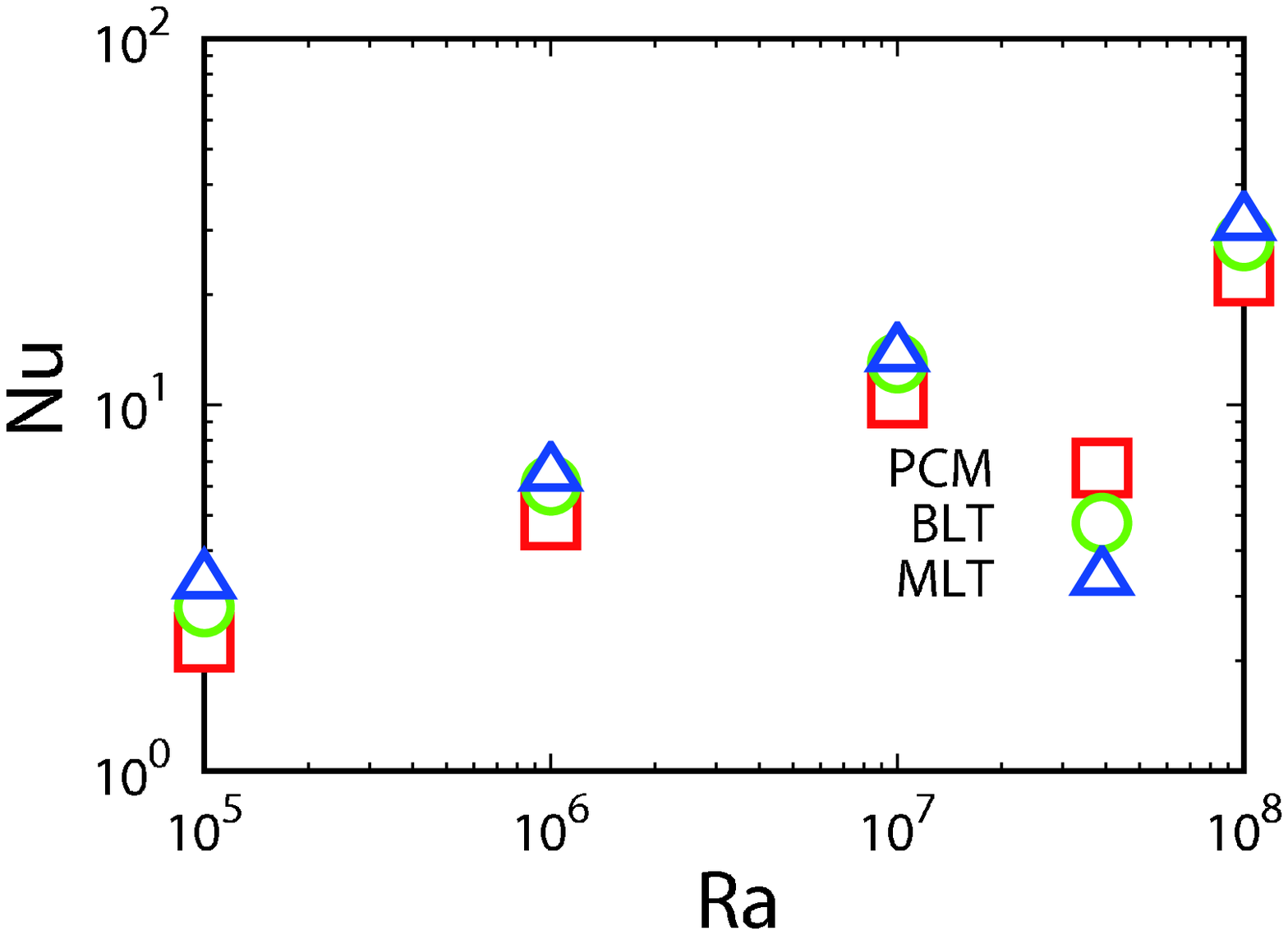} 
\end{flushleft}
\caption{Temporal averaged Nusselt numbers obtained by PCM, BLT and MLT models. The viscosity is changed from the bottom to the top with the ranges of (a) $\Delta \eta$= 10, (b) $10^2$, (c) $10^3$ and (d) $10^4$, respectively. }
\label{fig:nucompvvis}
\end{figure}

\begin{figure}[htbp]
\begin{flushleft}
(a)\\
\includegraphics[scale=0.3, angle=0]{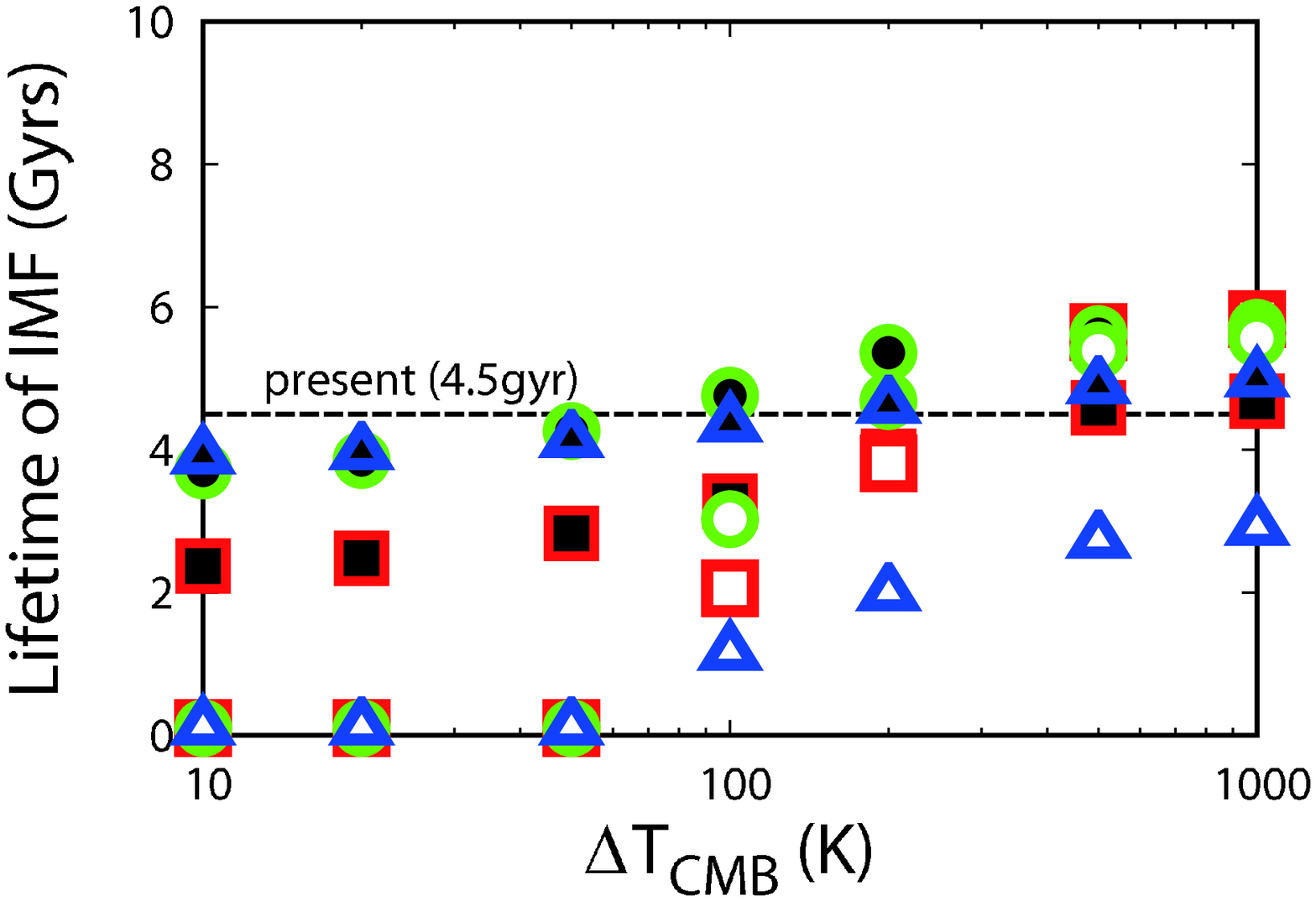}\\
(b)\\
\includegraphics[scale=0.3, angle=0]{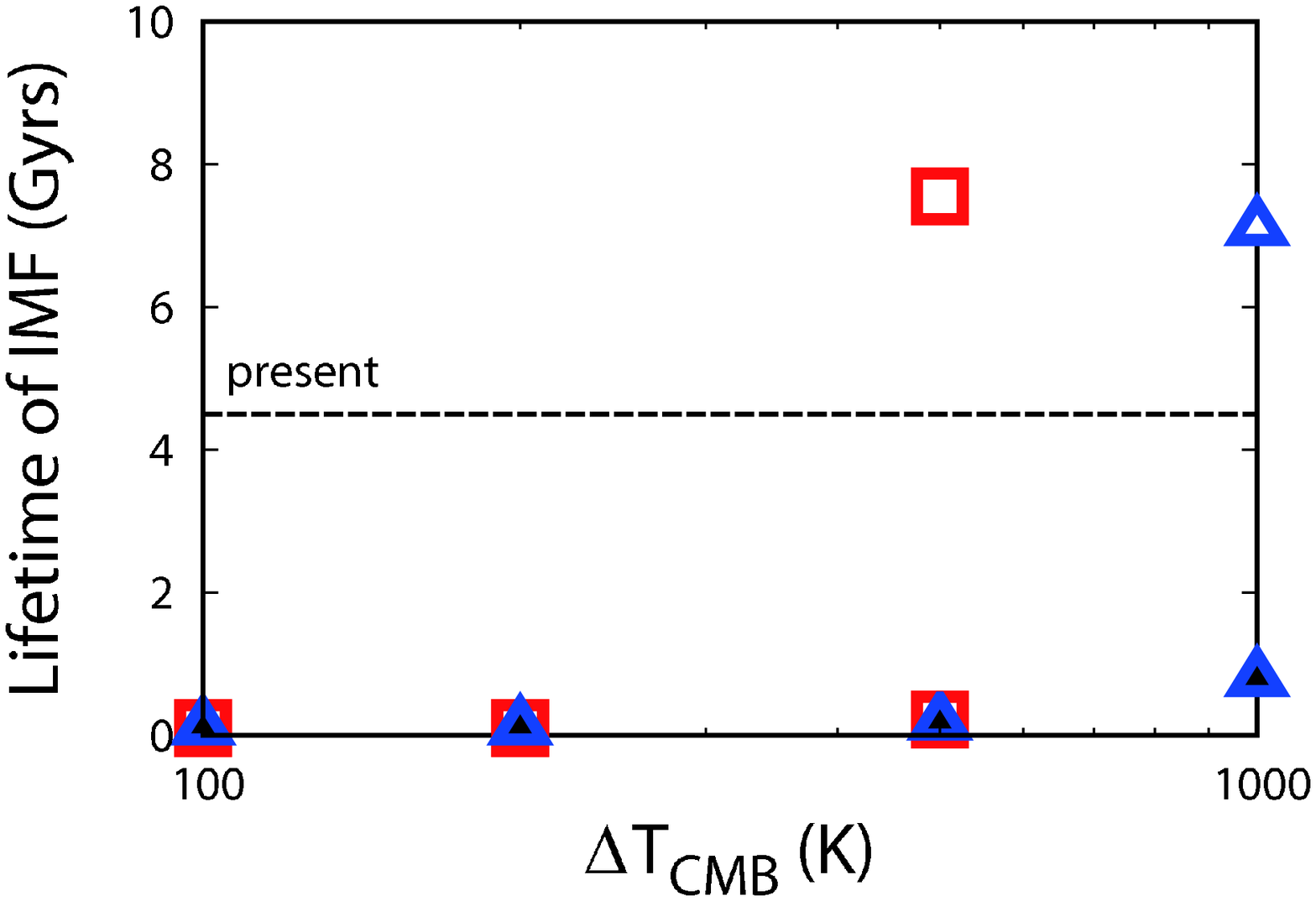}\\
(c)\\
\includegraphics[scale=0.3, angle=0]{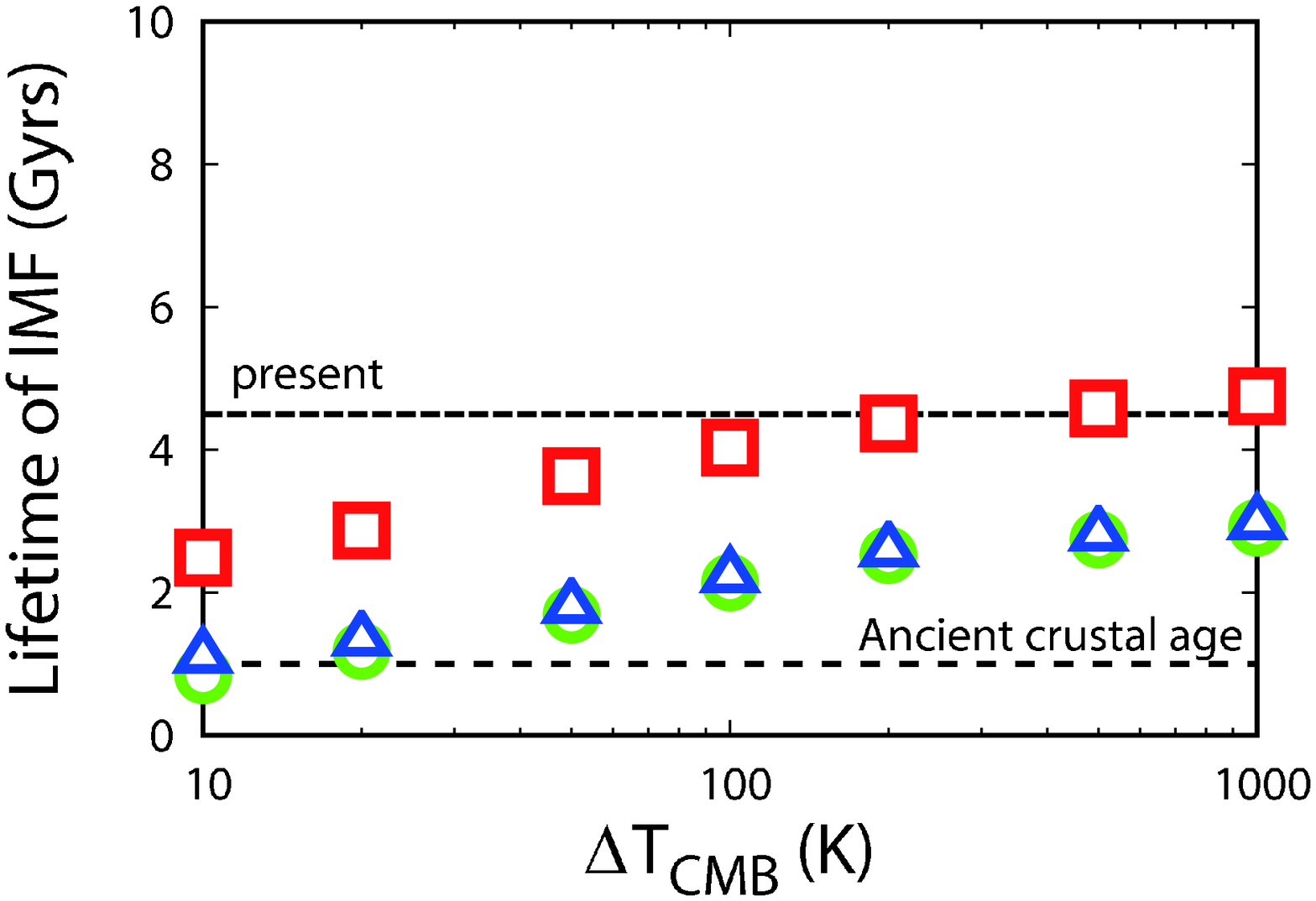}
\end{flushleft}
\caption{Lifetime of magnetic fields of (a) Mercury, (b) Venus and (c) Mars obtained through our simulations.
The free parameters are initial temperature gap at CMB $\Delta T_{\rm CMB}$ and 
initial impurity concentration in core  $x_0$, viscosity increase factor $\Delta \eta$, respectively.
(a)$\zeta_{m/c} = 3:7$, $T_{\rm surf} = 440$K, and squares, circles, and triangles are corresponding to models with $x_{0s} = 0.01, 0.05, $and 0.1, respectively.
The filled and open symbols represent models with the standard viscosity and a higher viscosity multiplied by $\Delta \eta = 100$.
(b)$\zeta_{m/c} = 7:3$, $T_{\rm surf} = 737$K and $x_0 = 0.1$.
The filled and open symbols represent the same meaning as is in the case of Mercury.
The square and triangles are models in which one- and two-layered mantle convection are assumed.
(c)$\zeta_{m/c} = 7:3$, $T_{\rm surf} = 210$K. squares, triangles and circles are corresponding to models with $x_0 = 0.1, 0.15$ and 0.2.
} 
\label{fig:ltmfss}
\end{figure}

\begin{figure}[htbp]
\begin{flushleft}
$1M_{\oplus}$\hspace{5cm} $2M_{\oplus}$\\
\includegraphics[scale=0.3, angle=0]{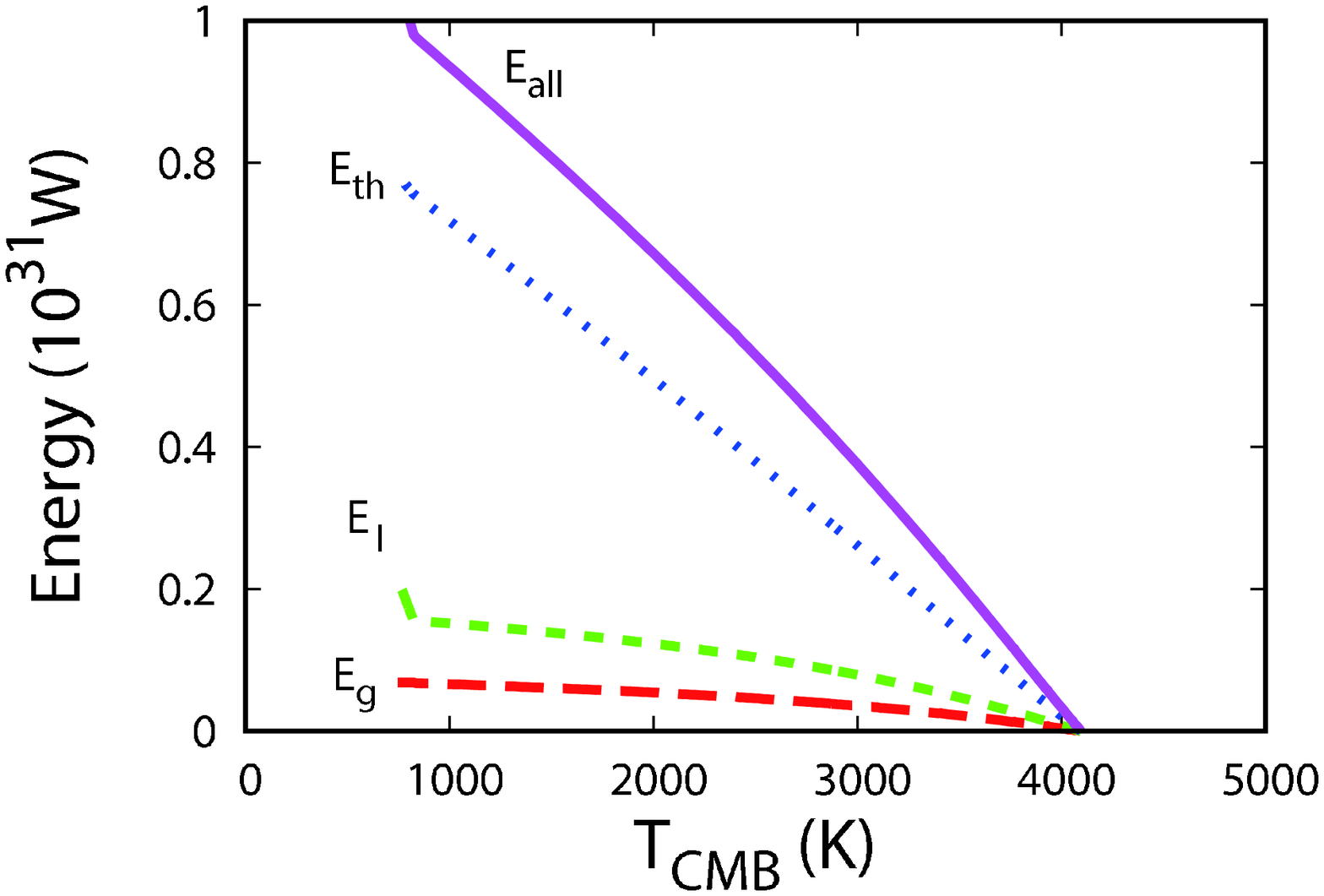}
\includegraphics[scale=0.3, angle=0]{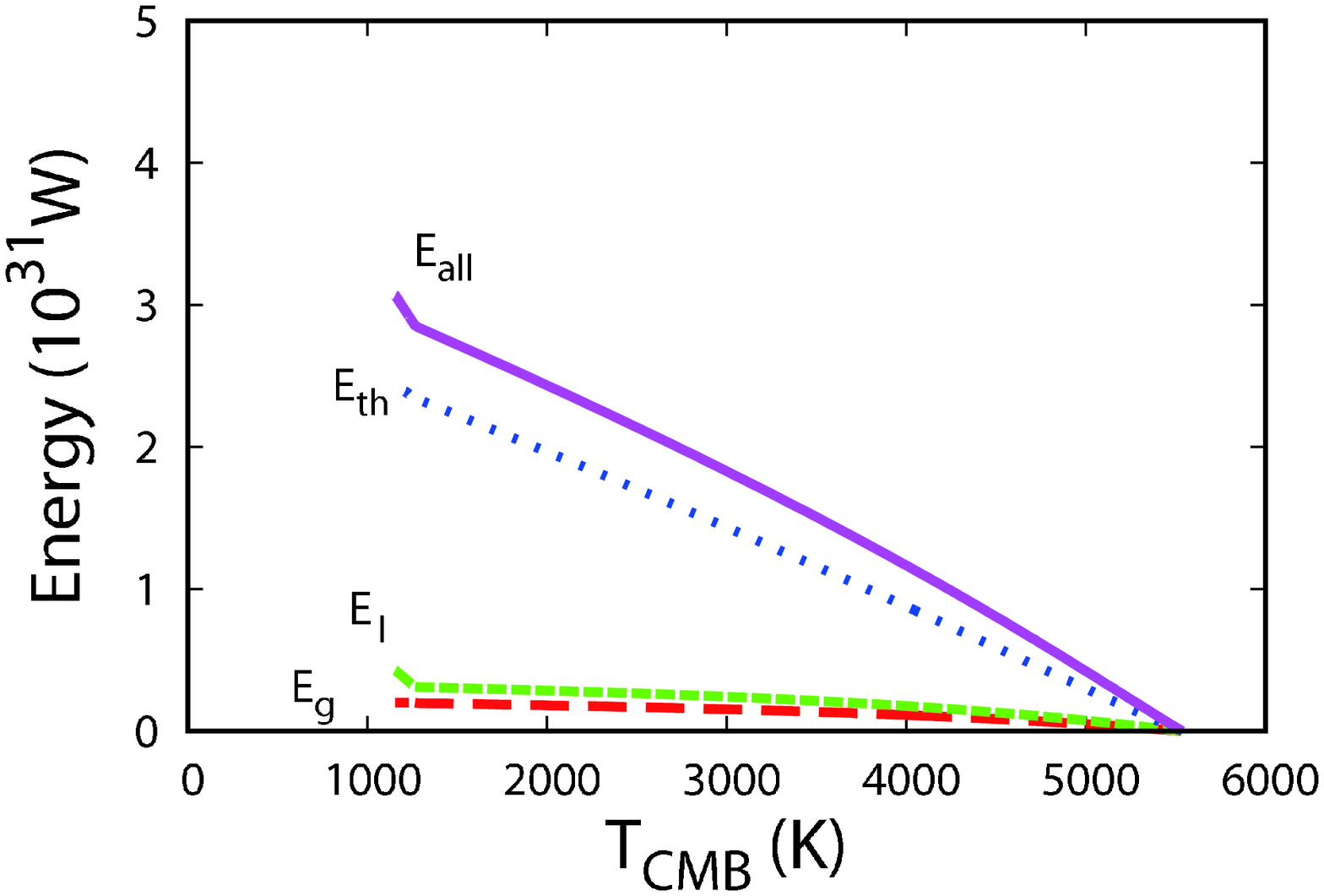}\\
$5M_{\oplus}$\hspace{5cm} $10M_{\oplus}$\\
\includegraphics[scale=0.3, angle=0]{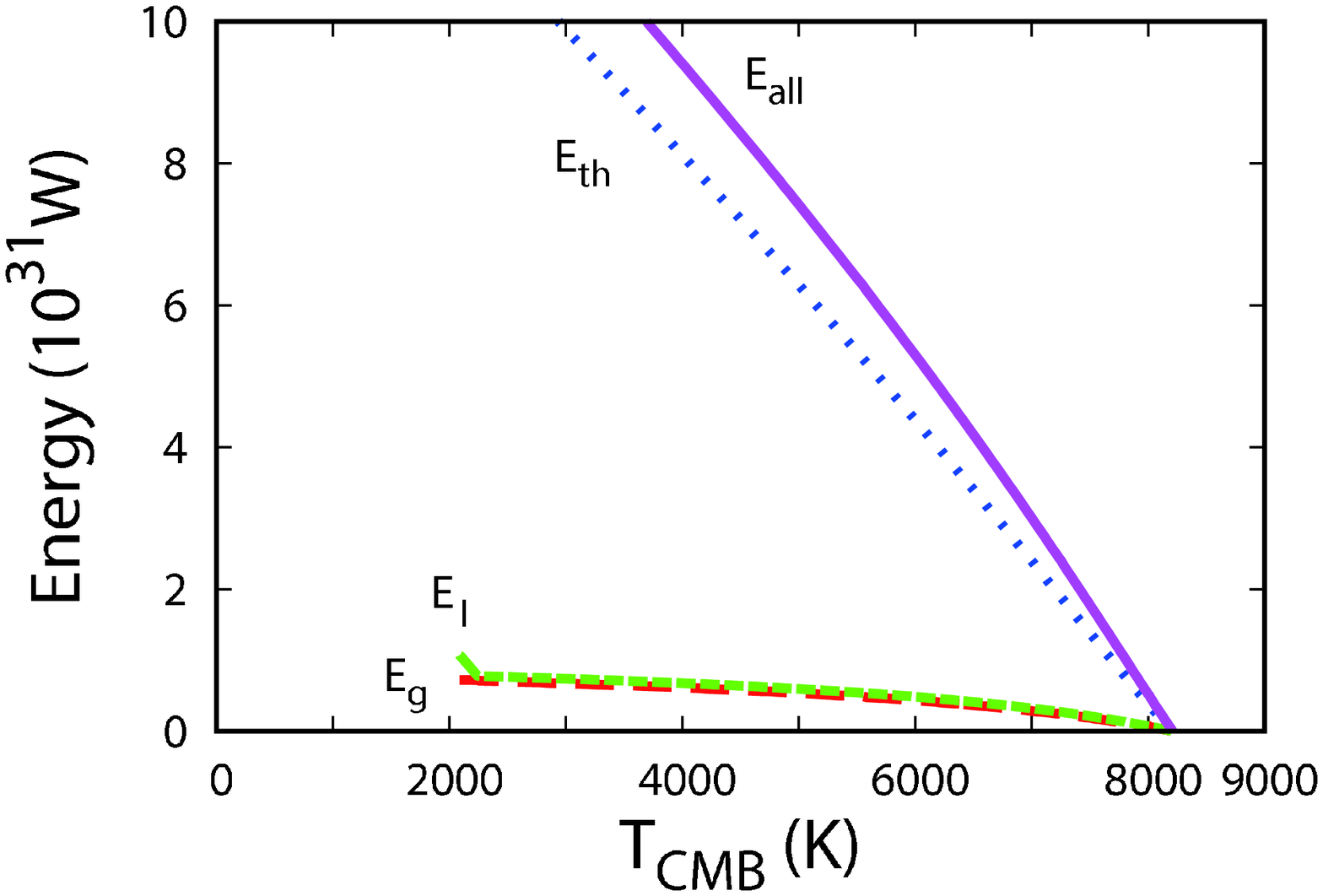}
\includegraphics[scale=0.3, angle=0]{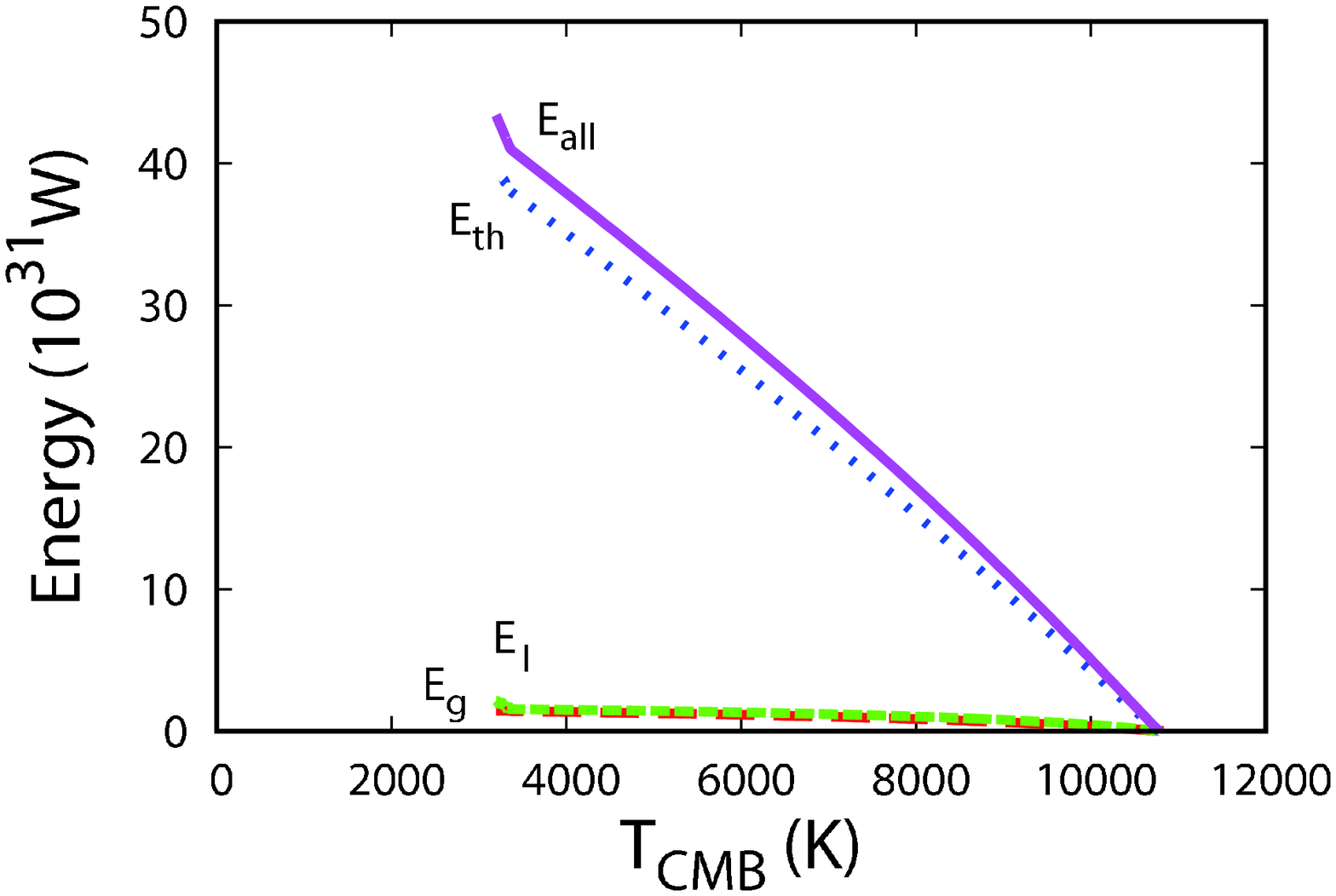}
\end{flushleft}
\caption{Individual core energies as a function of 
$T_{\rm CMB}$ for the nominal case with $M = 1, 2, 5, 10 M_{\oplus}$ and
$x_{\rm 0S} = 0.1$.
Dashed, dotted, and dot-dashed curves represent gravitational energy, 
latent heat, and thermal energy, respectively.
Solid curve shows the total energy. 
The gradient of total energy corresponds to 
the specific heat of the core.}
\label{core_energy}
\end{figure}

\end{document}